\documentclass[12pt,a4paper]{article}
\usepackage{jheppub}
\usepackage[T1]{fontenc}
\usepackage{slashed}
\usepackage{caption}
\usepackage{subcaption} 
\usepackage{float}

\usepackage{empheq}%

\usepackage{comment}
\usepackage{xcolor}
\usepackage{color,soul}

\usepackage{bbold}
\usepackage{graphicx}

\usepackage{hyperref}
\usepackage{mathtools}

\DeclarePairedDelimiter\floor{\lfloor}{\rfloor}

\usepackage{draft}
\newcommand{\ch}{\mathrm{ch}}
\renewcommand{\k}{k}
\newcommand{\DT}{\mathrm{DT}}

\newcommand{\MSW}{\mathrm{MSW}}
\renewcommand{\S}{\mathrm{S}}
\newcommand{\U}{\mathrm{U}}
\renewcommand{\L}{\mathrm{L}}
\newcommand{\R}{\mathrm{R}}
\newcommand{\SU}{\mathrm{SU}}

\usepackage{feynmp}

\everymath{\displaystyle}

\def\[{\left[}
\def\]{\right]}
\def\({\left(}
\def\){\right)}

\def\Tr{\mathrm{Tr}}

\usepackage{breqn}

\preprint{}

\title{Black hole/black ring transition}

\author{Indranil Halder and Ying-Hsuan Lin}

\affiliation{Jefferson Physical Laboratory, Harvard University, Cambridge, MA 02138, USA}

\abstract{We consider BPS states in M theory compactified on a Calabi-Yau threefold with one K\"ahler parameter, and investigate their relation to black holes and black rings. 
On the microscopic side, a characterization of the BPS degeneracy can be obtained from the non-perturbative part of the topological string free energy according to the Gopakumar-Vafa conjecture.
On the macroscopic side, the Wald entropy of relevant black objects in the four-dimensional IIA description is computed from the perturbative part of the same topological string free energy following the work of Cardoso-de Wit-Mohaupt and then lifted to five-dimensional M theory through the Gaiotto-Strominger-Yin correspondence.
For a given value of the M2 brane charge, we find that for small angular momenta, the microscopic BPS degeneracy matches the entropy of a rotating black hole, whereas an apparent transition occurs at a critical value of the angular momentum. Beyond this value, we find encouraging evidence that the microscopic counting is well approximated by the entropy of a black ring. We conclude by formulating a new puzzle involving the corrections due to degenerate worldsheet instantons.}

\begin{document}

\maketitle

\section{Introduction}

A microscopic explanation of the Bekenstein-Hawking entropy \cite{PhysRevD.7.2333, Hawking:1975vcx} of black holes is an important and fascinating question that any consistent quantum theory of gravity must address. A landmark achievement was made by Strominger and Vafa \cite{Strominger:1996sh} using the degrees of freedom on the constituent D-branes.\footnote{A separate development for understanding the entropy of a `small' black hole made up of fundamental strings originated in the work of Sen \cite{Sen:1995in}. For further developments on the subject and the current status, see \cite{Dabholkar:2004yr, Dabholkar:2004dq, Chen:2021dsw, Halder:2023nlp}.
}
The precise version of the argument relies highly on supersymmetry\footnote{This formalism is developed by a great amount over several years, see for instance \cite{Kinney:2005ej, Bhattacharya:2008zy, Benini:2015eyy, Cabo-Bizet:2018ehj, Choi:2018hmj} (also \cite{Chen:2023mbc, Chen:2023lzq}) and references therein. For a recent discussion on the microscopic origin of thermal entropy see \cite{Halder:2022ykw, Halder:2023adw}.}
, and at the time the only known supersymmetric black solutions to supergravity were single-centered black holes. The Cardy growth of the supersymmetric index of the theory on the brane reproduced to leading order (in the large charge limit) the Bekenstein-Hawking entropy of a five-dimensional single-centered black hole. Soon after, supersymmetric multi-centered 
black hole solutions were discovered \cite{Denef:2000nb, Bates:2003vx}.  A supergravity analysis shows these multi-centered black holes can sometimes entropically dominate over the single-centered ones with the same conserved charges \cite{Denef:2007vg}. It has been an open question since then if there is a precise microscopic counting in which we get the dominant contribution to entropy from both single- and multi-centered black holes with macroscopic horizons in different ranges of charges.

Consider M theory compactified on a Calabi-Yau threefold (CY$_3$). The microscopic entropy of the BPS solutions in 5d is captured by the non-perturbative part of the free energy of topological strings on the Calabi-Yau, following the Gopakumar-Vafa conjecture \cite{Gopakumar:1998ii, Gopakumar:1998jq}.\footnote{For further details of the conjecture see \cite{Katz:1999xq, Dedushenko:2014nya}.} We will investigate the existence of a transition between black holes and black rings in this context. 
In fact, under the Gaiotto-Strominger-Yin (GSY) lift \cite{Gaiotto:2005gf, Gaiotto:2005xt}, this transition is closely related to the aforementioned transition between 4d single- and multi-centered macroscopic black holes in IIA superstrings with $\mathcal{N}=2$ supersymmetry (in compactifications with higher supersymmetry, say compactification on $\mathrm{K3} \times \mathrm{T}^2$ \cite{Dijkgraaf:1996it, Shih:2005uc}, it is well known that the only black hole with macroscopic horizon that contributes to the suitable supersymmetric index is a single-centered one \cite{Dabholkar:2009dq, Chowdhury:2019mnb}). However, to compare the microscopic and macroscopic formulae, we need to be careful about the quantum corrections. The advantage of the GSY lift is the expectation that the BPS entropies of black solutions in 4d and 5d are the same, including perturbative quantum corrections. This allows us to systematically take into account corrections to the Bekenstein-Hawking entropy of black solutions in 5d in terms of the well-developed 4d formalism that we discuss below.  

Corrections to the Bekenstein-Hawking entropy of a black hole in four dimensions consist of various contributions. The first set of corrections come from the Wald entropy \cite{Wald:1993nt} of higher derivative terms in the Wilsonian effective action of massless fields---present due to integrating out other massive degrees of freedom. In the context of 4d $\mathcal{N}=2$ compactifications,
such contributions were systematically analyzed by Cardoso, de Wit, and Mohaupt in \cite{LopesCardoso:1998tkj, LopesCardoso:1999cv, LopesCardoso:1999xn}. Among these higher derivative terms, there is the $R\wedge R$ coupling that comes from the classical dimensional reduction of the $R^4$ term in the Wilsonian effective action of M theory \cite{Green:1997di, Kiritsis:1997em, Berkovits:1998ex}. All other higher derivative terms come from a degenerate limit of the contribution of  Euclidean M2 brane instantons wrapping a two-cycle in the Calabi-Yau and the M theory circle.\footnote{There are also `special' M2 brane instantons that contribute to the correction of the hypermultiplet metric \cite{Alexandrov:2013yva}. } In the language of topological string theory, 
these contributions are captured by the perturbative part of the free energy.  According to the Ooguri-Strominger-Vafa conjecture \cite{Ooguri:2004zv},\footnote{The conjecture has been sharpened over the years; for a more precise formulation see \cite{Denef:2007vg}.} apart from these higher derivative terms there also exist the contributions of non-degenerate M2 brane instantons and the one-loop fluctuations of massless fields in 4d that can be calculated systematically in Sen's formalism \cite{Sen:2008vm, Sen:2012kpz, Sen:2012cj}.\footnote{An alternative approach to calculating the entropy of an extremal (single-centered) black hole solution would be to perform the gravitational path integration in Sen's formalism beyond one-loop. Such a procedure has been carried out successfully for higher supersymmetric situations in \cite{Dabholkar:2010uh, Dabholkar:2011ec, Dabholkar:2014ema}.} 
In this work, by carefully analyzing the attractor equations \cite{Ferrara:1995ih, Strominger:1996kf}, we will show that for the D6-D2-D0 black hole with one unit of D6 charge, the volume of the Calabi-Yau at the horizon in string units is fixed to a large value (scales with the D2 charge) when the D0 charge is small.  The radius of the horizon is the same as the M theory circle, and the black hole is prone to all-order corrections from the degenerate worldsheet instantons. As a result of this phenomenon, the microscopic origin of this black hole has qualitatively new features compared to its higher-supersymmetric analogs (for $\mathcal{N}=4$ black holes these contributions vanish \cite{Harvey:1996ir, Dabholkar:2005dt}). Since the volume of the Calabi-Yau is large we can ignore the contribution of nontrivial M2 brane instantons. As a result, the entropy from the 4d point of view is captured by the perturbative part of the free energy of the topological strings, in sharp contrast to the microscopic calculation in 5d that gets contributions from the non-perturbative part only. This gives a nontrivial consistency check for the topological strings itself. 

As a concrete example, we consider the compactification on the quintic threefold. The calculation of Gopakumar-Vafa (GV) invariants\footnote{They determine the microscopic counting. When the Calabi-Yau is elliptic, they are related to F theoretic string \cite{Vafa:1997gr, Haghighat:2015ega, Haghighat:2013gba, Haghighat:2014vxa}.} on a compact manifold such as the quintic is a notoriously difficult task (more so because the quintic does not have any elliptic or K3 fibration---at least not before making a nontrivial geometric transition). A systematic procedure for the calculation of the GV invariants up to all genera is not known currently. The difficulty is related to the fact that the holomorphic anomaly equation \cite{Bershadsky:1993ta, Bershadsky:1993cx} satisfied by the free energy of topological strings requires suitable boundary conditions to give a unique solution. A procedure for calculating GV invariants up to genus $51$ for quintic was formulated in \cite{Huang:2006hq}, taking into account a set of physical boundary conditions at the orbifold point, the conifold point, and the large volume limit. However, the explicit information available in \cite{Huang:2006hq} for the GV invariants was up to much lower genera.
Therefore, we initially carried out the explicit calculation of GV invariants up to genus $49$ (available at \cite{WebUS}) to investigate the central questions of this work and make the main observations.\footnote{Section~\ref{Sec:Micro} was significantly inspired by \cite{Alexandrov:2023zjb}.} Later, after the appearance of \cite{Alexandrov:2023zjb, Web} which contains the state-of-the-art progress on the computation of GV invariants, we extended our analysis to make full use of the available data, and the results corroborated our earlier conclusions.

The numerical GV data allowed us to plot in figure~\ref{fgv} the BPS entropy in 5d at a fixed M2 brane charge $d$ as a function of left-moving angular momentum $j_{\L}$ (roughly speaking, the uplift from 4d to 5d maps the D4, D2, D0 charges to the M5, M2 charges, and the left-moving angular momentum).\footnote{The precise mapping also involves a shift in the M2 brane charge compared to the D2 brane charge that depends on the D4 brane charge.}
From figure~\ref{fgv}, it is clear that there is a sharp transition at $j_{\L} = j_{\L,c}(d)$. For small angular momentum $j_{\L}<j_{\L,c}$, we perform numerical extrapolation on the microscopic data to $d\to \infty$ in figure~\ref{BHexpol}, and notice that the results are well-approximated by the leading order entropy of the D6-D2-D0 black hole uplifted to 5d \cite{Breckenridge:1996is}. As part of figure~\ref{fig5}, we further justify this claim by showing that the microscopic curve obtained from the GV invariants (without any extrapolation) is close to the macroscopic entropy calculated in 4d, taking the contribution of the $R\wedge R$ term into account. This suggests that the re-summed contribution of the degenerate M2 brane instantons is small. At this stage, we do not have any deeper understanding of this fact. 

The understanding of the curve in figure~\ref{fgv} for larger values of the angular momentum $j_{\L}>j_{\L,c}$ is much more involved. 
The black ring under consideration is a multi-centered black hole in 4d made of two centers carrying one unit of D6 and D4-D2-D0 charge.  
In this work, we determine the range of charges for the existence of the bound state using the analysis of \cite{Denef:2007vg}. Within this domain, the black ring entropy has a local maximum as a function of the M5 brane parameter. In discussing the transition point above, we used the black ring whose M5 brane parameter $p$ was extremized to the local maximum at a given value of $d,j_{\L}$.
For the comparison to the microscopic curve, we focus on the black ring that maximizes the leading order Bekenstein-Hawking entropy at the critical angular momentum $j_{\L,c}(d) \sim d^{3/2}$, which determines the M5 brane parameter $p_c(d)\sim d^{1/2}$. In a similar way as the black hole, we take into account the corrections to the entropy due to the $R\wedge R$ term at fixed $p=p_c(d)$ as a function of $j_{\L},d$. 
The resulting black ring entropy obtained this way approximates the microscopic curve beyond $j_{\L}>j_{\L,c}$ reasonably well for a few Calabi-Yau compactifications.  
At this point, we note that the extremized value of M5 brane parameter $p_c(d)$ for the range of charges $d$ accessible to us (due to computational limitations) is small (smaller than one).\footnote{The extremized M5 brane parameter at larger angular momentum is even smaller. } This suggests that the supergravity plot we discussed here is again an extrapolation at best. Beyond the supergravity approximation, in the fully quantum string theory, $p$ is supposed to be an integer.

To take into account the quantization of the M5 brane charge $p$, we turn to the microscopic description \cite{Cyrier:2004hj} of black rings given by the Maldacena-Strominger-Witten (MSW) CFT \cite{Maldacena:1997de}.
The comparison of the entropy coming from GV invariants with that coming from the MSW index with $p=1$ shows a spectacular agreement.
This gives unequivocal evidence that black ring microstates indeed contribute to the 5d BPS index captured by the GV invariants, and that $p=1$ is a dominant contribution up to the M2 brane charge $d$ for which the GV data is available.  
It is unclear whether $p=1$ should remain a dominant contribution in the large $d$ limit, or whether $p>1$ black rings even contribute.
If all values of $p$ contribute, then for very large $d$, a better characterization of the black ring entropy would be an interpolation of the semi-classical entropy optimized over continuous (positive) values of $p$ for a range of $j_{\L}$, and the $p=1$ MSW entropy for $j_{\L}$ beyond the point where the optimal $p$ becomes $\cO(1)$.

The remainder of this paper is organized as follows.  In section~\ref{Sec:SUGRA}, we study the black hole/black ring transition within supergravity, using their Bekenstein-Hawking entropies.  In section~\ref{Sec:TOP}, we introduce the Gopakumar-Vafa invariants counting 5d BPS states, review how they are computed, and compare the corresponding microscopic entropy with supergravity.  In section~\ref{Sec:Wald}, we systematically study the Wald entropy as well as further corrections, first in the 4d language by way of the Ooguri-Strominger-Vafa conjecture, and then uplifted to 5d through the Gaiotto-Strominger-Yin correspondence.  In section~\ref{Sec:Micro}, we perform a microscopic/microscopic comparison between the GV invariants and the microstate counting of black rings with unit M5 brane charge.  Finally, section~\ref{Sec:Summary} ends with a summary and discussion of the open problem concerning degenerate instantons.

\paragraph{Note added:} After the completion of this work, we became aware that our findings in section~\ref{Sec:BMT} had been announced earlier \cite{Talk} and recently appeared in \cite{Alexandrov:2023ltz}.

\section{Black hole/black ring transition in supergravity}
\label{Sec:SUGRA}

In this section, we investigate the phase diagram of 5d minimal supergravity coming from M theory compactified on a Calabi-Yau (CY$_3$).
In the regime of large angular momentum, we will show that there is a transition (in the sense of entropic dominance) from a single-centered black hole with a spherical horizon (i.e.\ $\S^3$) to a black ring with a different horizon topology (more precisely $\S^2\times \S^1$).\footnote{See \cite{Horowitz:2017fyg, Kunduri:2014iga, Kunduri:2014kja} for similar discussions involving other horizon topologies. We thank Jorge E. Santos for pointing this out to us.
}
While this is certainly expected when the angular momentum is above the black hole extremality bound, the transition actually occurs slightly below it.\footnote{It is not completely unreasonable to wonder if there is a manifestation of this phenomena at low temperatures as a dynamical instability of the black hole. For a recent account of instability in the background of fast-rotating black holes (at high enough temperature) in AdS see \cite{Kim:2023sig} and references therein.  }

\subsection{Field content of 5d supergravity}

We work in units where the tension of an M2 brane is unity unless otherwise mentioned. There are eight real supercharges unbroken in this situation. We will describe the bosonic field content in the language of $d=5$ $\mathcal{N}=1$ supergravity at a generic point on the moduli space (at special points one might need to add additional matter fields that we will not discuss here). In eleven dimensions the only degrees of freedom are the metric $g_{\hat{\mu}\hat{\nu}}$ and the three form gauge field $A_{\hat{\mu}\hat{\nu} \hat{\rho}}$ ($\hat{\mu}=0,1,\dots, 10$). We decompose the coordinates into the 5d part $\mu=0,1,\dots, 4$ and the internal CY$_3$ part $i, \bar{i}=1,2,3$. Under the reduction to 5d, we obtain the 5d metric $g_{\mu \nu}$,  $h^{2,1}$ complex scalars $g_{i j}$, $h^{1,1}$ real scalars $g_{i \bar{j}}$, a real scalar $\Phi$ related by Hodge star to  $A_{\mu \nu \rho}$, one complex scalar $A_{i j k}=\epsilon_{ijk} c $, $h^{2,1}$ complex scalars $A_{i j \bar{k}}$, and $h^{1,1}$ real vectors $A_{\mu j \bar{k}}$. 

Before discussing the multiplet structure of these fields it is convenient to introduce an integer basis  $\alpha_A, A=1,2,.., h^{1,1}$ for H$^{1,1}$(CY$_3,\mathbb{Z})$. The K\"ahler class is harmonic and can be expanded as
\begin{equation}
J+iB=v^A \alpha_A,
\end{equation}
where $v^A$ are real scalar fields in 5d. 
The scalar field associated with the volume of the CY$_3$ is
  \begin{equation}
\mathcal{V}=\frac{C_{ABC}}{3!} v^A v^B v^C,
\end{equation}
where the intersection numbers are defined by
\begin{equation}
	C_{ABC}=\int_{CY_3}\alpha_A \wedge \alpha_B \wedge \alpha_C.
\end{equation}
In total we get $n_H=h^{2,1}+1$ hypermultiplets whose scalars are related to  $g_{i j}$, $A_{i j \bar{k}}$, $c$, $\Phi$,  $\mathcal{V}$.

The remaining $n_V=h^{1,1}-1$ scalars $\varphi^a$ related to the $g_{i \bar{j}}$ above are in the vector multiplet. Those can be parameterized as 
 \begin{equation}
 h^A(\varphi)=\frac{v^A}{v},~~~ \frac{C_{ABC}}{3!}h^A(\varphi) h^B(\varphi) h^C(\varphi)=1, ~~ v=(\mathcal{V})^{1/3},
\end{equation}
The abelian real vector fields in 5d coming from $A_{\mu j \bar{k}}$ can be also discussed by expanding the three form field strength  as
\begin{equation}
dA=dV^A \wedge \alpha_A.
\end{equation}
One of the field strengths is in the graviton multiplet along with $g_{\mu \nu}$ (enters in the SUSY variation of the gravitino) and given by
\begin{equation}
T=h_A dV^A, ~~~h_A=\frac{C_{ABC}}{3!}h^B h^C,
\end{equation}
and is called the graviphoton field strength (in general it is not gauge invariant),
and the remaining $n_V=h^{1,1}-1$ of them are in the vector multiplet mentioned above along with the $\varphi^a$s.

All the fields in the vector and hypermultiplets are uncharged under the $h^{1,1}$ vector fields mentioned here, and as a result, we get ungauged $d=5$ $\mathcal{N}=1$ supergravity.

\subsection{Entropy of the half-BPS black hole}\label{5dBMPV}

The Breckenridge-Myers-Peet-Vafa (BMPV) black hole \cite{Breckenridge:1996is} is a solution to 5d $\cN=1$ supergravity that preserves half the supersymmetries.
In this solution, only the fields in the graviton and vector multiplets are turned on  \cite{Kallosh:1996vy}.  The entropy of the black hole is given by 
\begin{equation}\label{BHsugra}
\begin{aligned}
	& S^{\mathrm{BH}}_0=2\pi \sqrt{Q^3-j_{\L}^2},\\
\end{aligned}
\end{equation}

where
\begin{equation}
	\begin{aligned}
	Q^{3/2}=\frac{C_{ABC}}{3!}y^A y^B y^C, ~~ \frac{C_{ABC}}{2!}y^Ay^B=d_C.
	\end{aligned}
\end{equation}
Here $y^A$ are related to the horizon values of the scalars in the vector multiplet $\varphi^a$, and fixed by the attractor mechanism  \cite{Ferrara:1995ih, Strominger:1996kf}. The black hole solution preserves half of the supersymmetries. Here $d_A$ is the charge with respect to $V^A$, normalized such that at the quantum level, it is integer-valued.
The flat 5d has a spatial rotation isometry of $\SU(2)_{\L} \times \SU(2)_{\R}$ that is broken down to $\U(1)_{\L}\times \SU(2)_{\R}$ by the black hole solution. And $j_{\L}$ is the half-integer quantized (in the full quantum theory) angular momentum for $\U(1)_{\L}$.

\subsection{Entropy of the half-BPS black ring}

Another half-BPS solution to 5d $\cN = 1$ supergravity is the Elvang-Emparan-Mateos-Reall/Bena-Warner/Gauntlett-Gutowski (EEMR/BW/GG) black ring \cite{Elvang:2004rt, Elvang:2004ds, Bena:2004de, Gauntlett:2004qy}, which has Bekenstein-Hawking entropy given by \cite{Cyrier:2004hj}
\begin{equation}\label{BRsugra}
	S_{0}^{\mathrm{BR}}=2\pi \sqrt{\frac{c_{\L}}{6} \hat{q}_0},
\end{equation}
where
\begin{equation}
	\begin{aligned}
	& c_{\L}= C_{ABC} p^A p^B p^C,
	\\
	& \hat{q}_0=-q_0+\frac{1}{2!}C^{AB}q_A q_B+\frac{c_{\L}}{24}, \quad C_{AB}=C_{ABC}p^C, \quad C^{AB}C_{BC}=\delta^A_C.\\
	\end{aligned}
\end{equation}
Here $p^A$ are the charges associated with M5 brane, and $q_0, q_A$ are related to charges mentioned above as follows
\begin{equation}
	q_0=2j_{\L}, ~~ d_A=q_A+ \frac{C_{ABC}}{2} p^B p^C.
\end{equation}

\subsection{Black hole/black ring transition on CY$_3$ with $h^{1,1}=1$}
\label{Sec:SUGRAComparison}

\begin{table} [t]
\centering
$$
\begin{array}{|c||c|c|c||c|c|}
\hline \mathrm{CY}_3 & ~~\chi~~ & ~~k~~ & ~~c_2~~ & g_{\mathrm{avail}} & d_{\mathrm{avail}} \\
\hline\hline
X_5(1^5) & -200 & 5 & 50 & 64 & 22 \\
X_6(1^4,2) & -204 & 3 & 42 & 49 & 15 \\
X_8(1^4,4) & -296 & 2 & 44 & 60 & 14 \\
X_{10}(1^3,2,5) & -288 & 1 & 34 & 56 & 10 \\
X_{4,3}(1^5,2) & -156 & 6 &  48 & 26 & 14 \\
X_{6,4}(1^3,2^2,3) & -156 & 2 & 32 & 17 & 7 \\
X_{3,3}(1^6) & -144 & 9 &  54 & 34 & 20 \\
X_{4,4}(1^4,2^2) & -144 & 4 & 40 & 34 & 14 \\
X_{6,6}(1^2,2^2,3^2) & -120 & 1 & 22 & 21 & 5 \\
X_{6,2}(1^5,3) & -256 & 4 & 52 & 49 & 17 \\
X_{4,2}(1^6) & -176 & 8 & 56 & 50 & 24 \\
X_{3,2,2}(1^7) & -144 & 12 & 60 & 14 & 13 \\
X_{2,2,2,2}(1^8) & -128 & 16 & 64 & 32 & 24 \\
\hline
\end{array}
$$
\caption{One-parameter Calabi-Yau threefolds studied in \cite{Huang:2007sb}.  The first column gives their definitions as complete intersections in weighted projective space; we refer to them by $X_\bullet$ without the weights.  The next three columns provide their basic topological data, including the Euler characteristic $\chi$, the self-intersection number $k$, and the second Chern class $c_2$.  The last two columns specify the maximal genus $g_{\mathrm{avail}}$ at which the topological string data is currently available from \cite{Alexandrov:2023zjb,Web}, and the corresponding maximal M2 brane charge $d_{\mathrm{avail}}$.
}
\label{Tab:CY}
\end{table}

Consider M theory compactified on a one-parameter CY$_3$ with $h^{1,1} = 1$.  Since the $A$ index only takes value 1, we define the shorthand
\ie
	p := p^1, \quad d:=d_1, \quad k := C_{111}.
\fe
Although not needed in the supergravity analysis of this section, we list in table~\ref{Tab:CY} the 13 one-parameter CY$_3$ studied in e.g.\ \cite{Huang:2006hq, Huang:2007sb, Alexandrov:2023zjb} and their relevant topological data, including the Euler character $\chi$ and second Chern class $c_2$.

\begin{figure}[ht]
	\centering
	\includegraphics[width=0.475\textwidth]{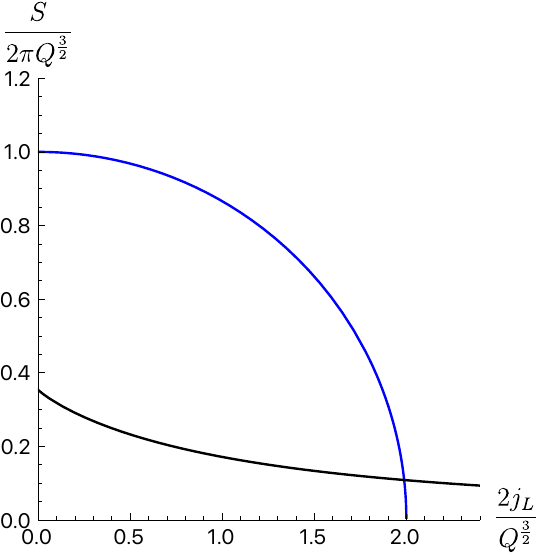}

	\caption{Entropy $S$ of the black hole (blue) and black ring (black) solutions in supergravity as a function of the $\SU(2)_{\L}$ angular momentum $j_{\L}$.  The black ring entropy is maximized over the M5 brane parameter $p>0$ at every value of $j_{\L}$ and does not vanish for arbitrary large $j_{\L}$.
	}
	\label{sugracurve}
\end{figure}

For such CY$_3$, the entropy of the black hole in (\ref{BHsugra}) simplifies to
\begin{equation}\label{BHentropy0}
	\begin{aligned}
		S_0^{\mathrm{BH}}=   2\pi \(\frac{2}{9k}\)^{\frac{1}{2}}d^{\frac{3}{2}} \(1-\frac{9}{32}y^2 \)^{\frac{1}{2}} , ~~ 
		y := \frac{4k^{\frac{1}{2}}}{d^{\frac{3}{2}}}j_{\L},
			\end{aligned}
\end{equation}
and the entropy of the black ring in (\ref{BRsugra}) simplifies to
\begin{equation}\label{BRentropy0}
	\begin{aligned}
		& S_0^{\mathrm{BR}}= 2\pi \(\frac{2}{9k}\)^{\frac{1}{2}}d^{\frac{3}{2}} \bigg(\frac{ x^2   \left(x^4-3 x^2-3 x y+3\right)}{8 } \bigg)^{\frac{1}{2}}, ~~ 
		x := \frac{k^{\frac{1}{2}}}{d^{\frac{1}{2}}} p.
	\end{aligned}
\end{equation}
From (\ref{BHentropy0}), (\ref{BRentropy0}), it is clear that the rescaled entropy 
\begin{equation}
	\begin{aligned}
		& s_0 := \frac{S_0}{Q(d)^{\frac{3}{2}}}  , ~~Q(d) := \(\frac{2}{9k}\)^{\frac{1}{3}}d
	\end{aligned}
\end{equation}
is independent of the geometric data ($k$) of the Calabi-Yau for both the black hole and black ring solutions. For the black hole, $s_0$ depends only on $y$, whereas for the black ring, $s_0$ depends on both $x,y$. For a given value of $y>0$, the scaled entropy of the black ring $s_0(x,y)$ has a local maximum as $x>0$ is varied, by which we
get a function of $y$ alone. The resulting curves are plotted in figure~\ref{sugracurve}. From the plot, one can see that for angular momentum
\ie\label{critialAM}
	j_{\L}>j_{\L,c}(d) \approx 0.99 \  Q(d)^{3/2},
\fe
the black ring is entropically dominant over the black hole. We emphasize that the critical value of the angular momentum above is universal among one parameter CY$_3$. We refer to this phenomenon as the black hole/black ring transition. In the rest of the paper, we will study this from the point of view of M theory.

\section{5d BPS indices from topological string theory}
\label{Sec:TOP}

Having established the existence of a black hole/black ring transition in supergravity, we can ask if this transition manifests in a quantum setting.  Had the entropic transition been for supergravity solutions in anti-de-Sitter space, the transition would have had ramifications in the dual conformal field theory.  In flat space, while the microstates describing a single black object admit a quantum description in terms of the worldvolume theory of the underlying brane, it is less clear whether the microstates of all objects (with the appropriate boundary conditions) can be captured by a coherent quantum theory.

Fortunately, a well-defined counting of microstates in flat space can be done by studying supersymmetric indices, where supersymmetry reduces the problem to geometric questions about the internal Calabi-Yau.
However, not all supersymmetric objects contribute to indices,\footnote{This phenomenon is well studied in 4d $\cN=4$ string theory (II on $\mathrm{K3} \times \mathrm{T}^2$), where none of the multi-centered black hole solutions (with at least one macroscopic horizon) contributes to the index (inverse Igusa cusp form) \cite{Dabholkar:2009dq, Chowdhury:2019mnb}. 
}
and thus it is unclear whether the black hole/black ring transition should manifest in the index, which is a central question we investigate.

The first two subsections are reviews, where we introduce 5d indices known as Gopakumar-Vafa (GV) invariants, and explain how they are computed using the machinery of topological string theory.  We then numerically compare the microscopic index with the macroscopic entropy, the results of which will motivate us to study in later sections the corrections to the semiclassical entropy from a macroscopic point of view, as well as the comparison of GV invariants with black ring indices for small charges.

\subsection{Gopakumar-Vafa conjecture}

Consider the worldvolume theory of an M2 brane wrapping 
two-cycle of class $[d]$ ( see section~\ref{5dBMPV} for the appropriate definition).
One obtains an effective supersymmetric quantum mechanics of a BPS particle of charge $d$ propagating in 5d (see \cite{Dedushenko:2014nya} for a careful discussion). 
These states can be further labeled by the representation $\[(j_{\L},j_{\R})\]$ under the spatial rotation group $\SU(2)_{\L} \times \SU(2)_{\R}$. Here both $j_{\L,\R}$ are taken to be half-integers. Abstractly, one can write the representation content as
\begin{equation}
	n_{d}^{j_{\L},j_{\R}}\[(j_{\L},j_{\R})\].
\end{equation}
Following \cite{Gopakumar:1998ii, Gopakumar:1998jq}, we can define a supersymmetric index by tracing over $\SU(2)_{\R}$ with signs, and further organize the results into $\SU(2)_{\L}$ representations to define the Gopakumar-Vafa (GV) invariants $n_{d}^g$, i.e.,
\begin{equation}
\begin{aligned}
		\sum_{j_{\L},j_{\R}} (-1)^{2j_{\R}}(2j_{\R}+1) n_{d}^{j_{\L},j_{\R}} \[(j_{\L})\] &=\sum_{g=0}^\infty n^g_{d} \ \[ \(\(\frac{1}{2}\)+ 2(0) \)^{\otimes(g+1)}\]\\
		&=\sum_{j_{\L}} \sum_{g=0}^\infty n^g_{d} \ {2(g+1)  \choose g+1+2j_{\L}} \ \[ \(j_{\L}\)\].
\end{aligned}
\end{equation}
The 5d supersymmetric index---a proxy for the BPS degeneracy at fixed $d, j_{\L}$---can be obtained from 
\begin{equation}\label{5ddegen}
	\Omega(j_{\L},d) = \sum_{g=0}^\infty n^g_{d} \ {2(g+1) \choose g+1+2j_{\L}}.
\end{equation}
Gopakumar and Vafa argued that the GV invariants $n^g_{d}$ can be computed by studying topological strings on the CY$_3$ (see \cite{Katz:1999xq} for more details and \cite{LopesCardoso:2012uug, Dedushenko:2014nya} for the subtleties involved).

For concreteness of discussion, let us focus on the quintic $X_5$ with $\chi =-200$, $k=5$, and $c_2=50$.
The GV conjecture can be expressed more explicitly as\footnote{ This expression is a formal expansion in $\lambda$ and does not have to be convergent.}
\begin{equation}\label{GV formula}
\begin{aligned}
	&\sum_{g=0}^\infty \ F^{(g)}(t, \bar{t}=-i\infty) \ \lambda^{2g-2}
	\\
	&=C_{3}(t)\lambda^{-2}+C_{1}(t)+\sum_{g=0}^\infty \sum_{k=1,d=0} ^\infty n_d^g \ \frac{1}{k} \bigg(2 \sin \frac{k \lambda}{2} \bigg)^{2g-2}\bigg(e^{2\pi i t }\bigg)^{k d}.
\end{aligned}
\end{equation}
Here $F^{(g)}$ is the free energy of topological strings at genus $g>0$ and $F^{(0)}$ is the prepotential (in a specific physical holomorphic gauge that we will discuss in detail later), $(t, \bar{t})$ keeps track of the K\"ahler moduli of the CY$_3$. Each $C_{n}$ is an $n$th order polynomial with the constant term set to $C_{n}(0)=0$ by absorbing it into $n_0^g$. All the terms with $d>0$ are coming from worldsheet instantons and would be referred to as the non-perturbative part and the rest of it would be called the perturbative part.

In IIA, $F^{(g)}$ for $g>1$ is expected to capture the coefficient in quantum effective action involving two anti-self dual gravitons and $2g-2$ anti-self dual graviphotons \cite{Antoniadis:1993ze, Antoniadis:1995zn}, so the right side of \eqref{GV formula} can be thought of as encoding the coefficients of the F terms \cite{Bershadsky:1993ta} in the Wilsonian effective action coming from integrating out a Euclidean M2 brane wrapping a holomorphic two-cycle in the CY$_3$ and the M theory circle. However, the conjecture relates them at the large volume limit. 

Physically, we expect the number of states to go to zero as we increase the angular momentum $j_{\L}$ for fixed mass $d$, and this expectation is met by the following bound on the GV invariants known as the Castelnuovo bound  \cite{Huang:2007sb}\footnote{The connection of GV invariants to Castelnuovo theory is discussed in \cite{Katz:1999xq}. It is proven recently in  \cite{Liu:2022agh} for the quintic and for all one parameter CY$_3$ (assuming the BMT inequality)  in \cite{Alexandrov:2023zjb}.}
\begin{equation}\label{CS bound}
\begin{aligned}
n^g_d=0, ~~ g>\frac{10+5d+d^2}{10}.
\end{aligned}
\end{equation}

\subsection{Computation of topological string free energy}

In this section, we review the method in \cite{Huang:2006hq} for computing the topological string free energy, %
which is based on the Bershadsky-Cecotti-Ooguri-Vafa (BCOV) holomorphic anomaly equation \cite{Bershadsky:1993ta, Bershadsky:1993cx} as well as \cite{Yamaguchi:2004bt, Aganagic:2006wq}, and focus on the quintic for concreteness. The reader who is familiar with the methodology, or is content with knowing the existence of an algorithm, can safely skip to the next subsection.

At genus zero, $F^{(0)}$ is determined in terms of the geometry of the moduli space. However, the K\"ahler moduli space parameterized by $t$ that we are after gets nontrivial $\alpha'$ corrections. On the other hand Mirror symmetry maps the K\"ahler structure moduli space to the complex structure moduli space parameterized by $z$ which is $\alpha'$ exact. Therefore we will study the topological strings on the mirror quintic and then we will use mirror map $t=t(z)$ to get the answer for the quintic. The coordinate $z$ is chosen such that at $z=0$ we have the Landau-Ginzburg orbifold description of the worldsheet CFT, at $z=1$ we encounter the conifold transition, and $z\to \infty$ is the large volume limit.

We rewrite the topological string free energy as 
\begin{equation}
\begin{aligned}
F^{(g)}(z,\bar{z})= \bigg(\frac{1-z}{5z} \omega_0^2(z)\bigg)^{g-1}\tilde{P}^{(g)}(z,\bar{z}), ~~~~ t(z)=\frac{1}{2\pi i} \frac{\omega_1(z)}{\omega_0(z)}.
\end{aligned}
\end{equation}
Here $\omega_0$ reflects the choice of the very specific holomorphic K\"ahler gauge (from the point of view of the worldsheet CFT of topological strings, there are nontrivial contact terms in the OPE at a generic point on the moduli space. Because of these contact terms, topological string free energy depends sensitively on the choice of the $\U(1)$ line bundle gauge field $A$ and the Zamolodchikov metric  $G$ on the moduli space. Once a holomorphic gauge is chosen for $A$, the answer still depends on the choice of the holomorphic gauge which can be further fixed at a given value of $\bar{t}=\bar{t}_0$ by demanding that the covariant derivative with respect to $A$ simplifies. It has been argued that this special choice is what is necessary to compare the results with the algebraic results on holomorphic curves). The mirror map can be calculated from the Picard-Fuchs equations at a point on the moduli space \cite{Candelas:1990rm}
\begin{equation}\label{PF equation}
\begin{aligned}
(\partial_u^4-e^{-u} L_1 L_2 L_3 L_4)\omega_n(z=e^u)=0, ~~ L_i=\partial_u-\frac{i}{5}.
\end{aligned}
\end{equation}
The boundary condition on $\omega $ is subtle and contains the choice of gauge. We will specify it at specific points below.

Near the large volume point $t=i\infty$, we have the following solutions
\begin{equation}\label{LV}
\begin{aligned}
\omega(z, \rho)=\sum_{n=0}^\infty \frac{\Gamma(5(n+\rho)+1)}{\Gamma(n+\rho+1)^5 (5^5 z)^{n+\rho}}, ~~ \omega_0(z)=\omega(z,0), ~~\omega_1(z)=\omega^{(0,1)}(z,0).
\end{aligned}
\end{equation}
The geometry of the moduli space in this holomorphic limit is determined by (the constant of proportionality would be unimportant for the calculation and we will not bother to fix it)
\begin{equation}\label{LVg}
\begin{aligned}
e^{-K} \propto \omega_0(z) ~~~~ G_{z\bar{z}} \propto \partial_z t(z).
\end{aligned}
\end{equation}

Near the orbifold point $t=0$, we have the following solutions (these are related to the solutions in the large volume limit but not exactly the same---in fact, their analytic continuations are related by a linear transformation)
\begin{equation}\label{orb}
\begin{aligned}
\omega_k(z)=25 z^{\frac{k+1}{5}} \sum _{n=0}^{\infty} \frac{\left(5^5 z\right)^n \left(\left(\frac{k+1}{5}\right)_n\right){}^5}{(k+1)_{5 n}},
\end{aligned}
\end{equation}
where $(k)_{n}$ is the Pochhammer symbol.
Again the geometry of the moduli space is given by (\ref{LVg}) with $\omega_{0}, \omega_{1}$ defined as in (\ref{orb}). 

A closed-form expression is not available near the conifold point $t=1$. We fix it by the following conditions (these are related to the solutions in the large volume limit and at the orbifold point presented earlier but not exactly the same)
\begin{equation}\label{coni}
\begin{aligned}
& \omega_k(z)=5^{1/2} \sum _{n=0}^{\infty} d_{k,n} (z-1)^n,\\
& d_{0,0}=1, ~~d_{0,1}=0, ~~d_{0,2}=0, ~~d_{0,3}=\frac{2}{625},\\
& d_{1,0}=0, ~~d_{1,1}=1, ~~d_{1,2}=-\frac{3}{10}, ~~d_{1,3}=\frac{11}{75}.\\
\end{aligned}
\end{equation}
Here (\ref{LVg})  remains valid.

Now we turn to state the results for the GV invariants. It will be convenient to define the following decomposition (at a generic point on the moduli space)
\begin{equation}\label{HolAm}
\begin{aligned}
&  \tilde{P}^{(0)}(z,\bar{z})=P^{(0)}(z,\bar{z}),  ~~~ X(z)= \frac{1}{1-z},\\
& \tilde{P}^{(g)}(z,\bar{z})=P^{(g)}(z,\bar{z})+\sum_{i=0}^{3(g-1)}a_{g,i} X(z)^i ~~ \text{ for }g\ge 1.
\end{aligned}
\end{equation}
The coefficients $a_{g,i}$ will be called the holomorphic ambiguity, the reason for which will be clear in a moment. The upper and lower limit on the sum has to do with the boundary conditions on the moduli space that we will discuss later.  We define a covariant derivative by the following formula \cite{Yamaguchi:2004bt}
\begin{equation}\label{CoVar}
\begin{aligned}
& P^{(g)}_{n+1}=z \partial_z P^{(g)}_{n}-(n(A_1+1)+(2-2g)(B_1-\frac{X}{2}))P^{(g)}_{n}, ~~ P^{(g)}_{0}=P^{(g)},\\
& A_p=\frac{(z \partial_z)^p G_{z\bar{z}}}{G_{z\bar{z}}}, ~~ B_p=\frac{(z \partial_z)^p e^{-K}}{e^{-K}}.
\end{aligned}
\end{equation}
At genus zero the data of GV invariants can be conveniently encoded into \cite{Yamaguchi:2004bt}
\begin{equation}\label{g0}
\begin{aligned}
& P^{(0)}_3=1.
\end{aligned}
\end{equation}

For higher genus non-holomorphic dependence of the topological string free energy is completely fixed by the holomorphic "anomaly" equations \cite{Bershadsky:1993cx}. Conceptually this can be understood as follows. $F^{(g)}(t, \bar{t})$ for the topological string B model can be written as an integration over the moduli space of genus $h$ Riemann surface just like the usual IIB string theory. An anti-holomorphic derivative ($\partial_{\bar{t}}$) of $F^{(g)}(t, \bar{t})$ inserts an operator in the path integral representation that is BRST exact and therefore gets non-zero contribution only from the boundary of the moduli space where one can factorize the answer through lower genus data. The equation is most conveniently expressed through Yamaguchi-Yau variables \cite{Yamaguchi:2004bt}
\begin{equation}\label{YY}
\begin{aligned}
A_1=v_1-2u-1, ~ B_1=u, ~ B_2=v_2+uv_1, ~ B_3=v_3+u(-v_2+(v_1-\frac{2}{5})X).
\end{aligned}
\end{equation}
All other $A_n, B_n$ can be  expressed as a polynomial of $A_1,B_{1,2,3},X$. For example
\begin{equation}\label{YY2}
\begin{aligned}
& A_2=-4 B_2-2 A_1 B_1-2B_1+2B_1^2-2A_1+2XB_1+XA_1+\frac{3}{5}X-1,\\
& B_4=2XB_3-\frac{7}{5}X B_2+\frac{2}{5}XB_1-\frac{24}{625}X.
\end{aligned}
\end{equation}
Also, note that 
\begin{equation}\label{YY3}
\begin{aligned}
& z\partial_z A_p=A_{p+1}-A_1A_p, ~~z\partial_z B_p=B_{p+1}-B_1B_p, ~~z\partial_z X=X(X-1).
\end{aligned}
\end{equation}

Genus one GV invariants can be obtained from 
\begin{equation}\label{g1}
\begin{aligned}
& P^{(1)}_1=-\frac{1}{2}A_1-\frac{31}{3}B_1+\frac{1}{12}(X-1)+\frac{5}{3}.
\end{aligned}
\end{equation}
The holomorphic anomaly equations for $g>1$ are given by 
\begin{equation}\label{HAE}
\begin{aligned}
& \frac{P^{(g-1)}_{2}+\sum_{r=1}^{g-1}P^{(r)}_{1}P^{(g-r)}_{1}}{2}=Q_0+uQ_1+u^2Q_2,\\
& \frac{\partial P^{(g)}_0}{\partial u}=0, ~~\frac{\partial P^{(g)}_0}{\partial v_1}=-Q_0, ~~\frac{\partial P^{(g)}_0}{\partial v_2}=Q_1-X Q_2, ~~\frac{\partial P^{(g)}_0}{\partial v_3}=Q_2.
\end{aligned}
\end{equation}
To use these equations one has to treat $u,v_{1,2,3},X$ as independent variables and $P^{(g)}_n(u, v_1,v_2,v_3,X)$ as a function of these variables. Now start from eq. (\ref{g0}) and eq. (\ref{g1}) and use (\ref{HAE}) recursively.  It is clear from this structure that the holomorphic anomaly terms parameterized by $a_{g,i}$ are not fixed in this procedure. 

There is no systematic procedure to fix the holomorphic ambiguity for all genera. However, from the additional input of boundary conditions at the orbifold point, the conifold point, and the large volume limit, we can fix them up to genus $g=51$ as follows.  The lower limit in holomorphic ambiguity sum (\ref{HolAm}) is constrained by the contribution of degenerate worldsheet instantons (these instantons are constant maps in the target space. These contributions can be calculated on the worldsheet for their relative simplicity) in the large volume limit \cite{Faber:1998gsw}
\begin{equation}\label{bLV}
\begin{aligned}
\lim_{z \to \infty} F^{(g)}=\frac{(-1)^{g-1}B_{2g}B_{2g-2}}{2g(2g-2)(2g-2)!}\frac{\chi}{2}.
\end{aligned}
\end{equation}
The upper limit is constrained by the knowledge of the gapless modes at the conifold point \cite{Strominger:1995cz, Vafa:1995ta}
In the IIA language, they come from a D2 brane's wrapping the $\S^2$ that shrinks as we approach the conifold point. At this point and until the new massless degrees of freedom are taken into account, the string background is expected to be singular. When the new massless degrees of freedom are taken into account, we go through a geometric transition to a different Calabi-Yau.
\begin{equation}\label{bC}
\begin{aligned}
\lim_{z \to 1} F^{(g)}=\frac{(-1)^{g-1} B_{2g}}{2g(2g-2)t^{2g-2}}+\mathcal{O}(1).
\end{aligned}
\end{equation}
This gives $2h-2$ conditions on $a_{g,i}$. 
More specifically this formula can be obtained by using the information of massless modes in the RHS of (\ref{GV formula}). On the other hand, we do not expect anything singular at the Gepner point 
\begin{equation}\label{bO}
\begin{aligned}
\lim_{z \to 0} F^{(g)}=\mathcal{O}(1).
\end{aligned}
\end{equation}
This gives $\lceil \frac{3}{5}(g-1) \rceil$ conditions on $a_{g,i}$. The number of unknown $a_{g,i}$ reduces to 
\begin{equation}\label{unka}
\begin{aligned}
(g-1)-\lceil \frac{3}{5}(g-1) \rceil=\floor*{\frac{2}{5}(g-1)} 
\end{aligned}
\end{equation}
The boundary conditions (\ref{bLV}), (\ref{bC}), (\ref{bO}) together with the knowledge of Castelnuovo bound (\ref{CS bound}) and explicit knowledge of $n^{51}_{20}=165$\footnote{Near the Castelnuovo bound some GV invariants are determined in the mathematics literature.} in principle determines the GV invariants up to genus $g=51$.\footnote{Note that the maximal genus for a non-zero GV invariant at $d=20$ is $g_{max}(20)=51$.} However, in practice, the evaluation is rather complex.

\paragraph{Note} As we computed the GV invariants for the quintic up to $g=49$, we came to learn of the beautiful recent work \cite{Alexandrov:2023zjb} that established new constraints on the GV invariants of various one-parameter CY$_3$, and pushed the determinable GV invariants beyond past limitations (in principle up to $g = 68$ for quintic).  Our results for quintic are in complete agreement with \cite{Alexandrov:2023zjb}. The details of our calculation can be found in appendix~\ref{aB} and on the website \cite{WebUS}.
For the subsequent numerical analysis, we use the data publicly available at \cite{Web}.

\subsection{Microscopic/macroscopic comparison}
\label{numana}

\begin{figure}[t]
	\centering
	\includegraphics[width=0.475\textwidth]{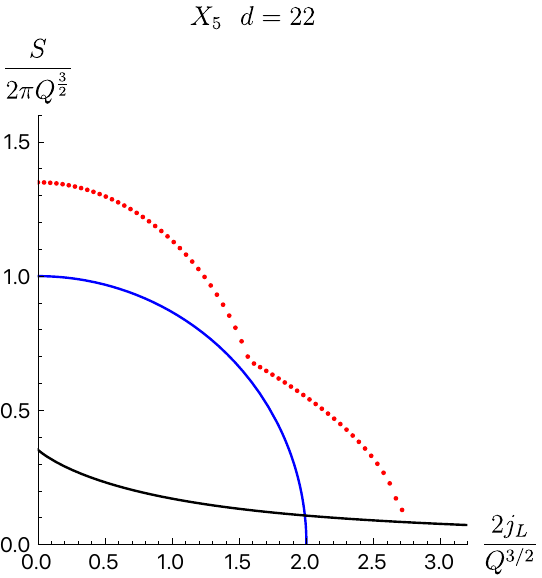}
	\quad 
	\includegraphics[width=0.475\textwidth]{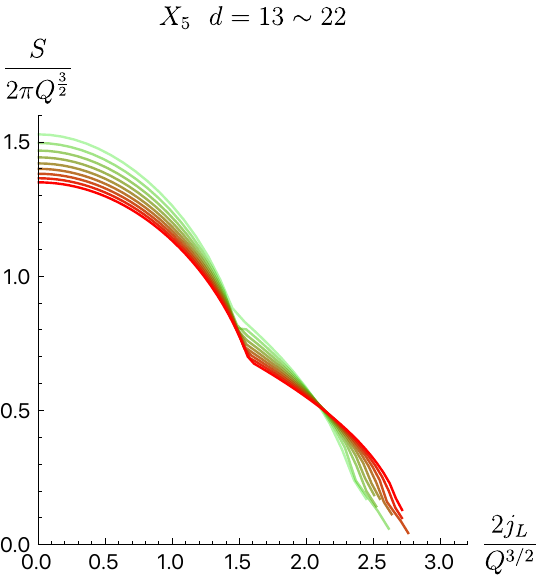}
	\caption{{\bf Left:} Entropy of BPS index ($S = \log|\Omega|$) obtained from the Gopakumar-Vafa invariants at $d=22$ (red), and compared with the entropy of the black hole (blue) and black ring (black) solutions in supergravity,
	plotted over the $\SU(2)_{\L}$ angular momentum $j_{\L}$. {\bf Right:} Entropy of BPS index as $d$ is increased from 13 to 22 (green to red).
	}
	\label{fgv}
\end{figure}

\begin{figure}[ht]
	\centering
	\includegraphics[width=0.45\textwidth]{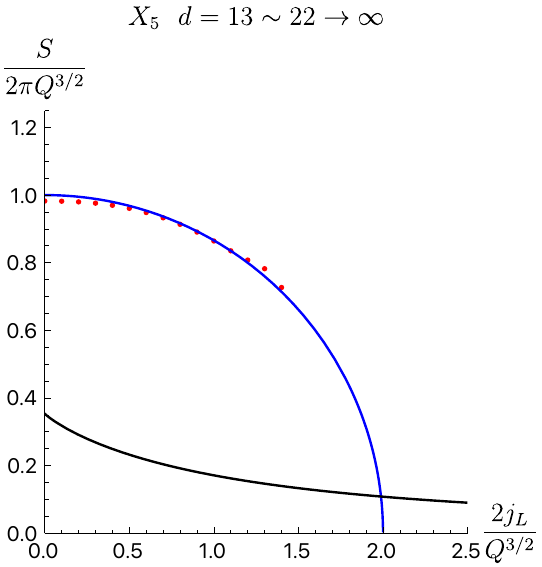}
	\vspace{-0.05in}
	\\
	\hrulefill
	\\
	~
	\vspace{-0.05in}
	\\
	\includegraphics[width=0.45\textwidth]{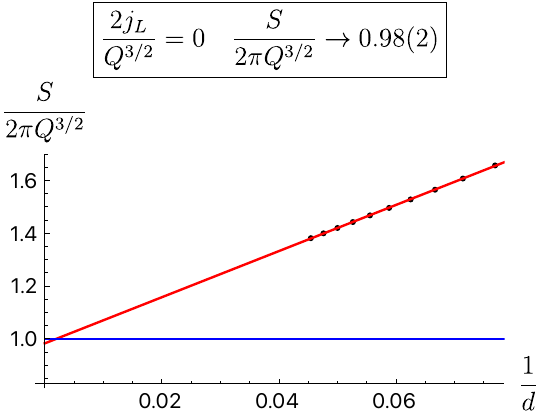}
	~~
	\includegraphics[width=0.45\textwidth]{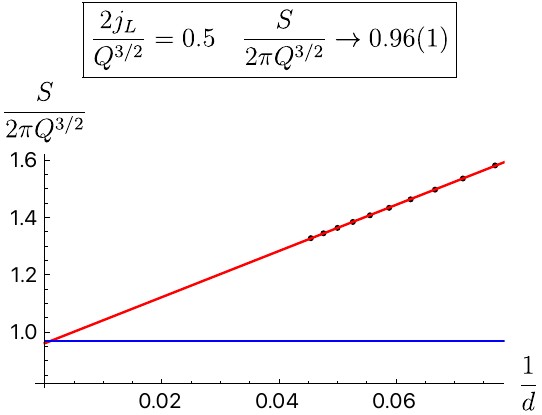}
	\\
	~\vspace{-0.1in}
	\\
	\includegraphics[width=0.45\textwidth]{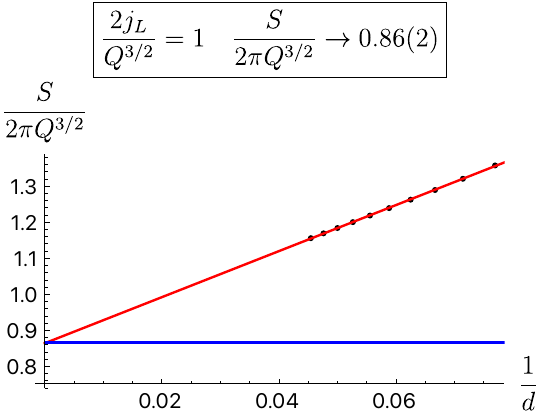}
	~~
	\includegraphics[width=0.45\textwidth]{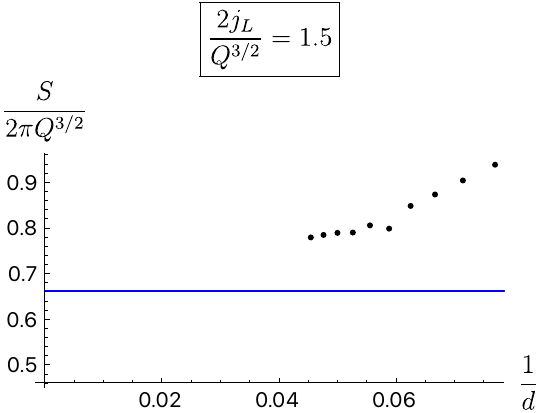}
	\\
	~\vspace{-0.1in}
	\\
  \caption{Entropy of BPS index ($S = \log|\Omega|$) for the quintic with M2 brane charge in the range $13 \le d \le 22$ (black), linearly extrapolated (in $1/d$) to $d \to \infty$ (red) whenever reasonable ($2j_{\L} \lesssim 1.5 Q^{\frac32}$), and compared with the black hole (blue) and black ring (black) entropies of supergravity.}
  \label{BHexpol}
  \end{figure}

We now perform a numerical study of the 5d BPS degeneracies \eqref{5ddegen} for M theory on quintic, using the GV invariants calculated in the previous section. For fixed M2 brane charge $d$, the entropy defined as
\ie 
	S(j_{\L}, d) := \log |\Omega(j_{\L}, d)|
\fe 
is plotted against the angular momentum $j_{\L}$ in figure~\ref{fgv}.   We pause for a moment to notice a few basic features of the graph. For a given value of $d$, there is a kink at a critical angular momentum $j_{\L,c}(d)$, which can be defined by maximizing over $j_{\L}$ the discrete second derivative
\ie 
	S''(j_{\L}, d) := \frac{S(j_{\L}+\frac12) + S(j_{\L}-\frac12) - 2 S(j_{\L})}{2}.
\fe
Thus we expect that the contribution to the entropy for angular momenta $j_{\L}<j_{\L,c}$ and $j_{\L}>j_{\L,c}$ are coming from different black objects from the macroscopic point of view. 
Is this the black hole/black ring transition?

A direct comparison of the BPS index entropy at $d=22$ with supergravity exhibits apparent discrepancy.  However, as we see on the right of figure~\ref{fgv}, the BPS index entropy continues to decrease as $d$ is increased and is far from stabilizing at $d=22$. 

\paragraph{Black hole}
For angular momenta below the transition point,
an agreement in the $d \to \infty$ limit seems plausible.
In fact, near $j_{\L} = 0$, this agreement has already been established by careful numerical extrapolation using the Richardson transform \cite{Huang:2007sb}.
In figure~\ref{BHexpol}, we perform linear extrapolations in $1/d$ to $d \to \infty$ for a range of $j_{\L}/Q^{\frac32}$ below the transition point, and find good agreement with supergravity.  To be precise, since the GV invariants define $\Omega(j_{\L}, d)$ for half-integral $j_{\L}$, in order to fix $j_{\L}/Q^{\frac32}$, we perform linear interpolations on $S = \log|\Omega|$ to define the entropy at non-half-integral values of $j_{\L}$.\footnote{By contrast, in \cite{Huang:2007sb} the extrapolation to $d \to \infty$ was performed for fixed $j_{\L}=0$.  By experimentation, we find that in the present limit of fixing $j_{\L}/Q^{\frac32}$, the Richardson transform or other sophisticated methods do not outperform a simple linear extrapolation.
}

The strength of the linear relationship between $S(d)$ and $1/d$ as shown in figure~\ref{BHexpol} suggests that the leading order correction to supergravity takes the form
\begin{equation}
	\begin{aligned}
		S^{\mathrm{BH}}=   S_0^{\mathrm{BH}} & \bigg(1+ \frac{c}{ d}\bigg), ~~ c>0.
			\end{aligned}
\end{equation}
This is the kind of correction that would arise from the dimensional reduction of the higher derivative terms $(\Tr{R^2})^2$ and $\Tr{R^4}$ in M theory.  
Although such corrections have been studied in the past \cite{LopesCardoso:1998tkj, LopesCardoso:1999cv, LopesCardoso:1999xn}, the structure of these corrections has not been carefully analyzed to the full extent possible.
We devote the entirety of section~\ref{Sec:Wald} to expounding the details of these corrections.

\paragraph{Black ring}  

\begin{figure}[t]
\centering
\includegraphics[width=0.475\textwidth]{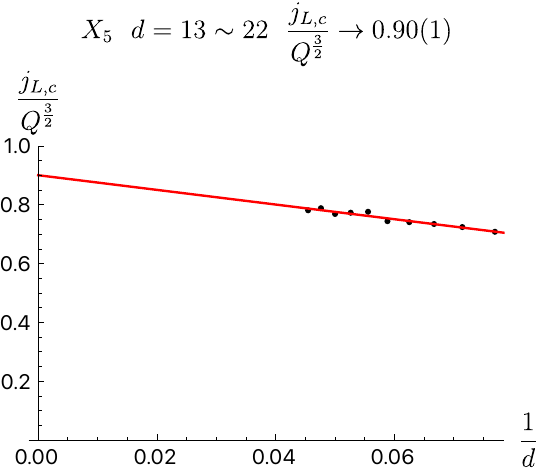}
\caption{Linear extrapolation of the critical angular momentum to M2 brane charge $d \to \infty$ (red) using quintic data ranging from $d = 13$ to 22 (black).
}
\label{fjc}
\end{figure}

Now we focus our attention on the transition point. The critical angular momentum is plotted in figure~\ref{fjc} as a function of $1/d$. Again we use a linear fit to give the extrapolated value for\footnote{We expect this extrapolated critical value of $j_c(d)$ to be universal among one parameter CY$_3$.} $$j_c(d\to \infty)= 0.9 Q(d)^{3/2}.$$ By comparing this value with the discussion of black hole/black ring transition in section~\ref{Sec:SUGRAComparison} (see (\ref{critialAM})), we conclude that it is possible that the black object dominating the entropy in the microscopic plot for angular momentum larger than the transition point is a black ring. However, there is no direct evidence for the index entropy to agree with the black ring entropy in the $d \to \infty$ limit.   %
We work under the hypothesis that it counts black rings.
In the next few sections, we will analyze in detail the following two viable scenarios: 
\begin{enumerate}
	\item The leading contribution to entropy is coming from a black ring whose M5 brane parameter  $p$ is macroscopic (scales as $d^{\frac12}$).  In this situation, we will carefully discuss the corrections to Bekenstein-Hawking entropy similar to the black hole above.
 
	\item The leading contribution to entropy is coming from a black ring whose $p$ is microscopic, invalidating the large $d$ analysis. 
	In section~\ref{Sec:MSW}, we will compare the tail with the exact black ring microstate counting for $p=1$, and find surprisingly good agreement for small values of $d$.  
\end{enumerate}

\section{Wald entropy and quantum corrections in 4d}
\label{Sec:Wald}

In 5d supergravity there is no systematic understanding of the superspace formulation for the higher derivative terms that are relevant for the corrections to the Bekenstein-Hawking entropy. However, the BPS black solutions in 5d can be considered as the Gaiotto-Strominger-Yin uplift \cite{Gaiotto:2005gf, Gaiotto:2005xt}  of the solutions in 4d where we have a much better understanding of the superspace formulation.

We turn to describe the uplift very briefly. Consider a configuration in IIA with $p^0$ D6 branes wrapping the Calabi-Yau. This configuration in M theory language corresponds to a factorized geometry involving the Calabi-Yau and a Taub-NUT space. The Taub-NUT radius $R_{TN}$ in this example is determined by the 10d IIA string coupling. At strong coupling, $R_{TN}$ becomes large in string units, and then for radial distance in Taub-NUT much smaller compared to $R_{TN}$ we have an effective flat Minkowski space for $p^0=1$. On the other hand, for weak string coupling in the region of Taub-NUT at distances much larger compared to $R_{TN}$, we have an effective 4d IIA description. The radius of the Taub-NUT is related to an asymptotic value of the moduli in the IIA language, and due to the attractor mechanism, the leading order entropy of the BPS black hole is independent of the asymptotic moduli. It is expected that this fact survives beyond leading order. This allows us to calculate the quantum corrected entropy of the black holes in 5d in terms of that in 4d. 
 Keeping this in mind, we first discuss the quantum correction in 4d following the Ooguri-Strominger-Vafa conjecture \cite{Ooguri:2004zv} and delay the detailed mapping of solutions to section \ref{4d/5d}.

\subsection{Ooguri-Strominger-Vafa conjecture}

In this section, we consider extremal black holes in 4d $\mathcal{N}=2$ supergravity obtained by compactifying M theory on $\S^1 \times$CY$_3$. This can be thought of as a compactification of the 5d $\mathcal{N}=1$ supergravity we discussed in the previous section. The compactification gives rise to an additional Kaluza-Klein gauge field $a$. The gauge fields in 4d can be related to the 5d ones as follows 
 \begin{equation}
A^C=V^C-\alpha^C a, ~~ A^0=-a,
\end{equation}
where $\alpha^I=V^I_{\S^1}$ is the component of 5d gauge fields in the M theory circle direction. Among $h^{1,1}+1$ gauge fields $A^\Lambda$, one combination becomes the 4d graviphoton field (we will describe it in detail in the later part of this section) and each of the remaining $h^{1,1}$ is in a vector multiplet. Along with the gauge fields, a scalar field is generated $e^\sigma = e^{\gamma} v$  where $e^\gamma $ keeps track of the radius of the $\S^1$ circle in M theory units.\footnote{Note that the 4d string coupling is given by $g_{s,4}=v^{-\frac{3}{2}}$ and it is in the hypermultiplet in 4d. Just as in 5d, in 4d we still have $h^{2,1}+1 $ hypermultiplets.} The complex scalar fields in the $h^{1,1}$ vector multiplets can be parameterized as
follows
 \begin{equation}\label{moduli}
t^C=\alpha^C+i e^\sigma h^C=\frac{X^C}{X^0}.
\end{equation}

Now we consider extremal black holes carrying magnetic and electric charges given by $(p^\Lambda, q_\Lambda)$. We associate the magnetic charges $p^0, p^A$ with the D6 and D4 branes, respectively, and the electric charges  $q_0, q_A$ with the D0 and D2 branes. In our convention, all charges are integer quantized. The leading order Bekenstein-Hawking entropy of the black hole may be written as \cite{Ferrara:1995ih, Strominger:1996kf} 
\begin{equation}
	S_{0}^{\mathrm{BH}}=\frac{\pi}{4} C\bar{C} e^{-K(X,\bar{X})},
\end{equation}
where the K\"ahler potential is determined from the holomorphic prepotential $F_0$  by\footnote{Here sum is over $\{\Lambda\}=0,\{A\}$. We are using the notation $\bar{f}(\bar{X})=(f(X))^*$.} 
\ie
	& e^{-K(X,\bar{X})}=i (\bar{X}^\Lambda F_{0,\Lambda}(X)-X^\Lambda \bar{F}_{0,\Lambda}(\bar{X})),
	\\
	& F_{0,\Lambda}(X)=\frac{\partial F_0(X)}{\partial X^\Lambda}, ~~ \bar{F}_{0,\Lambda}(\bar{X})=\frac{\partial \bar{F}_0(\bar{X})}{\partial \bar{X}^\Lambda}.
\fe
The moduli fields in the vector multiplet at the horizon are determined entirely in terms of the charges by the attractor mechanism (the BPS entropy does not depend on the fields in the hypermultiplet) 
\begin{equation}
	p^\Lambda=\Re(CX^\Lambda), ~~ q_\Lambda=\Re(F_{0,\Lambda}(CX)).
\end{equation}
Clearly, this determines $X^\Lambda, C$ only up to K\"ahler gauge transformations\footnote{The K\"ahler potential transforms as $K \to K-f(X)-\bar{f}(\bar{X})$.}
\begin{equation}
	X^\Lambda \to e^{f(X)}X^\Lambda, ~~ C \to e^{-f(X)}C.
\end{equation}
The entropy can be written in a very suggestive thermodynamic form following Ooguri-Strominger-Vafa (OSV) \cite{Ooguri:2004zv}
\begin{equation}
	S_0^{\mathrm{BH}}(\phi,p)=\mathcal{F}_0(\phi, p)-\phi^\Lambda \frac{\partial}{\partial \phi^\Lambda}\mathcal{F}_0(\phi, p),
\end{equation}
where we used the following definition for the electric potential $\phi^\Lambda$ 
\begin{equation}
	\begin{aligned}
		C X^\Lambda=p^\Lambda+ \frac{i}{\pi}\phi^\Lambda,~~ \mathcal{F}_0(\phi, p)=-\pi \Im(F_0(CX)).
	\end{aligned}
\end{equation}
Note that here $\phi^\Lambda$ is a real number and we can determine the electric charge in terms of it as follows,\footnote{Compared with~\cite{Denef:2007vg}, the parameter $\phi(\mathrm{here})=-2\pi \phi(\mathrm{there})$.} $$q_\Lambda= -\frac{\partial}{\partial \phi^\Lambda}\mathcal{F}_0(\phi, p).$$ This thermodynamic form suggests that we consider the partition function given by
\begin{equation}\label{pf0}
\begin{aligned}
		& Z^{\mathrm{BH}}_0(\phi, p) = \sum_{q}\Omega_0(q,p)e^{-\phi^\Lambda q_\Lambda}=e^{\mathcal{F}_0(\phi, p)}, \\
		& S^{\mathrm{BH}}_0(\phi, p)=\log Z^{\mathrm{BH}}_0(\phi, p) - \frac{\partial}{\partial \beta} \log Z^{\mathrm{BH}}_0(\beta \phi, p)|_{\beta=1}.
\end{aligned}
\end{equation}
The prepotential can be calculated from the topological string amplitude in holomorphic gauge %
\begin{equation}\label{pp0}
	\begin{aligned}
		F_0(CX)=-\frac{2i}{\pi}F_{top,0}(g_{\mathrm{top}},t), ~~ t^A= \frac{p^A+i\frac{\phi^A}{\pi}}{p^0++i\frac{\phi^0}{\pi}}, ~~ g_{\mathrm{top}}=-\frac{4\pi i}{p^0++i\frac{\phi^0}{\pi}},
	\end{aligned}
\end{equation}
with
\begin{equation}
	\begin{aligned}
	F_{top,0}( g_{\mathrm{top}},t)= &\frac{1}{g_{\mathrm{top}}^2}\(-(2\pi i)^3 \frac{C_{ABC}}{3!} t^A t^B t^C \).
	\end{aligned}
\end{equation}

The OSV conjecture \cite{Ooguri:2004zv} (here we present a refined formulation following Denef-Moore \cite{Denef:2007vg}) is the statement that the \emph{exact} degeneracy of the black hole  can be calculated from a formula  analogous to (\ref{pf0}), (\ref{pp0}) (near large volume limit of the Calabi-Yau threefold)
\begin{equation}
\begin{aligned}\label{4ddegen}
		& \Omega(q,p)=\int \frac{d\phi}{2\pi} \ e^{\phi^\Lambda q_\Lambda} \mu (\phi,p)  |Z_{\mathrm{top}}(g_{\mathrm{top}},t)|^2,
\end{aligned}
\end{equation}
with 
\begin{equation}
	\begin{aligned}
		t^A= \frac{p^A+i\frac{\phi^A}{\pi}}{p^0++i\frac{\phi^0}{\pi}}, ~~ g_{\mathrm{top}}=-\frac{4\pi i}{p^0++i\frac{\phi^0}{\pi}}.
	\end{aligned}
\end{equation}
Here the partition function of the topological strings is given by %
\begin{equation}\label{toppf}
	\begin{aligned}
	Z_{\mathrm{top}}(g_{\mathrm{top}},t) &= e^{F_{\mathrm{top}} (g_{\mathrm{top}},t)}=Z_{\mathrm{local}}(g_{\mathrm{top}},t)Z^0(g_{\mathrm{top}})Z'(g_{\mathrm{top}},e^{2\pi i t}),\\
	F_{\mathrm{local}}(g_{\mathrm{top}},t) &= \log Z_{\mathrm{local}}(g_{\mathrm{top}},t)
	\\
	&=\frac{1}{g_{\mathrm{top}}^2}\(-(2\pi i)^3 \frac{C_{ABC}}{3!} t^A t^B t^C \)-\frac{(2\pi i)}{24}c_{2A}t^A.
	\end{aligned}
\end{equation}
The perturbative contribution $F_{\mathrm{local}}(g_{\mathrm{top}},t)$ is the same as $C_{3,1}$ in (\ref{GV formula}). These contributions can be understood from the M theory point of view as follows: in 11d, lowest derivative order supergravity action \cite{Cremmer:1978km} gives the cubic term in $F_{\mathrm{local}}$ as expected. Apart from that, there is a specific $R^4$ term \cite{Green:1997di, Kiritsis:1997em, Berkovits:1998ex} that is required by the consistency of supersymmetry and anomaly cancellation \cite{Duff:1995wd, Vafa:1995fj}. Upon dimensional reduction of M theory to 4d IIA, this contains the $R\wedge R$ coupling (the two other factors of $R$ are in the Calabi-Yau directions and captured by $c_{2}$ above) that contributes to the linear term in $F_{\mathrm{local}}$ \cite{LopesCardoso:1998tkj}.

Other contributions from degenerate worldsheet instantons are given by (see \cite{Dabholkar:2005dt} for more details)\footnote{ These are the same as the ones in (\ref{bLV}) with the identification $\lambda = -i g_{\mathrm{top}}$.
}
\begin{equation}\label{topKK}
	\begin{aligned}
		Z^0(g_{\mathrm{top}})= e^{\frac{\chi}{2} \left(\frac{\zeta(3)}{g_{\mathrm{top}}^2} +\sum_{g=2}^\infty K_{g}(-g_{\mathrm{top}}^2)^{g-1}\right)}, ~~ K_g=\frac{(-1)^{g-1} B_{2 g} B_{2 g-2}}{2 g (2 g-2) \Gamma (2 g-1)}.
	\end{aligned}
\end{equation}
On the other hand, $Z'(g_{\mathrm{top}},e^{2\pi i t})$ comes from nontrivial worldsheet instanton contributions in topological strings. For this paper, we will always stay in the domain where the nontrivial worldsheet instanton corrections are negligible.
The precise form of the measure is not known, but the following form was suggested in \cite{Denef:2007vg} and found to be consistent with the analysis of \cite{Sen:2012kpz}:
\ie
	\mu(\phi,p) &= \text{constant} \times \bigg| \bigg( \frac{g_{\mathrm{top}}}{2\pi} \bigg)^{\frac{\chi}{24}-1}  \bigg|^2 e^{-K_{\mathrm{top}}},
 \\
 e^{-K_{\mathrm{top}}} &= \frac{2}{\pi}(\bar{X}^\Lambda F_{top,\Lambda}((g_{\mathrm{top}},t))+X^\Lambda \bar{F}_{top,\Lambda}((\bar{g}_{\mathrm{top}},\bar{t}))).
\fe

The simplest way of calculating the degeneracy as in (\ref{4ddegen}) would be to perform a saddle point approximation taking the contribution only from the $F_{\mathrm{local}}(g_{\mathrm{top}},t)$ term (we will come to the validity of such an approximation in details later in this section). The saddle point value of the charges of the black hole is given by
\begin{equation}
	q_\Lambda=-\frac{\partial}{\partial\phi^\Lambda}\log Z_{\mathrm{BH}}(\phi,p).
\end{equation}
The entropy is given by a thermodynamic formula
\begin{equation}
	S=\log Z_{\mathrm{BH}}(\phi,p)-\frac{\partial}{\partial \beta}\log Z_{\mathrm{BH}}(\beta\phi,p)|_{\beta=1},
\end{equation}
with
\begin{equation}
	Z_{\mathrm{BH}}(\phi,p)=  |Z_{\mathrm{local}}(g_{\mathrm{top}},t)|^2.
\end{equation}

\subsection{Local contributions}

\subsubsection{D6-D2-D0 single-centered black hole}
In this subsection, we focus on the discussion of the entropy of a single-centered large black hole carrying one unit of D6 charge ($p^0 = 1$) and D2-D0 charges, but no D4 charge $(p^A = 0)$.
For notational convenience that will be closer to the 5d setting which we aim to discuss later, we introduce
\begin{equation}\label{conPar}
	\phi^0=\frac{\omega}{2}, ~~ q_0 = 2 j_{\L}.
\end{equation}
 This is motivated by the fact under the 4d/5d uplift the Taub-NUT circle plays the role of the M theory circle and the momentum on the circle naturally keeps track of the D0 brane charge in the IIA description.
In terms of these parameters, we have
\begin{equation}
	t^A=\frac{2\phi^A}{\omega-2\pi i}  , ~~~g_{\mathrm{top}}=-\frac{8\pi^2}{\omega-2\pi i}.
\end{equation} 
As mentioned earlier, for the saddle point evaluation of the entropy, we approximate the topological string free energy by
\begin{equation}
	\begin{aligned}
	F_{\mathrm{top}}(t^A, g_{\mathrm{top}})= &\frac{1}{g_{\mathrm{top}}^2}\(-(2\pi i)^3 \frac{C_{ABC}}{3!} t^A t^B t^C\)-\frac{(2\pi i)}{24}c_{2A}t^A\\
		=& \frac{1}{\pi \mu}\(i\frac{C_{ABC}}{3!} \phi^A \phi^B \phi^C \)-\frac{i\pi}{6\mu}c_{2A}\phi^A,\\
	\Re(F_{\mathrm{top}}(t^A, g_{\mathrm{top}}))	=& \frac{1}{((\frac{\omega}{2\pi})^2+1)} \(-\frac{1}{2\pi^2} \frac{C_{ABC}}{3!}  \phi^A \phi^B \phi^C+\frac{1}{12}c_{2A}\phi^A\).
	\end{aligned}
\end{equation}
Here we have treated $\omega, \phi$ as real numbers while taking complex conjugate. Since we will be dealing with only Calabi-Yau compactification with one K\"ahler parameter in this paper, for simplicity we will denote  $\phi:=\phi^1$, $q:=q_1$. The black hole partition function is given by
\begin{equation}
	\begin{aligned}
		\log Z_{\mathrm{BH}}(\phi,\omega)=
		&  \frac{1}{((\frac{\omega}{2\pi})^2+1)} \(-\frac{1}{\pi^2} \frac{k}{6} \phi^3+\frac{1}{6}c_{2}\phi\),\\
	\end{aligned}
\end{equation}
and the charges evaluate to have the following simple expressions
\begin{equation}\label{BHsp}
\begin{aligned}
	q=
		& -\frac{1}{((\frac{\omega}{2\pi})^2+1)} \(-\frac{1}{\pi^2} \frac{k}{2} \phi^2+\frac{1}{6}c_{2}\),\\
	j_{\L}=
		&  \frac{\frac{\omega}{2\pi}}{\pi((\frac{\omega}{2\pi})^2+1)^2} \(-\frac{1}{\pi^2} \frac{k}{6} \phi^3+\frac{1}{6}c_{2}\phi\).
\end{aligned}
\end{equation}
We pause for a moment to note that when both the charges $q, j_{\L}$ are large, $|\phi/\omega|$ is a large number. This justifies the assumption that we can ignore nontrivial worldsheet instanton effects in this domain. In addition, in this limit, $g_{\mathrm{top}}$ is a small number and therefore we can ignore most of the degenerate worldsheet instantons from higher genera.\footnote{Strictly speaking, we cannot ignore the effect of the genus-zero degenerate instantons, or the effect of the measure factor in the OSV formula. The first effect is important only when we are too close to the extremality. Both these effects will be numerically sub-dominant in the parameter range where we will use the formula below.
}
These discussions are much more subtle when we are looking at the limit of small angular momenta. This issue will be discussed in detail later in this section. By inverting the expression of the charges, we can write down the entropy as a series expansion in $$Q:=\(\tfrac{2}{9k}\)^{1/3} q$$ as follows\footnote{At this stage, we do not have a clear logic for this particular choice of the expansion parameters. The explicit comparison of BPS entropy with the GV data will serve as an indirect justification. We thank Cumrun Vafa for an insightful discussion on this topic. Further details of the formula will appear in \cite{NewPaper}.} 
\begin{equation}\label{entropyD6D2D0}
	\begin{aligned}
		S &=  2\pi Q^{\frac{3}{2}} \(1-\frac{j_{\L}^2}{Q^3}\)^{\frac{1}{2}}\(1+ \sum_{m=1}^{\infty}  \frac{d^{(0)}_m}{Q^m}\(1+\sum_{n} d^{(1)}_{m,n} \frac{j_{\L}^2}{Q^3}\(1-\frac{j_{\L}^2}{Q^3}\)^{n}\)\).
	\end{aligned}
\end{equation}
where the first few coefficients are explicitly computed (by expanding around extremality, i.e., $j_{\L}^2\sim Q^3$) to be
\begin{equation}
\begin{aligned}
	& d^{(0)}_1=\(\frac{2}{9k}\)^{1/3} \frac{c_2}{4}, ~~~ d^{(0)}_2=-\frac{1}{2}\(\(\frac{2}{9k}\)^{1/3} \frac{c_2}{4}\)^2.%
\end{aligned}
\end{equation}

The corrections associated with $d^{(0)}_m$ do not vanish at zero angular momentum. Whereas the individual corrections associated with $d^{(1)}_{m,n}$ remain small for slowly rotating black holes and in the extremal limit. 
The value of $d^{(0)}_1$ in the formula above is the same as in \cite{Guica:2005ig}. Other corrections have their origin in the fact that the topological string K\"ahler parameter gets various $Q^{-1}$ corrections.

\subsubsection{D6, D4-D2-D0 multi-centered black hole}

First, we consider a single-centered black hole carrying D4-D2-D0 charge, but no D6 charge ($p^0 = 0$).
We will again use (\ref{conPar}) and work with one parameter CY$_3$, but now we have (as before $p := p^1$)
\begin{equation}
	\begin{aligned}
		t= \frac{p+i\frac{\phi}{\pi}}{i\frac{\omega}{2\pi}}, ~~ g_{\mathrm{top}}=-\frac{4\pi }{ \frac{\omega}{2\pi}},
	\end{aligned}
\end{equation}
which gives
\begin{equation}
	\begin{aligned}
	\Re(F_{\mathrm{top}}(t, g_{\mathrm{top}}))=\frac{k p }{\omega} \phi^2-\frac{\pi^2}{3\omega}(k p^3+c_2 p).
	\end{aligned}
\end{equation}
It immediately follows that the entropy is given by \cite{Dabholkar:2005dt}
\begin{equation}
	\begin{aligned}
		& S= 2\pi \( \frac{(k p^3+c_2 p)
}{6} (-q_0+ \frac{1}{2kp}q^2 ) \)^{\frac{1}{2}}.
	\end{aligned}
\end{equation}
This formula coincides with the Cardy-type argument in MSW CFT \cite{Maldacena:1997de}.
In the large $p$ limit, the expression for the entropy above reduces to the more general formula in \cite{Gaiotto:2005xt} when $p_0$ is set to zero. The whole effect of the Wald entropy correction was to shift $kp^3 \to kp^3+c_2p$. This is very different from the case considered previously with $p^0 \neq 0$.

In this paper, we will be interested in the D6, D4-D2-D0 multi-centered black hole with one unit of D6 charge. 
Now we turn to the discussion of the wall of marginal stability for this bound state following \cite{Denef:2007vg, Andriyash:2010qv}. We demand the distance between the centers to be positive and finite 
\begin{equation}\label{distance}
	\begin{aligned}
		& |x_{1,2}|=\frac{\langle \Gamma_1, \Gamma_2 \rangle}{2 \Im{(Z_1 \bar{Z}_2)}}|Z_1+Z_2|, ~~ \langle \Gamma, \Gamma' \rangle= -\Gamma^0 \Gamma'_0+\Gamma^A \Gamma'_{A}+\Gamma_0 \Gamma^{'0}-\Gamma_A \Gamma^{'A}.
	\end{aligned}
\end{equation}
Here $\Gamma=(p^\Lambda,q_\Lambda)$ is the charge vector and the central charge of single-center is given by (in large volume limit, we set the $B$ field to zero)
\begin{equation}\label{central charge}
	\begin{aligned}
		Z&=\frac{1}{6}k (B+iJ)^3 p^0-\frac{1}{2}k (B+iJ)^2 p^1+q_1(B+iJ)-q_0 
		\\
		&= Z_{D6} + Z_{D4-D2-D0},
		\\
		Z_{D6}&=-i \frac{k}{6}J^3, \quad Z_{D4-D2-D0}=\frac{1}{2}k J^2 p^1+q_1iJ-q_0.
	\end{aligned}
\end{equation}
This suggests that the bound state exists for the following choice of signs
\begin{equation}\label{BR sign}
	\begin{aligned}
		& p^0=1, ~~p^1>0,~~q_1>0,~~q_0>0.
	\end{aligned}
\end{equation}
In this work, we will talk about the bound state only in this domain. For the charges in this domain, the 4d solution when uplifted to 5d is certainly healthy. We propose this is the physical domain of the uplifted black ring solution (as far as its contribution to the GV formula is concerned). Keeping in mind the identification $q_0=2 j_{\L}$, based on the Cardy type argument, in  \cite{Cyrier:2004hj} it is advocated that in addition, one should take into account the zero point shift producing 
\begin{equation}\label{entropy1}
	\begin{aligned}
		& S= 2\pi \( \frac{(k p^3+c_2 p)
}{6} \(-q_0+ \frac{1}{2kp}q^2 +\frac{(k p^3+c_2 p)
}{24} \) \)^{\frac{1}{2}}.
	\end{aligned}
\end{equation}

\subsection{Loop corrections}
\label{puzzle}

We pause for a moment to specialize to the background that is suitable for the 4d/5d lift of the BMPV black hole 
\begin{equation}
	p^A=0, ~~ p^0=1, ~~ \phi^0=\frac{\omega}{2}, ~~ q_0=2 j_{\L}.
\end{equation}
This gives the following mapping of parameters
\begin{equation}
	t^A=\frac{2\phi^A}{\omega-2\pi i},~~ g_{\mathrm{top}}=-\frac{8\pi^2}{\omega-2\pi i}.
\end{equation}
The chemical potential $\omega$ is conjugate to $j_{\L}$ (and $\phi^A$ is conjugate to $q_A$). Note that the topological string coupling is not small in this setting for small $\omega$
\begin{equation}\label{strong}
	g_{\mathrm{top}}(\omega=0)=-4\pi i, ~~ \implies |g_{\mathrm{top}}(\omega=0)|>1.
\end{equation}
 Therefore to discuss slowly rotating BMPV black holes we need to sum over all order contributions of the topological strings amplitudes. In particular, we need to resum (\ref{topKK}). 
 
 Now we explain this from a more physical point of view as follows. We focus on the case $\omega=0$  and make some order of magnitude estimate using the attractor equations (\ref{BHsp}), (\ref{moduli}). Before we do that we want to restore the units for this section from M theory units to string units (see appendix \ref{AA} for more details). Note that so far we have been working in M theory units in which we have set
 \begin{equation}
	l_{11}=g_{s}^{\frac{1}{3}} l_s=1.
\end{equation}
 Say $L$ is the typical length scale of the Calabi-Yau, i.e.\ $V_{IIA}=L^6$. Therefore after putting back the units, we get 
 \begin{equation}
e^\sigma = v e^\gamma =\frac{L^2}{l_{s}^2}.
\end{equation}
 Near the horizon to the leading order in large D2 brane charge (from the solution of saddle point equations) we get 
 \begin{equation}
	\frac{L^2}{l_s^2}\approx \Im{(t)} \approx \phi \approx q^{\frac{1}{2}}.
\end{equation}
In 4d, the radius of the horizon  $r_H$ of the black hole when  $\omega=0$ is determined by\footnote{We thank Xi Yin for pointing this out, and Andrew Strominger for an insightful discussion.}
 \begin{equation}\label{rbh}
	\frac{1}{l_4^2} \approx \frac{1}{g_{s}^2 \ l_s^8}L(\infty)^6 \approx \frac{1}{g_{s}^2 \ l_s^8}(L e^{-\frac{\phi_d}{3}})^6 , ~~ \frac{r^2_{\mathrm{BH}}}{l_4^2} \approx q^{\frac{3}{2}} \implies r_{\mathrm{BH}} \approx l_s \ g_{s}e^{\phi_d} =R_M.
\end{equation}
 The black hole has a very tiny horizon---it is of the same order as the M theory circle! We can summarize this as follows. The black hole has a large horizon in 4d Planck units but since the volume of the Calabi-Yau in string units is fixed by the attractor mechanism to scale with the entropy, from 10d units the radius of the horizon is too small. Therefore it is highly quantum. 
 
 The contributions in (\ref{topKK}) are precisely the Wald entropy coming from higher derivative F-terms in the 4d Wilsonian effective action whose bosonic part contains $$  R_-^2 W_-^{2g-2}.$$ Here $R_-$ is the anti-self dual curvature, and $W_-$ is the anti-self dual gravi-photon field strength.   Now we want to estimate the order of magnitude of this term entirely from the 10d IIA perspective when  $\omega=0$, i.e.\ for the D6-D2 system.  The attractor value of the gravi-photon field strength is given by \cite{LopesCardoso:1999cv}\footnote{We used the following mapping of fields: 
 \begin{equation}
     \begin{aligned}
         C(\text{here})=-2i \bar{Z}e^{\frac{K}{2} } (\text{there}), ~~ W_-^2(\text{here})=\hat{A}(\text{there}).
     \end{aligned}
 \end{equation}
 }
   \begin{equation}
	\begin{aligned}
	W_-^2=\frac{64}{l_4^2Q^2} \approx \frac{1}{r_{\mathrm{BH}}^2}\approx \frac{1}{l_s^2 (g_s e^{\phi_d})^2} , ~~ S_{\mathrm{BH}}= \pi |Q|^2 \approx q^{\frac{3}{2}}.
	\end{aligned}
\end{equation}
 Here we used the result in (\ref{rbh}). Since from 10d IIA perspective, the term  $$  R_-^2 W_-^{2g-2}$$ originates exactly in genus $g$, as a result, we note that the effective dilaton dependence of the term is independent of the genus $g$. 
 
 We conclude that it is necessary to resum $Z_{GW}^0$ and re-expand around (\ref{strong}). A resummation is proposed in appendix~A of \cite{Dabholkar:2005dt}, assuming a certain choice of the contour (at this point without further physical inputs we do not have any justification for any particular choice) in terms of the MacMahon function:
 \begin{equation}
	f(\lambda)=\sum_{n=1}^{\infty} n \log(1-e^{i n \lambda}), ~~ \lambda = -i g_{\mathrm{top}}.
\end{equation}
Note that the value (\ref{strong}) is on the boundary of the domain of convergence and therefore there might be new non-analytic terms in $g_{\mathrm{top}}$ in this limit. We will leave a careful analysis of such questions to future work.

 Similar conclusions apply to the slowly rotating EEMR/BW/GG black ring.

\subsection{Gaiotto-Strominger-Yin uplift to 5d}\label{4d/5d}

In this subsection, we will compare the result for the entropy obtained on the macroscopic side around the large $d$ limit with $c_2$ corrections, and then compare it to the microscopic answer obtained through the GV invariants. This process has two limitations.  The first is that it is valid for large $d$, but here we are in a range of $d$ that is comparable to geometric data of the Calabi-Yau.  Secondly, as we have discussed extensively before, at subleading orders in $1/d$, there are an infinite number of quantum corrections that are naively large term by term (we do not know their resummed contribution at this stage).  Despite these issues, we will show that in four different Calabi-Yau compactifications, our large $d$ results are very close to the microscopic index.

The D6-D2-D0 black hole in 4d uplifts to BMPV black hole \cite{Breckenridge:1996is} solution in 5d---the D0 brane charge becomes the angular momentum and D2 brane charge becomes the M2 brane charge in 5d. 
\begin{equation}
	q_0=2j_{\L}, ~~ d_A=q_A.
\end{equation}
With $c_2$ corrections the entropy of BMPV black hole for one parameter Calabi-Yau takes the form given in (\ref{entropyD6D2D0})\footnote{At this point, it is natural to question how reasonable it is to keep the $\frac{1}{d^2}$ correction while ignoring other corrections of the form (at $j_{\L}=0$) $$-\log \(\frac{2}{9k}\)^{\frac{1}{3}}d$$  as expected from the 5d analysis of \cite{Sen:2012cj}. One way to justify this for the quintic is to note that we are interested in the range where $d \approx 20, c_2=50, k=5$, and the numerical value of the $\log$ term is smaller. As we mentioned previously, other $\mathcal{O}(d^0)$ contributions from degenerate instantons are also not understood for the BMPV black hole.}
\begin{equation}\label{BHentropy}
	\begin{aligned}
		S^{\mathrm{BH}}_{\mathrm{OSV}}=   2\pi \(\frac{2}{9k}\)^{\frac{1}{2}}d^{\frac{3}{2}} \(1-\frac{9}{32}y^2 \)^{\frac{1}{2}} & \bigg(1+ \frac{c_2}{4 d}- \frac{1}{2}\(\ \frac{c_2}{4d}\)^2\bigg),
		~~
		y := \frac{4k^{\frac{1}{2}}}{d^{\frac{3}{2}}}j_{\L}.
			\end{aligned}
\end{equation}
The expression above comes from 4d OSV type formalism discussed previously. On the other hand, in 5d supergravity superspace formalism is not well understood and a clear understanding of supersymmetry-protected terms in supergravity is not known presently. However, one can still consider certain $R^2$ corrections in five dimensions producing a different formula for the correction to the black hole entropy \cite{Cyrier:2004hj, Castro:2007ci}
\begin{equation}\label{BHentropySUGRA}
	\begin{aligned}
		S^{\mathrm{BH}}_{\mathrm{SUGRA}}=   2\pi \(\frac{2}{9k}\)^{\frac{1}{2}}d^{\frac{3}{2}} \(1-\frac{9}{32}y^2 \)^{\frac{1}{2}} & \bigg(1+\frac{3c_2}{16 d}+\frac{9 c_2 y^2}{512 d}\bigg)
	\end{aligned}
\end{equation}
In our plots we will compare both these formulas with the microscopic calculation. 

The multi-centered bound state of D6 and D4-D2-D0 black hole in 4d uplifts to the EEMR/BW/GG black ring in 5d for which the mapping of the charges is a little different
\begin{equation}
	q_0=2j_{\L}, ~~ d_A=q_A+ \frac{C_{ABC}}{2} p^B p^C.
\end{equation}
The entropy of the EEMR/BW/GG black ring solution \cite{Elvang:2004rt, Elvang:2004ds, Bena:2004de, Gauntlett:2004qy} for one parameter Calabi-Yau in terms of 5d parameters (\ref{entropy1}) takes the following form 
\begin{equation}
	\label{BRentropy}
	\begin{aligned}
		& \hspace{-.05in} S^{\mathrm{BR}}_{\mathrm{OSV}} = 2\pi \(\frac{2}{9k}\)^{\frac{1}{2}} {d}^{\frac{3}{2}} \bigg[ \frac{\left( x^2+\frac{c_2}{d}\right) \left(x^4-3 x^2-3 x y+3+\frac{c_2}{4d} x^2 \right)}{8} \bigg]^{\frac{1}{2}},
		~ x := \frac{k^{\frac{1}{2}}}{d^{\frac{1}{2}}} p.
	\end{aligned}
\end{equation}
Similar to the discussion of the black hole, from 5d supergravity one gets a different formula for the correction to the black ring entropy \cite{Bena:2005ae, Castro:2007ci} - essentially equivalent to dropping the zero point shift in (\ref{entropy1}) 
\begin{equation}
	\label{BRentropySUGRA}
	\begin{aligned}
		& \hspace{-.05in} S^{\mathrm{BR}}_{\mathrm{SUGRA}} = 2\pi \(\frac{2}{9k}\)^{\frac{1}{2}} {d}^{\frac{3}{2}} \bigg[ \frac{\left( x^2+\frac{c_2}{d}\right) \left(\frac{3}{4}x^4-3 x^2-3 x y+3 \right)}{8} \bigg]^{\frac{1}{2}},
		~ x := \frac{k^{\frac{1}{2}}}{d^{\frac{1}{2}}} p.
	\end{aligned}
\end{equation}

Our microscopic calculation does not keep track of the M5 brane parameter $p$. For $d$ as large as the currently available GV data, extremizing the black ring entropy \eqref{BRentropy} over (positive and continuous) $p$ for individual values of $j_{\L}$ results in small values of $p$ especially when $j_{\L}$ is large. Therefore, this procedure, while reasonable for much larger $d$, does not provide a meaningful comparison to the available GV data. We need some additional rules for fixing this parameter.
\begin{figure}[ht]
	\centering
	\includegraphics[width=0.475\textwidth]{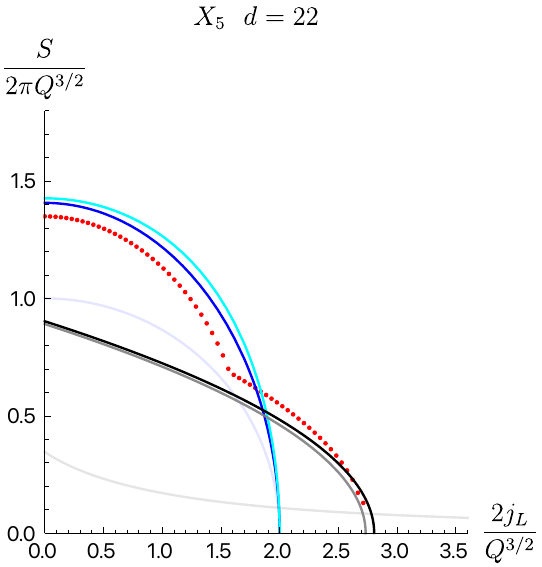}
	~
	\includegraphics[width=0.475\textwidth]{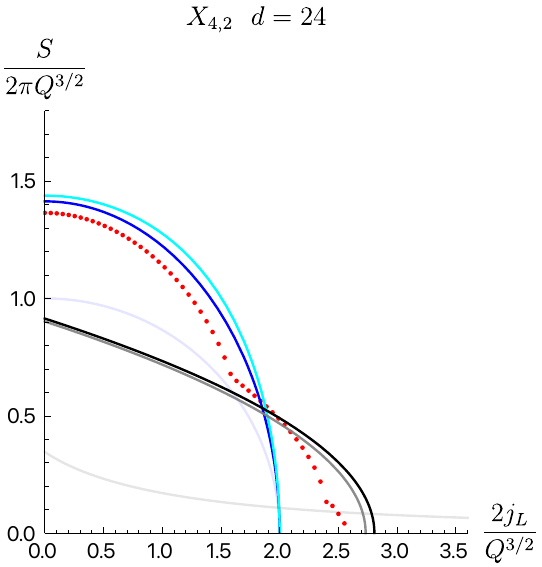}
	\\
	~
	\\
	\includegraphics[width=0.475\textwidth]{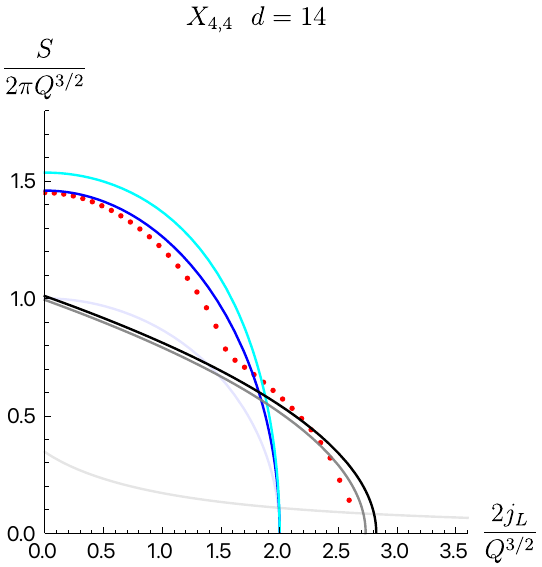}
	~
	\includegraphics[width=0.475\textwidth]{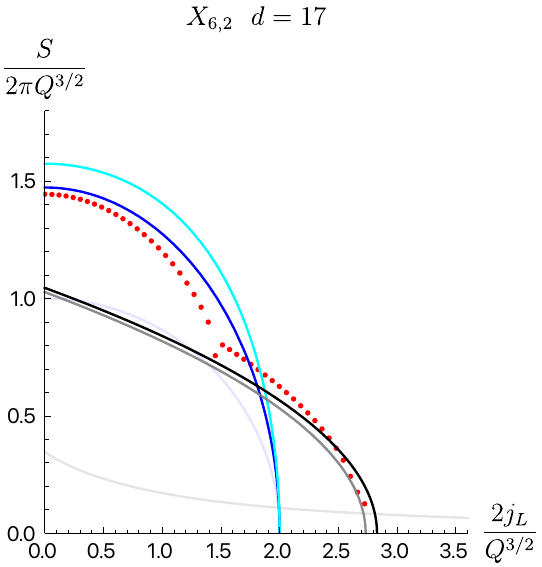}
	\\
	~
	\vspace{-0.1in}
	\\
	\caption{Entropy over $\SU(2)_{\L}$ angular momentum $j_{\L}$ for several one-parameter CY$_3$. {\bf Red:} $\log|\Omega|$ where $\Omega$ is the BPS index computed from the Gopakumar-Vafa invariants with the specified M2 brane charge $d$ 
	{\bf Blue:} black hole entropy obtained from 4d OSV given by \eqref{BHentropy} 
 {\bf Cyan:} black hole entropy obtained from 5d supergravity given by \eqref{BHentropySUGRA} {\bf Black:} black ring entropy obtained from 4d OSV \eqref{BRentropy} for a fixed M5 brane parameter (extremized at the critical angular momentum) {\bf Grey:} black ring entropy obtained from 5d supergravity \eqref{BRentropySUGRA} for a fixed M5 brane parameter (extremized at the critical angular momentum) {\bf Faded: }Bekenstein-Hawking entropy plotted before in figure \ref{fgv}.}
  \label{fig5}
  \end{figure}
A natural proposal might be the following: 
For a given $d$ from the plot of the Gopakumar-Vafa curve one can see that there is a critical angular momentum $j_{\L,c}(d)$ at which there is a sharp transition (studied in detail for quintic in the previous section). We will maximize the leading order (in large $d$, i.e.\ $c_2\to 0$) entropy of the black ring at $j_{\L}=j_{\L,c}(d)$ subjected to (\ref{BR sign}) to obtain $p=p_c(d)$. 
While comparing to the microscopic calculation, we will add $c_2$ corrections using (\ref{BRentropy}) with $p=p_c(d)$ fixed.  For $X_5, X_{4,2}, X_{4,4}, X_{6,2}$ we compare resulting plots in figure~\ref{fig5}
below for the maximal available value of $d$, and find that the numerical curves are close to each other. However, for the small range of $d$ that we can discuss, the extremized value of $p$ is pretty small---near or smaller than one. This suggests that the discussion here in terms of supergravity is at best an extrapolation.  To take this fact into account, in the next section, we discuss the comparison of the GV calculation with the MSW index which captures the microstates of the black ring, and discuss why the contribution to the tail of the GV curve might arise solely from bound states based on a mathematical conjecture.

\section{Microscopic observations and a quantum transition}
\label{Sec:Micro}

Whereas 5d BPS states in Calabi-Yau compactification of M theory are counted by the Gopakumar-Vafa invariants, 4d BPS states in that of IIA (in the large $B$ field chamber) are counted by the rank-one Donaldson-Thomas (DT) invariants \cite{Thomas:1998uj,Donaldson:1996kp}.  The latter in the case of unit D6 charge is related to the former by a mathematical conjecture \cite{Maulik:2003rzb,Maulik:2004txy} that was motivated by \cite{Iqbal:2003ds} and physically justified by \cite{Dijkgraaf:2006um} using the GSY 4d/5d lift \cite{Gaiotto:2005gf}, as will be reviewed in section~\ref{Sec:Gas}.
It was argued in \cite{Cyrier:2004hj} that a microscopic description of 5d black ring microstates 
is provided by the Maldacena-Strominger-Witten (MSW) CFT \cite{Maldacena:1997de}, which arises in the infrared limit of the worldvolume theory of M5 branes wrapping a 4-cycle (ample divisor) of the Calabi-Yau.  
Therefore, the comparison of the tail in the 5d entropy with the black ring entropy becomes, at the microscopic level, the comparison of the GV invariants with the MSW indices.
In section~\ref{Sec:MSW}, by invoking the 4d/5d lift and wall-crossing, we propose an approximate relation \eqref{Approx} between the two with the latter having unit D4 charge,
and find strong numerical evidence for the proposal.
In section~\ref{Sec:BMT}, we observe that the mathematical Bayer-Macr\`i-Toda (BMT) inequality \cite{bayer2011bridgeland,bayer2016space}, a necessary condition for the stability (with respect to a reduced central charge function) of 4d BPS states everywhere in the K\"ahler moduli space, coincides under the 4d/5d lift with the extremality bound of 5d BMPV black holes.\footnote{Given the GSY correspondence \cite{Gaiotto:2005xt} between 5d and 4d solutions \cite{Shmakova:1996nz}, the 5d black hole extremality bound is equivalent to a 4d bound on the D2 and D0 charges for the existence of single-centered macroscopic black hole solutions  (with unit D6 charge) to the attractor equations \cite{Strominger:1996kf, Ferrara:1995ih}. }

As before, we consider 
one-parameter ($h^{1,1} = 1$) Calabi-Yau threefolds $X$ whose divisor 
has self-intersection number $\k \in \bZ$ and second Chern class $c_2$.

\subsection{Free gas of M2 branes and the DT/GV relation}
\label{Sec:Gas}

We first review an exact relation between the 4d and 5d BPS counting as well as the physical interpretation, setting the stage for later subsections. 

\paragraph{DT/GV relation}

Given the GSY 4d/5d lift \cite{Gaiotto:2005gf}, one may expect that the 4d counting of DT invariants is related to the 5d counting of GV invariants in a deterministic fashion.
Indeed, such a relation 
was conjectured \cite{
Maulik:2003rzb, Maulik:2004txy} even before the GSY discovery, and can be stated as follows.\footnote{Mathematicians formulated a conjectural relation between Donaldson-Thomas and Gromov-Witten invariants, the latter of which can be transformed into the Gopakumar-Vafa invariants.}
The generating function of DT invariants
\ie 
	Z_\DT(y, q) := \sum_{Q,m} \DT(Q,m) y^Q q^m
\fe 
admits an expression in terms of the GV invariants $n_d^g$ as 
\ie\label{DT/GV}
	Z_\DT(y, q) &= M(-q)^{\chi} \prod_{d>0} \prod_{k>0} (1-(-q)^k y^d)^{k n_d^0}
	\\
	& \times \prod_{d>0} \prod_{g>0} \prod_{\ell=0}^{2g-2} (1-(-q)^{g-\ell-1} y^d)^{(-1)^{g+\ell} 
	\binom{2g-2}{\ell}
	n_d^g},
\fe 
where $\textstyle M(q) := \prod_{n>0} (1-q^n)^{-n}$ is the MacMahon function, and $\chi$ is the Euler characteristic of $X$.
The formula \eqref{DT/GV} admits a rewriting as\footnote{
	The second line of \eqref{DT/GV} can be rewritten through combinatoric identities as
	\ie
		\prod_{d>0} \prod_{\ell} (1-(-q)^\ell y^d)^{c_d^\ell}
		= 
		\prod_{d>0} \prod_m \prod_{n>0} (1-(-q)^{m+n} y^d)^{n \tilde N_d^m},
	\fe 
	where
	\ie 
		c_d^\ell &:= - (-1)^\ell \sum_{g>|\ell|}
		\binom{2g-2}{g-\ell-1} n_d^g = \sum_{n=1}^\infty n \tilde N_d^{|\ell|-n},
		\quad
		\tilde N_d^m := (-1)^m \sum_{g>0} \binom{2g}{g-m} n_d^g,
	\fe 
	with symmetries $c_d^\ell = c_d^{-\ell}$ and $\tilde N_d^m = \tilde N_d^{-m}$.
}
\ie\label{Gas}
	Z_\DT(y, q)
	&= 
	M(-q)^{\chi} \prod_{d>0} \prod_m \prod_{n>0} (1-(-q)^{m+n} y^d)^{n N_d^m}
\fe 
with
\ie\label{GasN}
	N_d^m &:= (-1)^m \sum_{g\ge0} \binom{2g}{g-m} n_d^g,
\fe 
which has the symmetry $N_d^m = N_d^{-m}$.  The MacMahon factor can also be absorbed into the triple-product formula 
\ie\label{Gas}
	Z_\DT(y, q)
	&= 
	\prod_{d\ge0} \prod_m \prod_{n>0} (1-(-q)^{m+n} y^d)^{n N_d^m}
\fe 
with $n_0^g = - \chi \delta_0^g$ (and hence $N_0^m = - \chi \delta_0^m$).

\paragraph{Free M2 gas}

It was suggested in \cite{Dijkgraaf:2006um} that the DT/GV relation \eqref{DT/GV} or \eqref{Gas} can be physically justified by a free M2 gas picture,
the basic idea of which is the following.  Suppose one already knows the full collection $\cC$ of 5d single-particle M2-brane bound states on flat spatial $\bR^4$, where `single-particle' (`irreducible' in the words of \cite{Dijkgraaf:2006um}) means a state with a normalizable wave function as opposed to a multi-particle scattering state which is $\delta$-function normalizable.  Now, instead of $\bR^4$ let us consider spatial Taub-NUT, which is a circle fibration over $\bR^3$ such that the circle shrinks at the origin and asymptotes to a fixed radius $R_{TN}$ at infinity.  Near the origin, the spatial section of Taub-NUT looks like $\bR^4$, so we can imagine putting inside it particles belonging to $\cC$.  Since Taub-NUT with large flux acts like a box preventing the particles from forming scattering states (the importance of large flux was later pointed out in \cite{Denef:2007vg, Aganagic:2009kf}), if we put multiple particles inside, we expect them to form normalizable `multi-particle' bound states which become scattering states in the $R_{TN} \to \infty$ limit as the Taub-NUT circle opens up.  

While the actual (unsigned) degeneracies have a complicated dependence on the Taub-NUT circle size $R_{TN}$,
supersymmetric indices are protected and have no $R_{TN}$ dependence.
In the small $R_{TN}$ limit, we get the index of 4d D6-D2-D0 bound states with unit D6 charge, whereas in the large $R_{TN}$ limit, we get the index of a free gas of M2-branes, counting all subsets of $\cC$.  
Since the indices of 4d BPS states in the chamber of large $B$ field give the DT invariants, and the indices of $\cC$ give the GV invariants, it is intuitively clear that DT is a plethystic exponential of GV.
The precise relation is \eqref{Gas}, where $m+1$ and $n-1$ are interpreted as the intrinsic and orbital angular momenta, respectively, of each 5d BPS particle \cite{Dijkgraaf:2006um}.

\paragraph{Quantum GSY lift}

According to the free gas picture \cite{Dijkgraaf:2006um}, when series expanding the DT/GV relation \eqref{Gas}, the terms with coefficients linear in $N_d^m$ give the indices of single-particle states, i.e.\  
\ie\label{DTSingle}
	Z_\DT(y, q) = - \sum_{d>0} \sum_m \sum_{n>0} {n N_d^m} (-q)^{m+n} y^d
	+ \text{(multi-particles)}.
\fe 
Moreover, the terms with $n=1$ count the states with no orbital angular momenta,
\ie\label{DTSingleNoOrbital}
	Z_\DT(y, q) = - \sum_{d>0} \sum_m N_d^m (-q)^{m+1} y^d + \text{(multi-particles)} + \text{(orbital)}.
\fe 
For negative values of $m$, this reads
\ie\label{OmegaDT}
	\DT(Q,m) &= (-1)^{m-1} N_Q^{m-1} + \text{(multi-particles)} + \text{(orbital)}
	\\
	&= n_Q^{1-m} + \text{(multi-particles)} + \text{(orbital)} + \text{(intrinsic)}.
\fe 
We will also be interested in the 5d index defined in \eqref{5ddegen}, which can be related to the above by
\ie\label{OmegaN}
	\Omega(j_{\L}, d) = (-1)^{2j_{\L}-1} N_d^{2j_{\L}-1} + \text{(5d supersymmetry descendants)}.
\fe

\paragraph{5d black hole extremality bound}

A key role in the following subsections is played by the extremality bound of a 5d BMPV black hole \cite{Breckenridge:1996is}.  Let us recall this bound, and translate it to 4d single-particle language.
The extremality bound on $\Omega(j_{\L}, d)$ can be read off from \eqref{BHsugra}, giving
\ie 
	(2j_{\L})^2 \le \frac{8d^3}{9k}.
\fe
In terms of the counting of 5d supersymmetry multiplets by $N_d^m$, this translates via \eqref{OmegaN} to
\ie 
	(m+1)^2 \le \frac{8d^3}{9k}.
\fe 
And in terms of the 4d counting by $\DT(Q,m)$, this translates via \eqref{OmegaDT} to
\ie\label{DTExtremalBound}
	(-m+2)^2 \le \frac{8Q^3}{9k}.
\fe

\subsection{Black ring microstates with unit M5 charge}
\label{Sec:MSW}

Black rings enter the story as their microscopic index is related to DT via wall-crossing, which can then be lifted to a relation with GV.  Of course, underlying this somewhat roundabout argument is the physical expectation that the GV tail counts black ring microstates.

\paragraph{Black ring microstates and the MSW CFT}

In \cite{Cyrier:2004hj}, the authors constructed 5d BPS rotating cosmic closed strings by wrapping M5 branes on four-cycles (ample divisor) of Calabi-Yau threefolds.  On the one hand, the macroscopic limit of this cosmic string becomes the EEMR/BW/GG black ring \cite{Elvang:2004rt, Elvang:2004ds, Bena:2004de, Gauntlett:2004qy}; on the other hand, the low energy limit of the M5 brane worldvolume theory is the Maldacena-Strominger-Witten (MSW) CFT \cite{Maldacena:1997de}.
Hence, the MSW CFT describes the microscopic constituents of a black ring.

Originally, the MSW CFT was formulated to describe D4-D2-D0 bound states in Calabi-Yau compactifications of IIA string theory.  In particular, the generating function of indices is (encoded in) the modified index of the MSW CFT.   The relation between the D4-D2-D0 context and the black ring context was later elucidated by the GSY 4d/5d lift \cite{Gaiotto:2005gf, Gaiotto:2005xt}: by adding a single D6 brane and considering two-centered bound states, with the D6 at one center and the D4-D2-D0 at the other, the resulting 4d configuration lifts to a 5d black ring.

\paragraph{Wall-crossing and a GV/MSW proposal}

Suppose there is a wall of marginal stability for a D6-D2-D0 bound state (counted by a rank-one DT invariant) to decay into a single D6 brane and a D4-D2-D0 bound state, where the extra D4 dissolves into background flux that shifts the asymptotic D2 and D0 charges.
Let $\MSW_{p,Q}(q_0)$ denote the MSW index with charge vector $\gamma = (0, p, Q, q_0)$.
For unit D4 charge $p=1$, the wall-crossing formula \cite{Joyce:2008pc,Kontsevich:2008fj} gives \cite{Alexandrov:2023zjb}\footnote{
	 Instead of DT, the wall-crossing analysis of \cite{Alexandrov:2023zjb} considers stable-pair invariants \cite{Pandharipande:2007qu}, also known as Pandharipande-Thomas (PT) invariants, whose generating function is related to that of DT by stripping of a MacMahon factor $M(-q)^{\chi}$.  As explained in \cite{Denef:2007vg}, the PT invariants count 4d BPS bound states in the chamber of large $B + iJ$ with $\arctan(J/B) = (2\pi/3)^+$, such that the D0-halos become unbound.  One could thus regard \eqref{DTMSW} as first crossing the wall at $\arctan(J/B) = 2\pi/3$ and then proceeding as in \cite{Alexandrov:2023zjb}.
}
\ie\label{DTMSW}
	\DT(Q,m) &= (-1)^{\chi(Q,m)} \chi(Q,m) \MSW_{1,Q}(\hat q_0(Q,m)) + \dotsb,
	\\
	\chi(Q,m) &:= m+Q-(\frac{k}{6}+\frac{c_2}{12}),
	\\
	\hat q_0(Q,m) &:= -m-\frac{Q^2}{2k}-\frac{Q}{2}+\frac{k+c_2}{24}.
\fe 
According to \cite{Alexandrov:2023zjb}, when $k \mid d$ and at the maximal value of $j_{\L}$ (saturating the Castelnuovo bound), the above equation is exact with no additional $\cdots$ corrections.
For notational convenience let us define
\ie 
	\MSW{'}(Q,m) := (-1)^{\chi(Q,m)} \chi(Q,m) \MSW_{1,Q}(\hat q_0(Q,m)).
\fe 
We now make two assumptions, whose regime of validity will be empirically justified:
\begin{enumerate}
	\item The contributions from multi-particle states, $\SU(2)_{\L}$ descendants, and 5d supersymmetry descendants are subleading in \eqref{OmegaDT} and \eqref{OmegaN}.
	\item The wall-crossing contribution of a single D6 and D4-D2-D0 with unit D4 charge exists and dominates $\DT(Q,m)$ in \eqref{DTMSW}.
\end{enumerate}
Then combined with \eqref{OmegaDT} and \eqref{OmegaN}, we would have an approximate relation in the context of 5d counting\footnote{Here, $A \sim B$ means that $A/B \to 1$ in the $d \to \infty$ limit with $j_L/d^{3/2}$ held fixed.
}
\ie\label{Approx}
	& \Omega(j_{\L},d) \sim (-1)^{2j_{\L}-1} N_d^{2j_{\L}-1} \sim \MSW{'}(d, 2j_{\L}-2).
\fe

\paragraph{Numerical evidence}

\begin{figure}[h]
	\centering
	\includegraphics[width=0.475\textwidth]{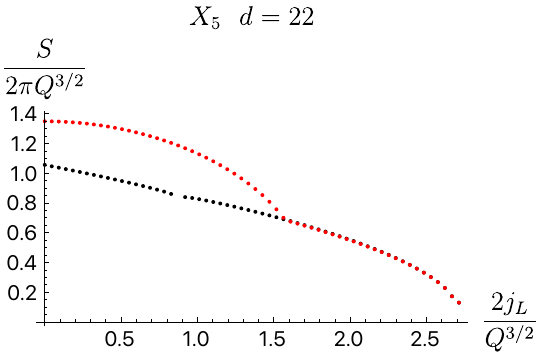} \quad
	\includegraphics[width=0.475\textwidth]{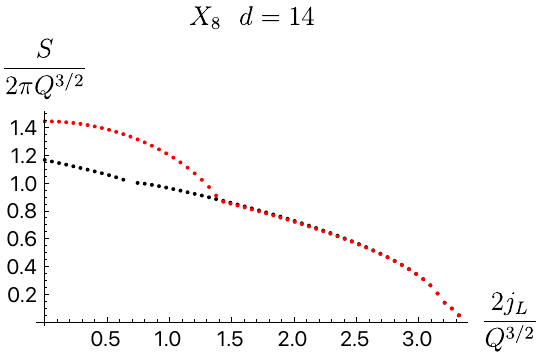}
	\\
	~
	\\
	\includegraphics[width=0.475\textwidth]{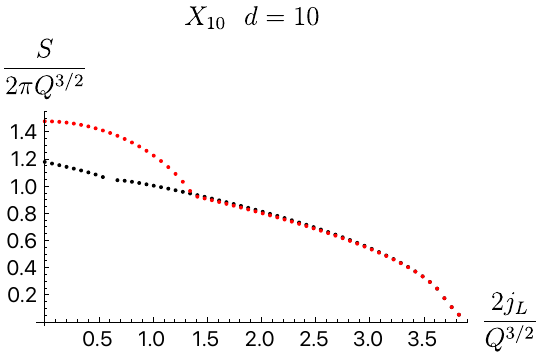} \quad
	\includegraphics[width=0.475\textwidth]{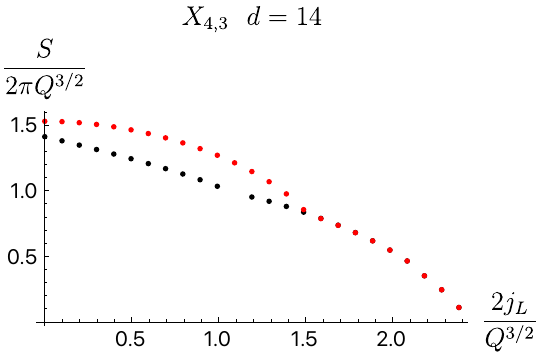}
	\\
	~
	\\
	\includegraphics[width=0.475\textwidth]{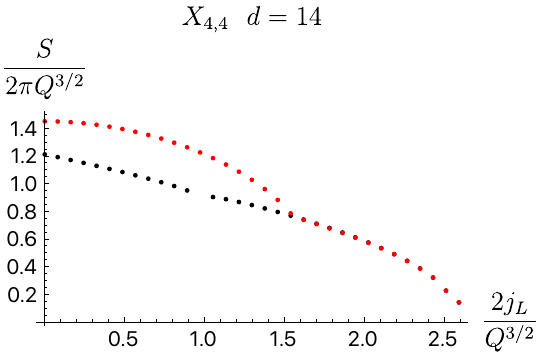} \quad
	\includegraphics[width=0.475\textwidth]{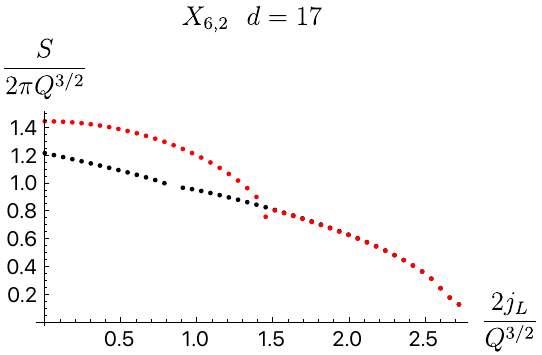}
	\\
	~
	\\
	\includegraphics[width=0.475\textwidth]{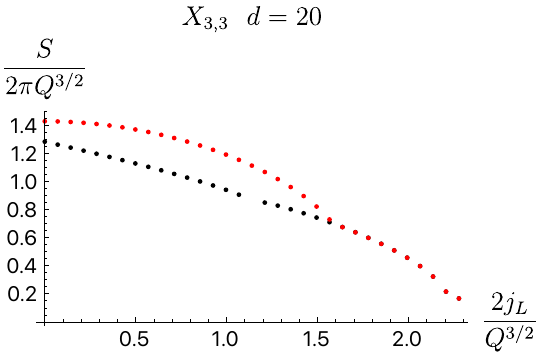} \quad
	\includegraphics[width=0.475\textwidth]{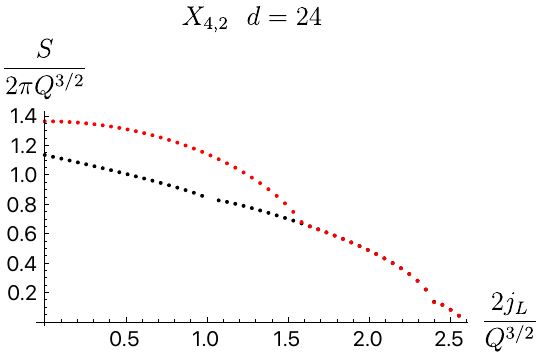}
	\\
	~
	\vspace{-0.1in}
	\\
	\caption{Juxtaposition of the 5d index entropy $S = \log|\Omega(j_{\L}, d)|$ in red with the Maldacena-Strominger-Witten entropy $S = \log|\MSW{'}(d, 2j_{\L}-2)|$ in black.}
	\label{Fig:MSW}
\end{figure}

In figure~\ref{Fig:MSW}, we compare the 5d index entropy 
$\log|\Omega(j_{\L}, d)|$, which is related to the GV invariants by \eqref{5ddegen}, with the Maldacena-Strominger-Witten entropy $\log|\MSW{'}(d, 2j_{\L}-2)|$ for several one-parameter Calabi-Yau threefolds, using the data and expressions of \cite{Alexandrov:2022pgd, Alexandrov:2023zjb, Web}. 
We find surprisingly good agreement for large enough angular momentum $j_{\L}$ even for small $d$.\footnote{The `missing' black dot in each plot is due to the vanishing of $\chi(Q,m)$.}  Notice that for $X_{42}$, even the kink at the far right matches.\footnote{
	 The comparison of $\log|N_d^m|$ with $\log|\MSW'(d,m-1)|$ yields a similar degree of agreement, while the comparison of $\log|\mathrm{PT}(Q,m)|$ with $\log|\MSW'(Q,-m)|$ seems to be even better.
}

\newpage 

\newpage

It would be interesting to understand how to account for the difference between $\Omega(j_{\L}, d)$ and $\MSW{'}(d, 2j_{\L}-2)$ in the tail region.  The 4d interpretation of this difference comes from crossing other walls \cite{Alexandrov:2023zjb}, the analysis of which is notoriously complicated, hence we hope that 5d microstate counting might offer a useful complementary perspective.  For instance, there can be contributions from black rings with non-unit D4 charges, but the MSW index in this case is not known (see \cite{Alexandrov:2022pgd, Alexandrov:2023zjb} for the current status).

\subsection{4d instability of 5d superextremal states
}
\label{Sec:BMT}

In this subsection, we make the observation that the GV invariants counting 5d BPS states with angular momenta beyond the 5d black hole extremality bound give rise to 4d BPS states that are stable in the large $B+iJ$ limit but become unbound in a chamber of the K\"ahler moduli space with respect to a `reduced' central charge function corresponding to what is known as a (very) weak stability condition.  
Although this stability condition is not the physical one away from large $J$, we argue in the end the conclusion should be valid in the large charge limit.

In the context of the black hole/black ring transition, we view this observation as a consistency check for the proposal that superextremal GV invariants count 5d black ring microstates.
Under the GSY 4d/5d correspondence \cite{Gaiotto:2005gf, Gaiotto:2005xt}, black rings descend to 4d two-centered bound states that can disassociate in certain chambers, whereas 5d black holes descend to 4d single-centered objects that should be stable everywhere.

\paragraph{Stability and wall-crossing}

The index counting of stable BPS states in Calabi-Yau compactifications of IIA string theory is a rich and intricate subject.  Inside the K\"ahler moduli space, the index remains constant in domains (chambers) but jumps across walls of marginal stability \cite{Gaiotto:2010okc, Joyce:2008pc, Kontsevich:2008fj, Gaiotto:2009hg}.  
For unit D6 charge $p^0 = 1$, an important chamber is that of large $J$ and $B$ with $B \gg J$, where BPS states are counted by the rank-one Donaldson-Thomas (DT) invariants \cite{Thomas:1998uj, Donaldson:1996kp}.
Given knowledge of the DT invariants, one can apply the celebrated wall-crossing formulae \cite{Joyce:2008pc, Kontsevich:2008fj} to compute the indices in various other chambers.  Conversely, starting from an empty chamber, one can cross walls to compute the DT invariants.

The stability of a BPS state is defined by a central charge function that maps charge vectors to complex numbers.  The central charge function of IIA on Calabi-Yau is extremely complicated at a generic point on the K\"ahler moduli space (as it depends on the exact prepotential), albeit simplifying in the large $J$ regime.  To circumvent this difficulty, mathematicians study not the physical central charge, but a space of central charge functions whose corresponding stability conditions satisfy a set of axioms formulated by Bridgeland \cite{bridgeland2007stability}.  The physical stability condition, known as $\Pi$-stability, is a particular map (sometimes an embedding) from the K\"ahler moduli space to the space of Bridgeland stability conditions.  Here is a key point: given a simpler `unphysical' map/embedding, and as long as it agrees with the physical one in the large $J$ limit, we can still use wall-crossing to compute the DT 
invariants starting from an empty chamber, as was done in \cite{Alexandrov:2023zjb} to formulate new constraints on GV invariants.  

In the following, we closely follow the discussions and conventions of \cite{Alexandrov:2023zjb}.

\paragraph{Central charge function}

For convenience, we first translate the charge (Mukai) vector $\gamma := (p^0, p^1, q_1, q_0)$ into the Chern characters (of line bundles describing D-brane bound states in the large $J$ limit) as follows:
\ie 
    C_0 = \k p^0, \quad C_1 = \k p^1, \quad C_2 = - q_1 - \frac{c_2}{24} p^0, \quad C_3 = q_0 - \frac{c_2}{24} p^1,
\fe 
and
\ie 
    \ch_0^b &= C_0, \quad \ch_1^b = C_1 - b C_0, \quad \ch_2^b = C_2 - b C_1 + \frac12 b^2 C_0, \quad
    \\
    \ch_3^b &= C_3 - b C_2 + \frac12 b^2 C_1 - \frac16 b^3 C_0.
\fe

\noindent The Bridgeland stability conditions are defined by a 4-parameter family of central charge functions
\ie 
    Z_{a,b,\alpha,\beta} := (- \ch_3^b + \beta \ch_2^b + \alpha \ch_1^b) + i (a \ch_2^b - \frac12 a^3 \ch_0^b),
\fe 
with the following important special cases:
\begin{itemize}
	\item The large $J$ limit of the physical central charge is given by (with $J^0$ and $e^{-J}$ corrections)
	\ie 
		a = \sqrt{\frac{J^2}{3} - \frac{c_2}{12k}}, \quad b = B, \quad \alpha = \frac{J^2}{2} - \frac{c_2}{24k}, \quad \beta = 0.
	\fe 
	\item A weak stability condition (which is located at the boundary of the space Bridgeland stability conditions) is defined with respect to a reduced central charge function
	\ie 
		Z_{a,b} := i Z_{a,b=0,\alpha=a^2,\beta=0} = - a \ch_2^b + \frac12 a^3 \ch_0^b + i a^2 \ch_1^b,
	\fe
	independent of the D0 charge $q_0$.
	A crucial point is the following: With the identification $a = J/\sqrt{3}$ and $b = B$, the weak stability condition agrees with the physical one in the large $J$ limit.
	In the following it is convenient to define $w := \tfrac12 (a^2+b^2)$; the meaningful region is $w \ge \tfrac12 b^2$.
\end{itemize}

\paragraph{Bayer-Macr\`i-Toda conjecture}

The Bayer-Macr\`i-Toda (BMT) conjecture \cite{bayer2011bridgeland,bayer2016space}, which has been proven by \cite{li2019stability} for the quintic threefold, states that
\ie\label{BMT}
    L_{b,w} := (C_1^2-2C_0C_2)w + (3C_0C_3-C_1C_2)b + (2C_2^2-3C_1C_3) \ge 0
\fe 
is a necessary condition for there to exist a weakly-semistable object of class $\gamma$.
In other words, for a fixed charge vector $\gamma$, the region on the side of the line $L_{b,w}$ that violates \eqref{BMT} is in an empty chamber.
It turns out that $C_1^2 - 2 C_0 C_2 > 0$, and hence the BMT inequality \eqref{BMT} is always satisfied for large enough K\"ahler modulus. 
If $L_{b,w}$ intersects $w = \tfrac12 b^2$ at two points, which happens for a positive discriminant
\ie\label{Discriminant}
    \Delta = (3C_0C_3 - C_1C_2)^2 - 2(C_1^2 - 2C_0C_2) (2C_2^2 - 3C_1C_3) > 0,
\fe 
then an empty chamber exists.
We now investigate the existence of empty chambers for BPS states counted by the DT invariants.

\paragraph{Donaldson-Thomas}

The rank-one DT invariants $\DT(Q,m)$ are defined with the charge assignments
\ie
	p^0 = 1, \quad p^1 = 0, \quad Q := q_1 + \frac{c_2}{24}, \quad m := -q_0.
\fe
The BMT line \eqref{BMT} and discriminant \eqref{Discriminant} read
\ie 
    L = -3 b \k  m+2 Q^2+2 \k  Q w,
    \quad 
    \Delta = 9 \k ^2 m^2-8 \k  Q^3,
\fe 
and the BMT conjecture states that an empty chamber
\ie 
    \frac12 b^2 \le w < \frac{3 b \k  m-2 Q^2}{2 \k Q}
\fe
exists when
\ie\label{mBound}
    m^2 > \frac{8Q^3}{9\k} > 0.
\fe

The largest value of $a = J/\sqrt{3}$ in the empty chamber is 
\ie 
    a_{\text{max}} = \frac{1}{2} \sqrt{\frac{9 m^2}{Q^2}-\frac{8 Q}{k}}.
\fe 
In the large $Q$ limit with $m/Q^{\frac32}$ fixed and \eqref{mBound} satisfied, the empty chamber contains a subregion with large $J \sim Q^{\frac12}$, such that the weak stability condition becomes a good approximation of the physical stability condition.  It is therefore reasonable to expect that at least for large charges, the existence of an empty chamber under weak stability implies the existence of an empty chamber under physical stability.

We now observe that the condition \eqref{mBound} for the existence of an empty chamber is, up to a constant shift, the complement of the 5d black hole extremality bound \eqref{DTExtremalBound}.
Hence, the BMT conjecture can be regarded as a microscopic version of the 5d black hole extremality bound, which lends credence to the conjecture's validity.  Conversely, the BMT conjecture provides a consistency check of our proposal that superextremal GV invariants count black ring microstates.

\section{On the contributions of degenerate instantons}
\label{Sec:Summary}

In this work, we noticed that the one-loop corrected Wald entropy for the black hole approximated the microscopic entropy coming from the topological strings well for smaller values of the angular momentum, whereas the individual contributions from higher-genus degenerate instantons were all of the same order, demanding a summation to all orders (see section~\ref{puzzle} for more details). In this section, we look into this puzzle from the viewpoint of a gravitational path integration of supergravity.

For supersymmetric extremal black holes with a near horizon AdS$_2$ factor, the entropy can be calculated by performing path integration over the fields localized in the near horizon region using Sen's formalism \cite{Sen:2008vm}\footnote{For a recent discussion of subtleties involved in the calculation see \cite{Sen:2023dps, Iliesiu:2022onk}.}. In this process,  $Z_{\mathrm{local}}$ comes from the Wald entropy associated with the classical reduction of the 5d (M theory on CY$_3$) Wilsonian effective action to 4d. The sub-leading contributions come from quantum fluctuations. It is tempting to identify the measure factor $\mu$ in the OSV formula as coming from the one-loop fluctuations of the massless fields in 4d \cite{Sen:2012kpz}. This fact has been established by the explicit one-loop localization of supergravity in \cite{Murthy:2015yfa, Jeon:2018kec} following the developments of \cite{Dabholkar:2010uh}. The origin of  $Z^0(g_{\mathrm{top}})$ is much more subtle; note that for higher supersymmetric compactifications ($\mathcal{N}=4,8$ in the 4d language), these contributions are not present \cite{Harvey:1996ir, Dabholkar:2005dt}. It has been observed in \cite{Dedushenko:2014nya} that one-loop contribution from the Kaluza-Klein (KK) modes of a massless charged hypermultiplet in 5d reproduces part of the structure in $Z^0(g_{\mathrm{top}})$. It is natural to wonder if in Taub-NUT space with one unit of D6 charge, a one-loop calculation of massless fields around BMPV black hole in 5d (similar to the ones done for the flat space in \cite{Sen:2012cj})\footnote{The results in \cite{Sen:2012cj} show that the functional form of the non-analytic logarithmic term in the entropy is sensitive to the way the angular momentum $j_{\L}$ is scaled compared to the charge $d$.} can be recast as a calculation of massless 4d modes along with the contribution of the KK modes on the Taub-NUT circle, effectively capturing the contribution of degenerate instantons.\footnote{We are grateful to Xi Yin for mentioning this possibility to us and Michael Green, Sameer Murthy for an extensive discussion on this topic.} In the context of the 4d/5d lift considered here, the KK modes on the Taub-NUT circle would be the KK modes on the M theory circle (for toroidal compactifications, such one-loop effects in M theory have been analyzed in \cite{Green:1997as, Green:1997me}). It would be very interesting to understand why these KK modes are not under control from the 4d point of view.

\bigskip

\section*{Acknowledgements}

We thank Xi Yin for the initial collaboration. We thank  Shiraz Minwalla, Cumrun Vafa, and Xi Yin for valuable feedback on a preliminary version of the manuscript. Also, we would like to thank Yiming Chen, Michael B. Green, Daniel Jafferis, Manki Kim, Neil D. Lambert, Jan Manschot, Samir D. Mathur, Sameer Murthy, Hirosi Ooguri, Fernando Quevedo, Harvey S. Reall, Jorge E. Santos, Andrew Strominger, and Zhi-Zhen Wang for helpful discussions. 
YL is supported by the Simons Collaboration Grant on the Non-Perturbative Bootstrap. IH is supported by the Harvard Quantum Initiative Fellowship. 

\appendix

\section{Length scales and units}\label{AA}

In this Appendix we will explicitly write down all the relevant length scales under compactification (we will not set any length scale to one) without being careful about order one factors. The length of the M theory in asymptotic infinity is 
\begin{equation}
	\begin{aligned}
		R= g_s l_s
	\end{aligned}
\end{equation}
Here $g_s$ is the 10d IIA string coupling (defined to be constant). 10d dilaton $\phi_d$ is defined with the convention that in asymptotic infinity $\phi_d(\infty)=0$. The length of the M theory circle at a generic point would be given by
\begin{equation}
	\begin{aligned}
		R_M= g_s l_s  \ e^{\frac{2}{3}\phi_d} \equiv l_{11} e^{\gamma},
	\end{aligned}
\end{equation}
where $l_{11}$ is the Planck scale in 11d.
The M2 brane tension will be given by
\begin{equation}
	\begin{aligned}
		T_{M2}= \frac{1}{l^3_{11}}=\frac{1}{g_s l_s^3} \implies l_{11}=l_s g_s^{\frac{1}{3}}
	\end{aligned}
\end{equation}
 \begin{equation}
	\begin{aligned}
		v^3=\frac{V_M}{l_{11}^6}
	\end{aligned}
\end{equation}
 On the other hand the IIA volume of the Calabi-Yau $V_{IIA}$ varies over the radial direction as 
  \begin{equation}
	\begin{aligned}
		V_{IIA}=V_M e^{2\phi_d}
	\end{aligned}
\end{equation}
 4d and 5d Planck scales are given by
 \begin{equation}
	\begin{aligned}
	\frac{1}{l^3_{5}}=\frac{1}{l_{11}^9} V_M, ~~ \frac{1}{l^2_{4}}=\frac{1}{l_{11}^9} R V_M \equiv \frac{1}{g_{s,4}^2 l_s^2} 
	\end{aligned}
\end{equation}
This determines the 4d string coupling to be the given just by the volume of $X$ in M theory units. 
\begin{equation}
	\begin{aligned}
		g_{s,4}=\frac{1}{v^{\frac{3}{2} } }
	\end{aligned}
\end{equation}

\section{GV invariants for quintic up to genus 4}\label{aB}

In this appendix, we will present some details (up to genus $4$---we necessarily need to rely on Castelnuovo bound to determine the holomorphic ambiguity for $g>3$) of the calculation of GV invariants for the quintic. For higher genera, the details are extremely complicated and we won't mention that here. The Mathematica file used for calculation up to genus $49$ can be found at \cite{WebUS}.

We will present results only in holomorphic limit at respective points on the moduli space.

\begin{dmath}[style={\small}]
 \tilde{P}^{(2)}=   -\frac{13}{288} v_1 X^2+\frac{1}{6} v_1^2 X-\frac{167}{720} v_1 X-\frac{475}{12} v_2 X+\frac{25 v_1^2}{24}+\frac{25}{6} v_1 v_2+\frac{350 v_3}{9}-\frac{5 v_1^3}{24}-\frac{625 v_2}{36}-\frac{625 v_1}{288}+\frac{X^3}{240}+\frac{41 X^2}{3600}-\frac{5759 X}{3600}+\frac{25}{144}
\end{dmath}

\begin{dmath}[style={\small}]
 \tilde{P}^{(3)}=   -\frac{19 v_1 X^5}{1440}+\frac{1307 v_1^2 X^4}{17280}+\frac{19 v_2 X^4}{1440}-\frac{1603 v_1 X^4}{43200}+\frac{22417 v_1^2 X^3}{86400}+\frac{61}{675} v_2 X^3-\frac{47}{720} v_3 X^3-\frac{1271 v_1 v_2 X^3}{8640}-\frac{2539 v_1^3 X^3}{10368}-\frac{4567 v_1 X^3}{43200}+\frac{35}{72} v_1^4 X^2+\frac{54497 v_1^2 X^2}{86400}-\frac{270997}{864} v_2^2 X^2+\frac{31}{108} v_1^2 v_2 X^2+\frac{611}{864} v_1 v_3 X^2-\frac{69457 v_2 X^2}{2700}-\frac{1927 v_3 X^2}{10800}-\frac{3977 v_1 X^2}{14400}-\frac{15797 v_1^3 X^2}{17280}-\frac{92027 v_1 v_2 X^2}{43200}-\frac{55}{96} v_1^5 X+\frac{799}{480} v_1^4 X+\frac{116687 v_1^2 X}{86400}+\frac{11381}{216} v_1 v_2^2 X+\frac{185}{144} v_1^3 v_2 X+\frac{19063 v_1^2 v_2 X}{2160}-\frac{47}{18} v_1^2 v_3 X+\frac{7849 v_1 v_3 X}{2160}+\frac{22325}{36} v_2 v_3 X+\frac{270673 v_3 X}{10800}-\frac{590273 v_2^2 X}{2160}-\frac{32359 v_2 X}{3600}-\frac{23059 v_1 X}{14400}-\frac{32063 v_1^3 X}{17280}-\frac{1292477 v_1 v_2 X}{43200}+\frac{5 v_1^6}{16}+\frac{625 v_1^4}{288}+\frac{1465}{72} v_1^2 v_2^2-\frac{5275}{216} v_1 v_2^2-\frac{15}{4} v_1^4 v_2+\frac{1625}{144} v_1^3 v_2-\frac{3125}{432} v_1^2 v_2+\frac{625 v_2}{864}+\frac{235}{72} v_1^3 v_3-\frac{1175}{72} v_1^2 v_3+\frac{29375}{864} v_1 v_3+\frac{29375}{108} v_2 v_3-\frac{1175}{18} v_1 v_2 v_3-\frac{8225 v_3^2}{27}-\frac{5225 v_2^3}{81}-\frac{125 v_1^5}{96}-\frac{14375 v_2^2}{288}-\frac{1175 v_3}{432}-\frac{15925 v_1 v_2}{1728}-\frac{15625 v_1^3}{10368}+\frac{X^6}{1008}+\frac{53 X^5}{25200}+\frac{377 X^4}{50400}+\frac{52187 X^3}{3543750}-\frac{56414003 X^2}{113400000}+\frac{27683 X}{226800}+\frac{125}{36288}
\end{dmath}

\begin{dmath}[style={\small}]
 \tilde{P}^{(4)}=  \frac{X^9}{1440}-\frac{6221 v_1 X^8}{604800}+\frac{1181 X^8}{1512000}+\frac{17063 v_1^2 X^7}{241920}+\frac{6221 v_2 X^7}{604800}-\frac{16949 v_1 X^7}{1008000}+\frac{42311 X^7}{15120000}+\frac{558667 v_1^2 X^6}{3628800}+\frac{18221 v_2 X^6}{432000}-\frac{16823 v_1 v_2 X^6}{120960}-\frac{16379 v_3 X^6}{604800}-\frac{187051 v_1^3 X^6}{622080}-\frac{2858929 v_1 X^6}{60480000}+\frac{32341651 X^6}{6804000000}+\frac{2235257 v_1^4 X^5}{2488320}+\frac{128269 v_1^2 X^5}{345600}+\frac{2369 v_2^2 X^5}{34560}+\frac{157183 v_1^2 v_2 X^5}{207360}+\frac{80293 v_2 X^5}{567000}+\frac{691 v_1 v_3 X^5}{1920}-\frac{31481 v_1^3 X^5}{38880}-\frac{86633 v_3 X^5}{1512000}-\frac{2488931 v_1 v_2 X^5}{3628800}-\frac{60903371 v_1 X^5}{680400000}+\frac{13084603 X^5}{1215000000}+\frac{8547583 v_1^4 X^4}{3110400}+\frac{19814171 v_1^2 X^4}{28350000}+\frac{568207 v_2^2 X^4}{518400}+\frac{459251 v_1^2 v_2 X^4}{103680}+\frac{8952953 v_2 X^4}{24300000}-\frac{1055}{512} v_1^2 v_3 X^4+\frac{87119 v_1 v_3 X^4}{86400}-\frac{551}{540} v_2 v_3 X^4-\frac{182341 v_1^5 X^4}{92160}-\frac{296461 v_1 v_2 X^4}{141750}-\frac{180817 v_1 v_2^2 X^4}{207360}-\frac{1285643 v_1^3 v_2 X^4}{622080}-\frac{13698191 v_1^3 X^4}{7776000}-\frac{10471823 v_3 X^4}{45360000}-\frac{1130133587 v_1 X^4}{6804000000}+\frac{2975053771 X^4}{170100000000}+\frac{66773 v_1^6 X^3}{20736}+\frac{11252893 v_1^4 X^3}{2073600}+\frac{2494373411 v_1^2 X^3}{2268000000}+\frac{2627 v_3^2 X^3}{2160}+\frac{25639 v_1^4 v_2 X^3}{10368}+\frac{10807753 v_1^2 v_2 X^3}{864000} +\frac{103577 v_1^3 v_3 X^3}{15552}+\frac{676451 v_1 v_3 X^3}{216000}+\frac{36409 v_1 v_2 v_3 X^3}{3240}-\frac{4207 v_1^2 v_2^2 X^3}{3456}-\frac{18479 v_2 v_3 X^3}{4050}-\frac{51059479 v_2^3 X^3}{12960}-\frac{202319 v_1^2 v_3 X^3}{28800}-\frac{2648369 v_1 v_2^2 X^3}{86400}-\frac{42062591 v_2^2 X^3}{86400}-\frac{5365157 v_1^3 v_2 X^3}{388800}-\frac{6523081 v_1^5 X^3}{1036800}-\frac{115074571 v_1^3 X^3}{38880000}-\frac{20093351 v_3 X^3}{42525000}-\frac{807655403 v_1 v_2 X^3}{113400000}-\frac{353743963 v_1 X^3}{1134000000}-\frac{21867787379 v_2 X^3}{1134000000}-\frac{20302003513 X^3}{85050000000}+\frac{33185 v_1^6 X^2}{3456}+\frac{569027287 v_1^4 X^2}{77760000}+\frac{7028005 v_1 v_2^3 X^2}{7776}+\frac{3475529731 v_1^2 X^2}{2268000000}+\frac{52255 v_1^3 v_2^2 X^2}{1728}+\frac{3186887 v_1^2 v_2^2 X^2}{25920}+\frac{107707 v_3^2 X^2}{32400}+\frac{29}{24} v_1^5 v_2 X^2+\frac{161347 v_1^4 v_2 X^2}{8640}+\frac{13676467 v_1^2 v_2 X^2}{405000}+\frac{159637 v_1^3 v_3 X^2}{6480}+\frac{15172127 v_2^2 v_3 X^2}{1296}+\frac{90142 v_1 v_3 X^2}{10125}+\frac{3184141 v_1 v_2 v_3 X^2}{32400}+\frac{15481423 v_2 v_3 X^2}{16200}+\frac{12739716247 v_3 X^2}{680400000}
\end{dmath}

\begin{dmath}
=\text{expression in previous page}-\frac{45343 v_1^2 v_2 v_3 X^2}{1296}-\frac{22799 v_1^4 v_3 X^2}{1728}-\frac{8549 v_1^7 X^2}{2304}-\frac{34151 v_1 v_3^2 X^2}{2592}-\frac{12484331 v_2^2 X^2}{32400}-\frac{998014159 v_2^3 X^2}{194400}-\frac{270035419 v_1 v_2^2 X^2}{518400}-\frac{18085807 v_1^3 v_2 X^2}{518400}-\frac{9206869 v_1^2 v_3 X^2}{518400}-\frac{37195201 v_1^5 X^2}{3456000}-\frac{835967159 v_1 v_2 X^2}{15120000}-\frac{1428996491 v_1^3 X^2}{388800000}-\frac{864671453 v_1 X^2}{648000000}-\frac{5197671079 v_2 X^2}{1134000000}+\frac{223340899 X^2}{2268000000}+\frac{1045}{384} v_1^8 X+\frac{145793 v_1^6 X}{11520}+\frac{345668753 v_1^4 X}{62208000}-\frac{1699379}{972} v_2^4 X+\frac{8645}{18} v_1^2 v_2^3 X
   -\frac{50485}{864} v_1^4 v_2^2 X+\frac{1129607 v_1^3 v_2^2 X}{4320}+\frac{2627}{54} v_1^2 v_3^2 X-\frac{1247825}{108} v_2 v_3^2 X+a_{4,1} X+\frac{6373 v_1 X}{120960}-\frac{365}{48} v_1^6 v_2 X+\frac{757 v_1^5 v_2 X}{4320}+\frac{188159 v_1^4 v_2 X}{5760}+\frac{130313 v_2 X}{90720}+\frac{745}{48} v_1^5 v_3 X+\frac{1286587 v_1^3 v_3 X}{25920}-\frac{680371}{324} v_1 v_2^2 v_3 X+\frac{33107443 v_2^2 v_3 X}{3240}+\frac{4967551 v_1 v_3 X}{86400}+\frac{35}{54} v_1^3 v_2 v_3 X+\frac{1553609 v_1 v_2 v_3 X}{1350}+\frac{1333633 v_2 v_3 X}{4050}-\frac{1365983 v_1^2 v_2 v_3 X}{3240}-\frac{192719 v_1^4 v_3 X}{4320}-\frac{51601 v_1^7 X}{5760}-\frac{438709 v_1 v_3^2 X}{6480}-\frac{9475661 v_1 v_2^3 X}{19440}-\frac{15128893 v_3^2 X}{32400}-\frac{6822121 v_1^2 v_2^2 X}{51840}-\frac{397608839 v_2^3 X}{194400}-\frac{105655439 v_1 v_2^2 X}{259200}-\frac{10744159 v_1^2 v_3 X}{259200}-\frac{3289267 v_1^2 v_2 X}{345600}-\frac{27925537 v_2^2 X}{518400}-\frac{20620903 v_1 v_2 X}{1209600}-\frac{3285931 v_1^3 X}{1555200}-\frac{17482339 v_1^5 X}{1728000}-\frac{52419967 v_1^3 v_2 X}{1944000}-\frac{810119 v_1^2 X}{2419200}-\frac{12103499 v_3 X}{2721600}+\frac{125 v_1^8}{32}+\frac{15625 v_1^6}{2592}+\frac{7315}{162} v_1 v_2^4+\frac{20755}{648} v_1^3 v_2^3+\frac{147875}{648} v_1^2 v_2^3+\frac{919450 v_3^3}{243}-\frac{7105}{288} v_1^5 v_2^2+\frac{625}{48} v_1^4 v_2^2+\frac{68125}{576} v_1^3 v_2^2+\frac{78125 v_2^2}{20736}-\frac{13135}{216} v_1^3 v_3^2+\frac{65675}{216} v_1^2 v_3^2-\frac{1641875}{324} v_2 v_3^2+\frac{65675}{54} v_1 v_2 v_3^2+\frac{65675 v_3^2}{1296}
\end{dmath}

\begin{dmath}
    =\text{expression in previous page}+\frac{695}{96} v_1^7 v_2-\frac{125}{6} v_1^6 v_2+\frac{3125}{216} v_1^5 v_2+\frac{39625 v_1^4 v_2}{2592}+\frac{15625 v_1^2 v_2}{20736}+\frac{3125 v_2}{108864}-\frac{2435}{288} v_1^6 v_3+\frac{625}{18} v_1^5 v_3-\frac{10625}{192} v_1^4 v_3+\frac{248875 v_1^3 v_3}{7776}+\frac{428450}{243} v_2^3 v_3+\frac{420625 v_1^2 v_3}{41472}-\frac{56315}{108} v_1^2 v_2^2 v_3+\frac{122525}{324} v_1 v_2^2 v_3+\frac{2549375 v_2^2 v_3}{1296}+\frac{3565}{36} v_1^4 v_2 v_3-\frac{29875}{108} v_1^3 v_2 v_3+\frac{115625 v_1^2 v_2 v_3}{1296}-\frac{6875}{216} v_2 v_3+\frac{175175}{432} v_1 v_2 v_3-\frac{726275 v_2^4}{972}-\frac{1105 v_1^9}{1152}-\frac{15625 v_1^7}{2304}-\frac{1178125 v_1 v_2^3}{2592}-\frac{371875 v_1^2 v_2^2}{2592}-\frac{1641875 v_1 v_3^2}{2592}-\frac{1617725 v_2^3}{7776}-\frac{15625 v_1 v_3}{10368}-\frac{125 v_1 v_2}{18144}-\frac{1853125 v_1 v_2^2}{41472}-\frac{2824375 v_1^3 v_2}{124416}-\frac{390625 v_1^5}{165888}-\frac{7375 v_3}{217728}+\frac{125}{435456}
\end{dmath}
These expressions are obtained using the following expression of holomorphic ambiguity (upto genus $4$)

\begin{equation}
	\begin{aligned}
		& a_{2,0}= \frac{25}{144}, \quad a_{2,3}= \frac{1}{240}, \quad a_{2,2}= \frac{41}{3600}, \quad a_{2,1}= -\frac{5759}{3600},
  \\& a_{3,0}= \frac{125}{36288}, \quad a_{3,6}= \frac{1}{1008}, \quad a_{3,5}= \frac{53}{25200}, \quad a_{3,4}= \frac{377}{50400},
  \\& a_{3,3}= \frac{52187}{3543750}, \quad a_{3,2}= -\frac{56414003}{113400000}, \quad a_{3,1}= \frac{27683}{226800},
  \\& a_{4,0}= \frac{125}{435456}, \quad a_{4,9}= \frac{1}{1440}, \quad a_{4,8}= \frac{1181}{1512000}, \quad a_{4,7}= \frac{42311}{15120000},
  \\& a_{4,6}= \frac{32341651}{6804000000}, \quad a_{4,5}= \frac{13084603}{1215000000}, \quad a_{4,4}= \frac{2975053771}{170100000000},
  \\& a_{4,3}= -\frac{20302003513}{85050000000}, \quad a_{4,2}= \frac{223340899}{2268000000}
	\end{aligned}
\end{equation}
Near the orbifold point, we have (up to higher order in $z$ corrections)
\begin{dmath}
    G_{z\bar{z}}=-\frac{110069 i z^{6/5}}{9072000 \pi }-\frac{52200923 i z^{11/5}}{6191640000 \pi }-\frac{1181622906671 i z^{16/5}}{182847974400000 \pi }-\frac{971860467770484913 i z^{21/5}}{185462700433920000000 \pi }-\frac{517712195376322778317819 i z^{26/5}}{117464655946827571200000000 \pi }-\frac{i}{10 \pi  z^{4/5}}-\frac{13 i \sqrt[5]{z}}{600 \pi }
\end{dmath}

\begin{dgroup}
    \begin{dmath}
    A_1(z)= \frac{220354390465111604731 z^6}{1323775600931204628480}+\frac{9128302521513673 z^5}{53413257724968960}+\frac{488137713751 z^4}{2764661372928}+\frac{2433517 z^3}{13208832}+\frac{3551 z^2}{18144}+\frac{13 z}{60}-\frac{4}{5}
\end{dmath}
\begin{dmath}
    A_2(z)= \frac{455465431612371468283 z^6}{500770375352264908800}+\frac{64957905930744067 z^5}{89022096208281600}+\frac{3745355101183 z^4}{6911653432320}+\frac{1385861 z^3}{4043520}+\frac{14201 z^2}{113400}-\frac{13 z}{100}+\frac{16}{25}
\end{dmath}
\begin{dmath}
    A_3(z)= \frac{1965068367346519805362051 z^6}{380585485267721330688000}+\frac{236047909509127831 z^5}{70280602269696000}+\frac{1277237750879 z^4}{664582060800}+\frac{1442437583 z^3}{1651104000}+\frac{40073 z^2}{162000}+\frac{169 z}{1500}-\frac{64}{125}
\end{dmath}
\end{dgroup}

\begin{dgroup}
    \begin{dmath}
    B_1(z)= \frac{118181324517177899 z^6}{83530422006632939520}+\frac{2237487657991 z^5}{1318845869752320}+\frac{144615385 z^4}{68263243776}+\frac{24833 z^3}{8805888}+\frac{17 z^2}{4032}+\frac{z}{120}+\frac{1}{5}
\end{dmath}
\begin{dmath}
    B_2(z)= \frac{3804394348955513741 z^6}{417652110033164697600}+\frac{60801809239057 z^5}{6594229348761600}+\frac{3203648143 z^4}{341316218880}+\frac{85051 z^3}{8805888}+\frac{1027 z^2}{100800}+\frac{7 z}{600}+\frac{1}{25}
\end{dmath}
\begin{dmath}
    B_3(z)= \frac{1301547134020985771 z^6}{22947918133690368000}+\frac{226966909677019 z^5}{4710163820544000}+\frac{67688041507 z^4}{1706581094400}+\frac{376837 z^3}{12096000}+\frac{1633 z^2}{72000}+\frac{43 z}{3000}+\frac{1}{125}
\end{dmath}
\end{dgroup}
The mirror map $s_0=2\pi i t(z)$ near the orbifold  is given by (up to higher order in $s_0$ corrections)
\begin{dmath}
    z=s_0^5 \left(-\frac{4685818879361815620241 s_0^{30}}{20937783610175266434023424}-\frac{15426361501608623 s_0^{25}}{33607621760550469632}-\frac{341636050705 s_0^{20}}{334524026124288}-\frac{6942139 s_0^{15}}{2615348736}-\frac{1193 s_0^{10}}{399168}-\frac{13 s_0^5}{72}+1\right)
\end{dmath}

Near the large volume limit, we have (up to higher order in $1/z$ corrections)
\begin{dmath}
    G_{z\bar{z}}=\frac{325267989289300474494993 i}{14551915228366851806640625 \pi  z^9}+\frac{587545043347548828192 i}{23283064365386962890625 \pi  z^8}+\frac{1079824013793424848 i}{37252902984619140625 \pi  z^7}+\frac{2031077058780702 i}{59604644775390625 \pi  z^6}+\frac{6315703313 i}{152587890625 \pi  z^5}+\frac{12901232 i}{244140625 \pi  z^4}+\frac{28713 i}{390625 \pi  z^3}+\frac{77 i}{625 \pi  z^2}+\frac{i}{2 \pi  z}
\end{dmath}

\begin{dgroup}
    \begin{dmath}
    A_1(z)= -\frac{91136}{390625 z^2}-\frac{54528844}{244140625 z^3}-\frac{32921029136}{152587890625 z^4}-\frac{2499906384811904}{11920928955078125 z^5}-\frac{1525393024559459176}{7450580596923828125 z^6}-\frac{933819657938692482104}{4656612873077392578125 z^7}-\frac{573110486047238165452576}{2910383045673370361328125 z^8}-\frac{154}{625 z}-1
\end{dmath}
\begin{dmath}
    A_2(z)= \frac{77652}{78125 z^2}+\frac{300714108}{244140625 z^3}+\frac{222626829264}{152587890625 z^4}+\frac{20009189497115328}{11920928955078125 z^5}+\frac{2818972488865881168}{1490116119384765625 z^6}+\frac{9778647325104080827032}{4656612873077392578125 z^7}+\frac{6704866032955695360371616}{2910383045673370361328125 z^8}+\frac{462}{625 z}+1
\end{dmath}
\begin{dmath}
    A_3(z)=-\frac{1327064}{390625 z^2}-\frac{1359282148}{244140625 z^3}-\frac{1252941937376}{152587890625 z^4}-\frac{134813955503205872}{11920928955078125 z^5}-\frac{110609405838188962216}{7450580596923828125 z^6}-\frac{17517486998759780560936}{931322574615478515625 z^7}-\frac{13498002552990569588470112}{582076609134674072265625 z^8}-\frac{1078}{625 z}-1
\end{dmath}
\end{dgroup}

\begin{dgroup}
    \begin{dmath}
    B_1(z)= -\frac{8496}{390625 z^2}-\frac{3723264}{244140625 z^3}-\frac{1795273056}{152587890625 z^4}-\frac{114279890313024}{11920928955078125 z^5}-\frac{60297588712958976}{7450580596923828125 z^6}-\frac{32617546341983714304}{4656612873077392578125 z^7}-\frac{17973379161831778554816}{2910383045673370361328125 z^8}-\frac{24}{625 z}
\end{dmath}
\begin{dmath}
    B_2(z)= \frac{17568}{390625 z^2}+\frac{463104}{9765625 z^3}+\frac{7431990912}{152587890625 z^4}+\frac{118015860527424}{2384185791015625 z^5}+\frac{372816963835435008}{7450580596923828125 z^6}+\frac{234830021433200630784}{4656612873077392578125 z^7}+\frac{147631116217732855017984}{2910383045673370361328125 z^8}+\frac{24}{625 z}
\end{dmath}
\begin{dmath}
    B_3(z)= -\frac{35712}{390625 z^2}-\frac{35358336}{244140625 z^3}-\frac{30244442112}{152587890625 z^4}-\frac{119942000161344}{476837158203125 z^5}-\frac{2271029905283733504}{7450580596923828125 z^6}-\frac{1667282909354805344256}{4656612873077392578125 z^7}-\frac{47875293873762326102016}{116415321826934814453125 z^8}-\frac{24}{625 z}
\end{dmath}
\end{dgroup}
The mirror map $q=e^{2\pi i t(z)}$ near the large volume limit  is given by (up to higher order in $q$ corrections)
\begin{dmath}
    z=\frac{3211453397717989716 q^6}{25}+\frac{487153794602541 q^5}{5}+\frac{251719793608904 q^4}{3125}+75834339 q^3+\frac{438412 q^2}{5}+\frac{3371 q}{25}+\frac{1}{3125 q}+\frac{154}{625}
\end{dmath}

Near the conifold point (up to higher order in $z-1$ corrections)
\begin{dmath}
    G_{z\bar{z}}=\frac{764048190197 i z^9}{7382812500000 \pi }-\frac{213525194443 i z^8}{205078125000 \pi }+\frac{968228100047 i z^7}{205078125000 \pi }-\frac{558790404511 i z^6}{43945312500 \pi }+\frac{1321459994197 i z^5}{58593750000 \pi }-\frac{67351692503 i z^4}{2441406250 \pi }+\frac{2081315276147 i z^3}{87890625000 \pi }-\frac{1461487713461 i z^2}{102539062500 \pi }+\frac{4914441099397 i z}{820312500000 \pi }-\frac{3603437486393 i}{1845703125000 \pi }
\end{dmath}

\begin{dgroup}
    \begin{dmath}
    A_1(z)=-\frac{1384560214737911 (z-1)^{10}}{1153564453125000}-\frac{862803763771 (z-1)^9}{439453125000}-\frac{69901289 (z-1)^8}{307617187500}+\frac{786241 (z-1)^7}{30761718750}+\frac{48046 (z-1)^6}{146484375}
	\qquad 
	-\frac{9614 (z-1)^5}{9765625}+\frac{109 (z-1)^4}{46875}	
	\qquad 
	-\frac{52 (z-1)^3}{9375}+\frac{2}{125} (z-1)^2-\frac{2 (z-1)}{25}-\frac{3}{5}
\end{dmath}
\begin{dmath}
    A_2(z)=-\frac{27584282576503483 (z-1)^{10}}{3204345703125000}-\frac{6302295161034667 (z-1)^9}{230712890625000}-\frac{90602200782679 (z-1)^8}{5126953125000}-\frac{273688951 (z-1)^7}{153808593750}+\frac{2953091 (z-1)^6}{1464843750}-\frac{340222 (z-1)^5}{146484375}+\frac{5336 (z-1)^4}{1953125}-\frac{152 (z-1)^3}{46875}+\frac{8 (z-1)^2}{3125}+\frac{6 (z-1)}{125}+\frac{7}{25}
\end{dmath}
\begin{dmath}
    A_3(z)=-\frac{3556357603096849823 (z-1)^{10}}{48065185546875000}-\frac{3025026699994756781 (z-1)^9}{9613037109375000}-\frac{517209274128583 (z-1)^8}{1373291015625}-\frac{543657766868221 (z-1)^7}{3845214843750}-\frac{1552212 (z-1)^6}{1220703125}+\frac{209153 (z-1)^5}{146484375}-\frac{48011 (z-1)^4}{29296875}+\frac{12632 (z-1)^3}{5859375}-\frac{86 (z-1)^2}{15625}+\frac{6 (z-1)}{3125}-\frac{3}{25}
\end{dmath}
\end{dgroup}

\begin{dgroup}
    \begin{dmath}
    B_1(z)= -\frac{50316142216889 (z-1)^{10}}{1153564453125000}-\frac{34402381991 (z-1)^9}{11535644531250}+\frac{2073387011 (z-1)^8}{615234375000}-\frac{59382199 (z-1)^7}{15380859375}+\frac{219619 (z-1)^6}{48828125}-\frac{156709 (z-1)^5}{29296875}+\frac{611 (z-1)^4}{93750}-\frac{76 (z-1)^3}{9375}+\frac{6}{625} (z-1)^2
\end{dmath}
\begin{dmath}
    B_2(z)= -\frac{526374297156437 (z-1)^{10}}{1201629638671875}-\frac{53437362851627 (z-1)^9}{115356445312500}+\frac{860715463 (z-1)^8}{2563476562500}-\frac{105071938 (z-1)^7}{384521484375}+\frac{111427 (z-1)^6}{732421875}+\frac{12617 (z-1)^5}{146484375}-\frac{3419 (z-1)^4}{5859375}+\frac{82 (z-1)^3}{46875}-\frac{16 (z-1)^2}{3125}+\frac{12 (z-1)}{625}
\end{dmath}
\begin{dmath}
    B_3(z)= -\frac{631476601118473739 (z-1)^{10}}{144195556640625000}-\frac{41093539953993277 (z-1)^9}{4806518554687500}-\frac{160213015272943 (z-1)^8}{38452148437500}+\frac{1738065043 (z-1)^7}{1922607421875}-\frac{7054558 (z-1)^6}{6103515625}+\frac{372779 (z-1)^5}{244140625}-\frac{61763 (z-1)^4}{29296875}+\frac{18154 (z-1)^3}{5859375}-\frac{78 (z-1)^2}{15625}+\frac{28 (z-1)}{3125}+\frac{12}{625}
\end{dmath}
\end{dgroup}
The mirror map $s_1(z)=2\pi i t$ near the conifold point  is given by (up to higher order in $s_1$ corrections)
\begin{dmath}
    z=1+\left(\frac{720943 s_1^9}{10937500000000}-\frac{99151 s_1^8}{708750000000}+\frac{90291 s_1^7}{43750000000}+\frac{10187 s_1^6}{3281250000}+\frac{13 s_1^5}{150000}+\frac{169 s_1^4}{375000}+\frac{s_1^3}{200}+\frac{s_1^2}{30}+\frac{3 s_1}{10}+1\right) s_1
\end{dmath}
The GV invariants $n^g_d$ are given by
 \begin{equation}
     \begin{aligned}
        & n^0_1= 2875,n^0_2= 609250, n^0_3= 317206375,n^0_4= 242467530000,
        \\& n^1_1= 0,n^1_2= 0,n^1_3= 609250,n^1_4= 3721431625,
        \\& n^2_1= 0,n^2_2= 0,n^2_3= 0,n^2_4= 534750,
        \\& n^3_1= 0,n^3_2= 0,n^3_3= 0,n^3_4= 8625,
        \\& n^4_1= 0,n^4_2= 0,n^4_3= 0,n^4_4= 0
     \end{aligned}
 \end{equation}

\newpage

\providecommand{\href}[2]{#2}\begingroup\raggedright\endgroup


\begin{thebibliography}{100}

\bibitem{PhysRevD.7.2333}
J.D.~Bekenstein, \emph{Black holes and entropy},
  \href{https://doi.org/10.1103/PhysRevD.7.2333}{\emph{Phys. Rev. D} {\bfseries
  7} (1973) 2333}.

\bibitem{Hawking:1975vcx}
S.W.~Hawking, \emph{{Particle Creation by Black Holes}},
  \href{https://doi.org/10.1007/BF02345020}{\emph{Commun. Math. Phys.}
  {\bfseries 43} (1975) 199}.

\bibitem{Strominger:1996sh}
A.~Strominger and C.~Vafa, \emph{{Microscopic origin of the Bekenstein-Hawking
  entropy}}, \href{https://doi.org/10.1016/0370-2693(96)00345-0}{\emph{Phys.
  Lett. B} {\bfseries 379} (1996) 99}
  [\href{https://arxiv.org/abs/hep-th/9601029}{{\ttfamily hep-th/9601029}}].

\bibitem{Sen:1995in}
A.~Sen, \emph{{Extremal black holes and elementary string states}},
  \href{https://doi.org/10.1142/S0217732395002234}{\emph{Mod. Phys. Lett. A}
  {\bfseries 10} (1995) 2081}
  [\href{https://arxiv.org/abs/hep-th/9504147}{{\ttfamily hep-th/9504147}}].

\bibitem{Dabholkar:2004yr}
A.~Dabholkar, \emph{{Exact counting of black hole microstates}},
  \href{https://doi.org/10.1103/PhysRevLett.94.241301}{\emph{Phys. Rev. Lett.}
  {\bfseries 94} (2005) 241301}
  [\href{https://arxiv.org/abs/hep-th/0409148}{{\ttfamily hep-th/0409148}}].

\bibitem{Dabholkar:2004dq}
A.~Dabholkar, R.~Kallosh and A.~Maloney, \emph{{A Stringy cloak for a classical
  singularity}},
  \href{https://doi.org/10.1088/1126-6708/2004/12/059}{\emph{JHEP} {\bfseries
  12} (2004) 059} [\href{https://arxiv.org/abs/hep-th/0410076}{{\ttfamily
  hep-th/0410076}}].

\bibitem{Chen:2021dsw}
Y.~Chen, J.~Maldacena and E.~Witten, \emph{{On the black hole/string
  transition}}, \href{https://doi.org/10.1007/JHEP01(2023)103}{\emph{JHEP}
  {\bfseries 01} (2023) 103}
  [\href{https://arxiv.org/abs/2109.08563}{{\ttfamily 2109.08563}}].

\bibitem{Halder:2023nlp}
I.~Halder and D.L.~Jafferis, \emph{{Double winding condensate CFT}},
  \href{https://arxiv.org/abs/2308.11702}{{\ttfamily 2308.11702}}.

\bibitem{Kinney:2005ej}
J.~Kinney, J.M.~Maldacena, S.~Minwalla and S.~Raju, \emph{{An Index for 4
  dimensional super conformal theories}},
  \href{https://doi.org/10.1007/s00220-007-0258-7}{\emph{Commun. Math. Phys.}
  {\bfseries 275} (2007) 209}
  [\href{https://arxiv.org/abs/hep-th/0510251}{{\ttfamily hep-th/0510251}}].

\bibitem{Bhattacharya:2008zy}
J.~Bhattacharya, S.~Bhattacharyya, S.~Minwalla and S.~Raju, \emph{{Indices for
  Superconformal Field Theories in 3,5 and 6 Dimensions}},
  \href{https://doi.org/10.1088/1126-6708/2008/02/064}{\emph{JHEP} {\bfseries
  02} (2008) 064} [\href{https://arxiv.org/abs/0801.1435}{{\ttfamily
  0801.1435}}].

\bibitem{Benini:2015eyy}
F.~Benini, K.~Hristov and A.~Zaffaroni, \emph{{Black hole microstates in
  AdS$_{4}$ from supersymmetric localization}},
  \href{https://doi.org/10.1007/JHEP05(2016)054}{\emph{JHEP} {\bfseries 05}
  (2016) 054} [\href{https://arxiv.org/abs/1511.04085}{{\ttfamily
  1511.04085}}].

\bibitem{Cabo-Bizet:2018ehj}
A.~Cabo-Bizet, D.~Cassani, D.~Martelli and S.~Murthy, \emph{{Microscopic origin
  of the Bekenstein-Hawking entropy of supersymmetric AdS$_{5}$ black holes}},
  \href{https://doi.org/10.1007/JHEP10(2019)062}{\emph{JHEP} {\bfseries 10}
  (2019) 062} [\href{https://arxiv.org/abs/1810.11442}{{\ttfamily
  1810.11442}}].

\bibitem{Choi:2018hmj}
S.~Choi, J.~Kim, S.~Kim and J.~Nahmgoong, \emph{{Large AdS black holes from
  QFT}},  \href{https://arxiv.org/abs/1810.12067}{{\ttfamily 1810.12067}}.

\bibitem{Chen:2023mbc}
Y.~Chen and G.J.~Turiaci, \emph{{Spin-Statistics for Black Hole Microstates}},
  \href{https://arxiv.org/abs/2309.03478}{{\ttfamily 2309.03478}}.

\bibitem{Chen:2023lzq}
Y.~Chen, M.~Heydeman, Y.~Wang and M.~Zhang, \emph{{Probing Supersymmetric Black
  Holes with Surface Defects}},
  \href{https://arxiv.org/abs/2306.05463}{{\ttfamily 2306.05463}}.

\bibitem{Halder:2022ykw}
I.~Halder, D.L.~Jafferis and D.K.~Kolchmeyer, \emph{{A duality in string theory
  on AdS$_{3}$}}, \href{https://doi.org/10.1007/JHEP07(2023)049}{\emph{JHEP}
  {\bfseries 07} (2023) 049}
  [\href{https://arxiv.org/abs/2208.00016}{{\ttfamily 2208.00016}}].

\bibitem{Halder:2023adw}
I.~Halder and D.L.~Jafferis, \emph{{Thermal Bekenstein-Hawking entropy from the
  worldsheet}},  \href{https://arxiv.org/abs/2310.02313}{{\ttfamily
  2310.02313}}.

\bibitem{Denef:2000nb}
F.~Denef, \emph{{Supergravity flows and D-brane stability}},
  \href{https://doi.org/10.1088/1126-6708/2000/08/050}{\emph{JHEP} {\bfseries
  08} (2000) 050} [\href{https://arxiv.org/abs/hep-th/0005049}{{\ttfamily
  hep-th/0005049}}].

\bibitem{Bates:2003vx}
B.~Bates and F.~Denef, \emph{{Exact solutions for supersymmetric stationary
  black hole composites}},
  \href{https://doi.org/10.1007/JHEP11(2011)127}{\emph{JHEP} {\bfseries 11}
  (2011) 127} [\href{https://arxiv.org/abs/hep-th/0304094}{{\ttfamily
  hep-th/0304094}}].

\bibitem{Denef:2007vg}
F.~Denef and G.W.~Moore, \emph{{Split states, entropy enigmas, holes and
  halos}}, \href{https://doi.org/10.1007/JHEP11(2011)129}{\emph{JHEP}
  {\bfseries 11} (2011) 129}
  [\href{https://arxiv.org/abs/hep-th/0702146}{{\ttfamily hep-th/0702146}}].

\bibitem{Gopakumar:1998ii}
R.~Gopakumar and C.~Vafa, \emph{{M theory and topological strings. 1.}},
  \href{https://arxiv.org/abs/hep-th/9809187}{{\ttfamily hep-th/9809187}}.

\bibitem{Gopakumar:1998jq}
R.~Gopakumar and C.~Vafa, \emph{{M theory and topological strings. 2.}},
  \href{https://arxiv.org/abs/hep-th/9812127}{{\ttfamily hep-th/9812127}}.

\bibitem{Katz:1999xq}
S.H.~Katz, A.~Klemm and C.~Vafa, \emph{{M theory, topological strings and
  spinning black holes}},
  \href{https://doi.org/10.4310/ATMP.1999.v3.n5.a6}{\emph{Adv. Theor. Math.
  Phys.} {\bfseries 3} (1999) 1445}
  [\href{https://arxiv.org/abs/hep-th/9910181}{{\ttfamily hep-th/9910181}}].

\bibitem{Dedushenko:2014nya}
M.~Dedushenko and E.~Witten, \emph{{Some Details On The Gopakumar-Vafa and
  Ooguri-Vafa Formulas}},
  \href{https://doi.org/10.4310/ATMP.2016.v20.n1.a1}{\emph{Adv. Theor. Math.
  Phys.} {\bfseries 20} (2016) 1}
  [\href{https://arxiv.org/abs/1411.7108}{{\ttfamily 1411.7108}}].

\bibitem{Gaiotto:2005gf}
D.~Gaiotto, A.~Strominger and X.~Yin, \emph{{New connections between 4-D and
  5-D black holes}},
  \href{https://doi.org/10.1088/1126-6708/2006/02/024}{\emph{JHEP} {\bfseries
  02} (2006) 024} [\href{https://arxiv.org/abs/hep-th/0503217}{{\ttfamily
  hep-th/0503217}}].

\bibitem{Gaiotto:2005xt}
D.~Gaiotto, A.~Strominger and X.~Yin, \emph{{5D black rings and 4D black
  holes}}, \href{https://doi.org/10.1088/1126-6708/2006/02/023}{\emph{JHEP}
  {\bfseries 02} (2006) 023}
  [\href{https://arxiv.org/abs/hep-th/0504126}{{\ttfamily hep-th/0504126}}].

\bibitem{Dijkgraaf:1996it}
R.~Dijkgraaf, E.P.~Verlinde and H.L.~Verlinde, \emph{{Counting dyons in N=4
  string theory}},
  \href{https://doi.org/10.1016/S0550-3213(96)00640-2}{\emph{Nucl. Phys. B}
  {\bfseries 484} (1997) 543}
  [\href{https://arxiv.org/abs/hep-th/9607026}{{\ttfamily hep-th/9607026}}].

\bibitem{Shih:2005uc}
D.~Shih, A.~Strominger and X.~Yin, \emph{{Recounting Dyons in N=4 string
  theory}}, \href{https://doi.org/10.1088/1126-6708/2006/10/087}{\emph{JHEP}
  {\bfseries 10} (2006) 087}
  [\href{https://arxiv.org/abs/hep-th/0505094}{{\ttfamily hep-th/0505094}}].

\bibitem{Dabholkar:2009dq}
A.~Dabholkar, M.~Guica, S.~Murthy and S.~Nampuri, \emph{{No entropy enigmas for
  N=4 dyons}}, \href{https://doi.org/10.1007/JHEP06(2010)007}{\emph{JHEP}
  {\bfseries 06} (2010) 007} [\href{https://arxiv.org/abs/0903.2481}{{\ttfamily
  0903.2481}}].

\bibitem{Chowdhury:2019mnb}
A.~Chowdhury, A.~Kidambi, S.~Murthy, V.~Reys and T.~Wrase, \emph{{Dyonic black
  hole degeneracies in $\mathcal{N} = 4$ string theory from Dabholkar-Harvey
  degeneracies}}, \href{https://doi.org/10.1007/JHEP10(2020)184}{\emph{JHEP}
  {\bfseries 10} (2020) 184}
  [\href{https://arxiv.org/abs/1912.06562}{{\ttfamily 1912.06562}}].

\bibitem{Wald:1993nt}
R.M.~Wald, \emph{{Black hole entropy is the Noether charge}},
  \href{https://doi.org/10.1103/PhysRevD.48.R3427}{\emph{Phys. Rev. D}
  {\bfseries 48} (1993) R3427}
  [\href{https://arxiv.org/abs/gr-qc/9307038}{{\ttfamily gr-qc/9307038}}].

\bibitem{LopesCardoso:1998tkj}
G.~Lopes~Cardoso, B.~de~Wit and T.~Mohaupt, \emph{{Corrections to macroscopic
  supersymmetric black hole entropy}},
  \href{https://doi.org/10.1016/S0370-2693(99)00227-0}{\emph{Phys. Lett. B}
  {\bfseries 451} (1999) 309}
  [\href{https://arxiv.org/abs/hep-th/9812082}{{\ttfamily hep-th/9812082}}].

\bibitem{LopesCardoso:1999cv}
G.~Lopes~Cardoso, B.~de~Wit and T.~Mohaupt, \emph{{Deviations from the area law
  for supersymmetric black holes}}, {\emph{Fortsch. Phys.} {\bfseries 48}
  (2000) 49} [\href{https://arxiv.org/abs/hep-th/9904005}{{\ttfamily
  hep-th/9904005}}].

\bibitem{LopesCardoso:1999xn}
G.~Lopes~Cardoso, B.~de~Wit and T.~Mohaupt, \emph{{Area law corrections from
  state counting and supergravity}},
  \href{https://doi.org/10.1088/0264-9381/17/5/310}{\emph{Class. Quant. Grav.}
  {\bfseries 17} (2000) 1007}
  [\href{https://arxiv.org/abs/hep-th/9910179}{{\ttfamily hep-th/9910179}}].

\bibitem{Green:1997di}
M.B.~Green and P.~Vanhove, \emph{{D instantons, strings and M theory}},
  \href{https://doi.org/10.1016/S0370-2693(97)00785-5}{\emph{Phys. Lett. B}
  {\bfseries 408} (1997) 122}
  [\href{https://arxiv.org/abs/hep-th/9704145}{{\ttfamily hep-th/9704145}}].

\bibitem{Kiritsis:1997em}
E.~Kiritsis and B.~Pioline, \emph{{On R**4 threshold corrections in IIb string
  theory and (p, q) string instantons}},
  \href{https://doi.org/10.1016/S0550-3213(97)00645-7}{\emph{Nucl. Phys. B}
  {\bfseries 508} (1997) 509}
  [\href{https://arxiv.org/abs/hep-th/9707018}{{\ttfamily hep-th/9707018}}].

\bibitem{Berkovits:1998ex}
N.~Berkovits and C.~Vafa, \emph{{Type IIB R**4 H**(4g-4) conjectures}},
  \href{https://doi.org/10.1016/S0550-3213(98)00475-1}{\emph{Nucl. Phys. B}
  {\bfseries 533} (1998) 181}
  [\href{https://arxiv.org/abs/hep-th/9803145}{{\ttfamily hep-th/9803145}}].

\bibitem{Alexandrov:2013yva}
S.~Alexandrov, J.~Manschot, D.~Persson and B.~Pioline, \emph{{Quantum
  hypermultiplet moduli spaces in N=2 string vacua: a review}}, {\emph{Proc.
  Symp. Pure Math.} {\bfseries 90} (2015) 181}
  [\href{https://arxiv.org/abs/1304.0766}{{\ttfamily 1304.0766}}].

\bibitem{Ooguri:2004zv}
H.~Ooguri, A.~Strominger and C.~Vafa, \emph{{Black hole attractors and the
  topological string}},
  \href{https://doi.org/10.1103/PhysRevD.70.106007}{\emph{Phys. Rev. D}
  {\bfseries 70} (2004) 106007}
  [\href{https://arxiv.org/abs/hep-th/0405146}{{\ttfamily hep-th/0405146}}].

\bibitem{Sen:2008vm}
A.~Sen, \emph{{Quantum Entropy Function from AdS(2)/CFT(1) Correspondence}},
  \href{https://doi.org/10.1142/S0217751X09045893}{\emph{Int. J. Mod. Phys. A}
  {\bfseries 24} (2009) 4225}
  [\href{https://arxiv.org/abs/0809.3304}{{\ttfamily 0809.3304}}].

\bibitem{Sen:2012kpz}
A.~Sen, \emph{{Logarithmic Corrections to N=2 Black Hole Entropy: An Infrared
  Window into the Microstates}},
  \href{https://doi.org/10.1007/s10714-012-1336-5}{\emph{Gen. Rel. Grav.}
  {\bfseries 44} (2012) 1207}
  [\href{https://arxiv.org/abs/1108.3842}{{\ttfamily 1108.3842}}].

\bibitem{Sen:2012cj}
A.~Sen, \emph{{Logarithmic Corrections to Rotating Extremal Black Hole Entropy
  in Four and Five Dimensions}},
  \href{https://doi.org/10.1007/s10714-012-1373-0}{\emph{Gen. Rel. Grav.}
  {\bfseries 44} (2012) 1947}
  [\href{https://arxiv.org/abs/1109.3706}{{\ttfamily 1109.3706}}].

\bibitem{Dabholkar:2010uh}
A.~Dabholkar, J.~Gomes and S.~Murthy, \emph{{Quantum black holes, localization
  and the topological string}},
  \href{https://doi.org/10.1007/JHEP06(2011)019}{\emph{JHEP} {\bfseries 06}
  (2011) 019} [\href{https://arxiv.org/abs/1012.0265}{{\ttfamily 1012.0265}}].

\bibitem{Dabholkar:2011ec}
A.~Dabholkar, J.~Gomes and S.~Murthy, \emph{{Localization \& Exact
  Holography}}, \href{https://doi.org/10.1007/JHEP04(2013)062}{\emph{JHEP}
  {\bfseries 04} (2013) 062} [\href{https://arxiv.org/abs/1111.1161}{{\ttfamily
  1111.1161}}].

\bibitem{Dabholkar:2014ema}
A.~Dabholkar, J.~Gomes and S.~Murthy, \emph{{Nonperturbative black hole entropy
  and Kloosterman sums}},
  \href{https://doi.org/10.1007/JHEP03(2015)074}{\emph{JHEP} {\bfseries 03}
  (2015) 074} [\href{https://arxiv.org/abs/1404.0033}{{\ttfamily 1404.0033}}].

\bibitem{Ferrara:1995ih}
S.~Ferrara, R.~Kallosh and A.~Strominger, \emph{{N=2 extremal black holes}},
  \href{https://doi.org/10.1103/PhysRevD.52.R5412}{\emph{Phys. Rev. D}
  {\bfseries 52} (1995) R5412}
  [\href{https://arxiv.org/abs/hep-th/9508072}{{\ttfamily hep-th/9508072}}].

\bibitem{Strominger:1996kf}
A.~Strominger, \emph{{Macroscopic entropy of N=2 extremal black holes}},
  \href{https://doi.org/10.1016/0370-2693(96)00711-3}{\emph{Phys. Lett. B}
  {\bfseries 383} (1996) 39}
  [\href{https://arxiv.org/abs/hep-th/9602111}{{\ttfamily hep-th/9602111}}].

\bibitem{Harvey:1996ir}
J.A.~Harvey and G.W.~Moore, \emph{{Five-brane instantons and R**2 couplings in
  N=4 string theory}},
  \href{https://doi.org/10.1103/PhysRevD.57.2323}{\emph{Phys. Rev. D}
  {\bfseries 57} (1998) 2323}
  [\href{https://arxiv.org/abs/hep-th/9610237}{{\ttfamily hep-th/9610237}}].

\bibitem{Dabholkar:2005dt}
A.~Dabholkar, F.~Denef, G.W.~Moore and B.~Pioline, \emph{{Precision counting of
  small black holes}},
  \href{https://doi.org/10.1088/1126-6708/2005/10/096}{\emph{JHEP} {\bfseries
  10} (2005) 096} [\href{https://arxiv.org/abs/hep-th/0507014}{{\ttfamily
  hep-th/0507014}}].

\bibitem{Vafa:1997gr}
C.~Vafa, \emph{{Black holes and Calabi-Yau threefolds}},
  \href{https://doi.org/10.4310/ATMP.1998.v2.n1.a8}{\emph{Adv. Theor. Math.
  Phys.} {\bfseries 2} (1998) 207}
  [\href{https://arxiv.org/abs/hep-th/9711067}{{\ttfamily hep-th/9711067}}].

\bibitem{Haghighat:2015ega}
B.~Haghighat, S.~Murthy, C.~Vafa and S.~Vandoren, \emph{{F-Theory, Spinning
  Black Holes and Multi-string Branches}},
  \href{https://doi.org/10.1007/JHEP01(2016)009}{\emph{JHEP} {\bfseries 01}
  (2016) 009} [\href{https://arxiv.org/abs/1509.00455}{{\ttfamily
  1509.00455}}].

\bibitem{Haghighat:2013gba}
B.~Haghighat, A.~Iqbal, C.~Koz\c{c}az, G.~Lockhart and C.~Vafa,
  \emph{{M-Strings}},
  \href{https://doi.org/10.1007/s00220-014-2139-1}{\emph{Commun. Math. Phys.}
  {\bfseries 334} (2015) 779}
  [\href{https://arxiv.org/abs/1305.6322}{{\ttfamily 1305.6322}}].

\bibitem{Haghighat:2014vxa}
B.~Haghighat, A.~Klemm, G.~Lockhart and C.~Vafa, \emph{{Strings of Minimal 6d
  SCFTs}}, \href{https://doi.org/10.1002/prop.201500014}{\emph{Fortsch. Phys.}
  {\bfseries 63} (2015) 294} [\href{https://arxiv.org/abs/1412.3152}{{\ttfamily
  1412.3152}}].

\bibitem{Bershadsky:1993ta}
M.~Bershadsky, S.~Cecotti, H.~Ooguri and C.~Vafa, \emph{{Holomorphic anomalies
  in topological field theories}},
  \href{https://doi.org/10.1016/0550-3213(93)90548-4}{\emph{Nucl. Phys. B}
  {\bfseries 405} (1993) 279}
  [\href{https://arxiv.org/abs/hep-th/9302103}{{\ttfamily hep-th/9302103}}].

\bibitem{Bershadsky:1993cx}
M.~Bershadsky, S.~Cecotti, H.~Ooguri and C.~Vafa, \emph{{Kodaira-Spencer theory
  of gravity and exact results for quantum string amplitudes}},
  \href{https://doi.org/10.1007/BF02099774}{\emph{Commun. Math. Phys.}
  {\bfseries 165} (1994) 311}
  [\href{https://arxiv.org/abs/hep-th/9309140}{{\ttfamily hep-th/9309140}}].

\bibitem{Huang:2006hq}
M.-x.~Huang, A.~Klemm and S.~Quackenbush, \emph{{Topological string theory on
  compact Calabi-Yau: Modularity and boundary conditions}},
  \href{https://doi.org/10.1007/978-3-540-68030-7_3}{\emph{Lect. Notes Phys.}
  {\bfseries 757} (2009) 45}
  [\href{https://arxiv.org/abs/hep-th/0612125}{{\ttfamily hep-th/0612125}}].

\bibitem{WebUS}
\url{https://github.com/I-Halder/Topological-Strings-on-Quintic}.

\bibitem{Alexandrov:2023zjb}
S.~Alexandrov, S.~Feyzbakhsh, A.~Klemm, B.~Pioline and T.~Schimannek,
  \emph{{Quantum geometry, stability and modularity}},
  \href{https://arxiv.org/abs/2301.08066}{{\ttfamily 2301.08066}}.

\bibitem{Web}
\url{http://www.th.physik.uni-bonn.de/Groups/Klemm/data.php}.

\bibitem{Breckenridge:1996is}
J.C.~Breckenridge, R.C.~Myers, A.W.~Peet and C.~Vafa, \emph{{D-branes and
  spinning black holes}},
  \href{https://doi.org/10.1016/S0370-2693(96)01460-8}{\emph{Phys. Lett. B}
  {\bfseries 391} (1997) 93}
  [\href{https://arxiv.org/abs/hep-th/9602065}{{\ttfamily hep-th/9602065}}].

\bibitem{Cyrier:2004hj}
M.~Cyrier, M.~Guica, D.~Mateos and A.~Strominger, \emph{{Microscopic entropy of
  the black ring}},
  \href{https://doi.org/10.1103/PhysRevLett.94.191601}{\emph{Phys. Rev. Lett.}
  {\bfseries 94} (2005) 191601}
  [\href{https://arxiv.org/abs/hep-th/0411187}{{\ttfamily hep-th/0411187}}].

\bibitem{Maldacena:1997de}
J.M.~Maldacena, A.~Strominger and E.~Witten, \emph{{Black hole entropy in M
  theory}}, \href{https://doi.org/10.1088/1126-6708/1997/12/002}{\emph{JHEP}
  {\bfseries 12} (1997) 002}
  [\href{https://arxiv.org/abs/hep-th/9711053}{{\ttfamily hep-th/9711053}}].

\bibitem{Talk}
B.~Pioline, ``Counting calabi-yau black holes with (mock) modular forms.''
  \href{https://www.lpthe.jussieu.fr/\~pioline/Seminars/sem_BPSMock_Brussels2023.pdf}{Talk
  at Joint Belgian Seminars, Bruxelles}, 3, 2023.

\bibitem{Alexandrov:2023ltz}
S.~Alexandrov, S.~Feyzbakhsh, A.~Klemm and B.~Pioline, \emph{{Quantum geometry
  and mock modularity}},  \href{https://arxiv.org/abs/2312.12629}{{\ttfamily
  2312.12629}}.

\bibitem{Horowitz:2017fyg}
G.T.~Horowitz, H.K.~Kunduri and J.~Lucietti, \emph{{Comments on Black Holes in
  Bubbling Spacetimes}},
  \href{https://doi.org/10.1007/JHEP06(2017)048}{\emph{JHEP} {\bfseries 06}
  (2017) 048} [\href{https://arxiv.org/abs/1704.04071}{{\ttfamily
  1704.04071}}].

\bibitem{Kunduri:2014iga}
H.K.~Kunduri and J.~Lucietti, \emph{{Black hole non-uniqueness via spacetime
  topology in five dimensions}},
  \href{https://doi.org/10.1007/JHEP10(2014)082}{\emph{JHEP} {\bfseries 10}
  (2014) 082} [\href{https://arxiv.org/abs/1407.8002}{{\ttfamily 1407.8002}}].

\bibitem{Kunduri:2014kja}
H.K.~Kunduri and J.~Lucietti, \emph{{Supersymmetric Black Holes with Lens-Space
  Topology}}, \href{https://doi.org/10.1103/PhysRevLett.113.211101}{\emph{Phys.
  Rev. Lett.} {\bfseries 113} (2014) 211101}
  [\href{https://arxiv.org/abs/1408.6083}{{\ttfamily 1408.6083}}].

\bibitem{Kim:2023sig}
S.~Kim, S.~Kundu, E.~Lee, J.~Lee, S.~Minwalla and C.~Patel, \emph{{'Grey
  Galaxies' as an endpoint of the Kerr-AdS superradiant instability}},
  \href{https://arxiv.org/abs/2305.08922}{{\ttfamily 2305.08922}}.

\bibitem{Kallosh:1996vy}
R.~Kallosh, A.~Rajaraman and W.K.~Wong, \emph{{Supersymmetric rotating black
  holes and attractors}},
  \href{https://doi.org/10.1103/PhysRevD.55.R3246}{\emph{Phys. Rev. D}
  {\bfseries 55} (1997) R3246}
  [\href{https://arxiv.org/abs/hep-th/9611094}{{\ttfamily hep-th/9611094}}].

\bibitem{Elvang:2004rt}
H.~Elvang, R.~Emparan, D.~Mateos and H.S.~Reall, \emph{{A Supersymmetric black
  ring}}, \href{https://doi.org/10.1103/PhysRevLett.93.211302}{\emph{Phys. Rev.
  Lett.} {\bfseries 93} (2004) 211302}
  [\href{https://arxiv.org/abs/hep-th/0407065}{{\ttfamily hep-th/0407065}}].

\bibitem{Elvang:2004ds}
H.~Elvang, R.~Emparan, D.~Mateos and H.S.~Reall, \emph{{Supersymmetric black
  rings and three-charge supertubes}},
  \href{https://doi.org/10.1103/PhysRevD.71.024033}{\emph{Phys. Rev. D}
  {\bfseries 71} (2005) 024033}
  [\href{https://arxiv.org/abs/hep-th/0408120}{{\ttfamily hep-th/0408120}}].

\bibitem{Bena:2004de}
I.~Bena and N.P.~Warner, \emph{{One ring to rule them all ... and in the
  darkness bind them?}},
  \href{https://doi.org/10.4310/ATMP.2005.v9.n5.a1}{\emph{Adv. Theor. Math.
  Phys.} {\bfseries 9} (2005) 667}
  [\href{https://arxiv.org/abs/hep-th/0408106}{{\ttfamily hep-th/0408106}}].

\bibitem{Gauntlett:2004qy}
J.P.~Gauntlett and J.B.~Gutowski, \emph{{General concentric black rings}},
  \href{https://doi.org/10.1103/PhysRevD.71.045002}{\emph{Phys. Rev. D}
  {\bfseries 71} (2005) 045002}
  [\href{https://arxiv.org/abs/hep-th/0408122}{{\ttfamily hep-th/0408122}}].

\bibitem{Huang:2007sb}
M.-x.~Huang, A.~Klemm, M.~Marino and A.~Tavanfar, \emph{{Black holes and large
  order quantum geometry}},
  \href{https://doi.org/10.1103/PhysRevD.79.066001}{\emph{Phys. Rev. D}
  {\bfseries 79} (2009) 066001}
  [\href{https://arxiv.org/abs/0704.2440}{{\ttfamily 0704.2440}}].

\bibitem{LopesCardoso:2012uug}
G.~Lopes~Cardoso, B.~de~Wit and S.~Mahapatra, \emph{{Non-holomorphic
  deformations of special geometry and their applications}},
  \href{https://doi.org/10.1007/978-3-319-00215-6_1}{\emph{Springer Proc.
  Phys.} {\bfseries 144} (2013) 1}
  [\href{https://arxiv.org/abs/1206.0577}{{\ttfamily 1206.0577}}].

\bibitem{Antoniadis:1993ze}
I.~Antoniadis, E.~Gava, K.S.~Narain and T.R.~Taylor, \emph{{Topological
  amplitudes in string theory}},
  \href{https://doi.org/10.1016/0550-3213(94)90617-3}{\emph{Nucl. Phys. B}
  {\bfseries 413} (1994) 162}
  [\href{https://arxiv.org/abs/hep-th/9307158}{{\ttfamily hep-th/9307158}}].

\bibitem{Antoniadis:1995zn}
I.~Antoniadis, E.~Gava, K.S.~Narain and T.R.~Taylor, \emph{{N=2 type II
  heterotic duality and higher derivative F terms}},
  \href{https://doi.org/10.1016/0550-3213(95)00467-7}{\emph{Nucl. Phys. B}
  {\bfseries 455} (1995) 109}
  [\href{https://arxiv.org/abs/hep-th/9507115}{{\ttfamily hep-th/9507115}}].

\bibitem{Liu:2022agh}
Z.~Liu and Y.~Ruan, \emph{{Castelnuovo bound and higher genus Gromov-Witten
  invariants of quintic 3-folds}},
  \href{https://arxiv.org/abs/2210.13411}{{\ttfamily 2210.13411}}.

\bibitem{Yamaguchi:2004bt}
S.~Yamaguchi and S.-T.~Yau, \emph{{Topological string partition functions as
  polynomials}},
  \href{https://doi.org/10.1088/1126-6708/2004/07/047}{\emph{JHEP} {\bfseries
  07} (2004) 047} [\href{https://arxiv.org/abs/hep-th/0406078}{{\ttfamily
  hep-th/0406078}}].

\bibitem{Aganagic:2006wq}
M.~Aganagic, V.~Bouchard and A.~Klemm, \emph{{Topological Strings and (Almost)
  Modular Forms}},
  \href{https://doi.org/10.1007/s00220-007-0383-3}{\emph{Commun. Math. Phys.}
  {\bfseries 277} (2008) 771}
  [\href{https://arxiv.org/abs/hep-th/0607100}{{\ttfamily hep-th/0607100}}].

\bibitem{Candelas:1990rm}
P.~Candelas, X.C.~De~La~Ossa, P.S.~Green and L.~Parkes, \emph{{A Pair of
  Calabi-Yau manifolds as an exactly soluble superconformal theory}},
  \href{https://doi.org/10.1016/0550-3213(91)90292-6}{\emph{Nucl. Phys. B}
  {\bfseries 359} (1991) 21}.

\bibitem{Faber:1998gsw}
C.~Faber and R.~Pandharipande, \emph{{Hodge integrals and Gromov-Witten
  theory}}, \href{https://doi.org/10.1007/s002229900028}{\emph{Inventiones
  mathematicae} {\bfseries 139} (2000) 173}
  [\href{https://arxiv.org/abs/math/9810173}{{\ttfamily math/9810173}}].

\bibitem{Strominger:1995cz}
A.~Strominger, \emph{{Massless black holes and conifolds in string theory}},
  \href{https://doi.org/10.1016/0550-3213(95)00287-3}{\emph{Nucl. Phys. B}
  {\bfseries 451} (1995) 96}
  [\href{https://arxiv.org/abs/hep-th/9504090}{{\ttfamily hep-th/9504090}}].

\bibitem{Vafa:1995ta}
C.~Vafa, \emph{{A Stringy test of the fate of the conifold}},
  \href{https://doi.org/10.1016/0550-3213(95)00279-2}{\emph{Nucl. Phys. B}
  {\bfseries 447} (1995) 252}
  [\href{https://arxiv.org/abs/hep-th/9505023}{{\ttfamily hep-th/9505023}}].

\bibitem{Cremmer:1978km}
E.~Cremmer, B.~Julia and J.~Scherk, \emph{{Supergravity Theory in
  Eleven-Dimensions}},
  \href{https://doi.org/10.1016/0370-2693(78)90894-8}{\emph{Phys. Lett. B}
  {\bfseries 76} (1978) 409}.

\bibitem{Duff:1995wd}
M.J.~Duff, J.T.~Liu and R.~Minasian, \emph{{Eleven-dimensional origin of
  string-string duality: A One loop test}},
  \href{https://doi.org/10.1016/0550-3213(95)00368-3}{\emph{Nucl. Phys. B}
  {\bfseries 452} (1995) 261}
  [\href{https://arxiv.org/abs/hep-th/9506126}{{\ttfamily hep-th/9506126}}].

\bibitem{Vafa:1995fj}
C.~Vafa and E.~Witten, \emph{{A One loop test of string duality}},
  \href{https://doi.org/10.1016/0550-3213(95)00280-6}{\emph{Nucl. Phys. B}
  {\bfseries 447} (1995) 261}
  [\href{https://arxiv.org/abs/hep-th/9505053}{{\ttfamily hep-th/9505053}}].

\bibitem{NewPaper}
I.~Halder, C.~Vafa and K.~Xu, \emph{{To appear}}.

\bibitem{Guica:2005ig}
M.~Guica, L.~Huang, W.~Li and A.~Strominger, \emph{{R**2 corrections for 5-D
  black holes and rings}},
  \href{https://doi.org/10.1088/1126-6708/2006/10/036}{\emph{JHEP} {\bfseries
  10} (2006) 036} [\href{https://arxiv.org/abs/hep-th/0505188}{{\ttfamily
  hep-th/0505188}}].

\bibitem{Andriyash:2010qv}
E.~Andriyash, F.~Denef, D.L.~Jafferis and G.W.~Moore, \emph{{Wall-crossing from
  supersymmetric galaxies}},
  \href{https://doi.org/10.1007/JHEP01(2012)115}{\emph{JHEP} {\bfseries 01}
  (2012) 115} [\href{https://arxiv.org/abs/1008.0030}{{\ttfamily 1008.0030}}].

\bibitem{Castro:2007ci}
A.~Castro, J.L.~Davis, P.~Kraus and F.~Larsen, \emph{{Precision Entropy of
  Spinning Black Holes}},
  \href{https://doi.org/10.1088/1126-6708/2007/09/003}{\emph{JHEP} {\bfseries
  09} (2007) 003} [\href{https://arxiv.org/abs/0705.1847}{{\ttfamily
  0705.1847}}].

\bibitem{Bena:2005ae}
I.~Bena and P.~Kraus, \emph{{R**2 corrections to black ring entropy}},
  \href{https://arxiv.org/abs/hep-th/0506015}{{\ttfamily hep-th/0506015}}.

\bibitem{Thomas:1998uj}
R.P.~Thomas, \emph{{A Holomorphic Casson invariant for Calabi-Yau three folds,
  and bundles on K3 fibrations}}, {\emph{J. Diff. Geom.} {\bfseries 54} (2000)
  367} [\href{https://arxiv.org/abs/math/9806111}{{\ttfamily math/9806111}}].

\bibitem{Donaldson:1996kp}
S.K.~Donaldson and R.P.~Thomas, \emph{{Gauge theory in higher dimensions}},  in
  \emph{{Conference on Geometric Issues in Foundations of Science in honor of
  Sir Roger Penrose's 65th Birthday}}, pp.~31--47, 6, 1996.

\bibitem{Maulik:2003rzb}
D.~Maulik, N.~Nekrasov, A.~Okounkov and R.~Pandharipande,
  \emph{{Gromov\textendash{}Witten theory and Donaldson\textendash{}Thomas
  theory, I}}, \href{https://doi.org/10.1112/S0010437X06002302}{\emph{Compos.
  Math.} {\bfseries 142} (2006) 1263}
  [\href{https://arxiv.org/abs/math/0312059}{{\ttfamily math/0312059}}].

\bibitem{Maulik:2004txy}
D.~Maulik, N.~Nekrasov, A.~Okounkov and R.~Pandharipande,
  \emph{{Gromov\textendash{}Witten theory and Donaldson\textendash{}Thomas
  theory, II}}, \href{https://doi.org/10.1112/S0010437X06002314}{\emph{Compos.
  Math.} {\bfseries 142} (2006) 1286}
  [\href{https://arxiv.org/abs/math/0406092}{{\ttfamily math/0406092}}].

\bibitem{Iqbal:2003ds}
A.~Iqbal, N.~Nekrasov, A.~Okounkov and C.~Vafa, \emph{{Quantum foam and
  topological strings}},
  \href{https://doi.org/10.1088/1126-6708/2008/04/011}{\emph{JHEP} {\bfseries
  04} (2008) 011} [\href{https://arxiv.org/abs/hep-th/0312022}{{\ttfamily
  hep-th/0312022}}].

\bibitem{Dijkgraaf:2006um}
R.~Dijkgraaf, C.~Vafa and E.~Verlinde, \emph{{M-theory and a topological string
  duality}},  \href{https://arxiv.org/abs/hep-th/0602087}{{\ttfamily
  hep-th/0602087}}.

\bibitem{bayer2011bridgeland}
A.~Bayer, E.~Macr{\`\i} and Y.~Toda, \emph{Bridgeland stability conditions on
  threefolds i: Bogomolov-gieseker type inequalities}, {\emph{arXiv preprint
  arXiv:1103.5010} (2011) }.

\bibitem{bayer2016space}
A.~Bayer, E.~Macr{\`\i} and P.~Stellari, \emph{The space of stability
  conditions on abelian threefolds, and on some calabi-yau threefolds},
  {\emph{Inventiones mathematicae} {\bfseries 206} (2016) 869}.

\bibitem{Shmakova:1996nz}
M.~Shmakova, \emph{{Calabi-Yau black holes}},
  \href{https://doi.org/10.1103/PhysRevD.56.R540}{\emph{Phys. Rev. D}
  {\bfseries 56} (1997) 540}
  [\href{https://arxiv.org/abs/hep-th/9612076}{{\ttfamily hep-th/9612076}}].

\bibitem{Aganagic:2009kf}
M.~Aganagic, H.~Ooguri, C.~Vafa and M.~Yamazaki, \emph{{Wall Crossing and
  M-theory}}, {\emph{Publ. Res. Inst. Math. Sci. Kyoto} {\bfseries 47} (2011)
  569} [\href{https://arxiv.org/abs/0908.1194}{{\ttfamily 0908.1194}}].

\bibitem{Joyce:2008pc}
D.~Joyce and Y.~Song, \emph{{A Theory of generalized Donaldson-Thomas
  invariants}},  \href{https://arxiv.org/abs/0810.5645}{{\ttfamily 0810.5645}}.

\bibitem{Kontsevich:2008fj}
M.~Kontsevich and Y.~Soibelman, \emph{{Stability structures, motivic
  Donaldson-Thomas invariants and cluster transformations}},
  \href{https://arxiv.org/abs/0811.2435}{{\ttfamily 0811.2435}}.

\bibitem{Pandharipande:2007qu}
R.~Pandharipande and R.P.~Thomas, \emph{{Stable pairs and BPS invariants}},
  \href{https://doi.org/10.1090/S0894-0347-09-00646-8}{\emph{Electron. Res.
  Announ. AMS} {\bfseries 23} (2010) 267}
  [\href{https://arxiv.org/abs/0711.3899}{{\ttfamily 0711.3899}}].

\bibitem{Alexandrov:2022pgd}
S.~Alexandrov, N.~Gaddam, J.~Manschot and B.~Pioline, \emph{{Modular bootstrap
  for D4-D2-D0 indices on compact Calabi-Yau threefolds}},
  \href{https://arxiv.org/abs/2204.02207}{{\ttfamily 2204.02207}}.

\bibitem{Gaiotto:2010okc}
D.~Gaiotto, G.W.~Moore and A.~Neitzke, \emph{{Four-dimensional wall-crossing
  via three-dimensional field theory}},
  \href{https://doi.org/10.1007/s00220-010-1071-2}{\emph{Commun. Math. Phys.}
  {\bfseries 299} (2010) 163}
  [\href{https://arxiv.org/abs/0807.4723}{{\ttfamily 0807.4723}}].

\bibitem{Gaiotto:2009hg}
D.~Gaiotto, G.W.~Moore and A.~Neitzke, \emph{{Wall-crossing, Hitchin systems,
  and the WKB approximation}},
  \href{https://doi.org/10.1016/j.aim.2012.09.027}{\emph{Adv. Math.} {\bfseries
  234} (2013) 239} [\href{https://arxiv.org/abs/0907.3987}{{\ttfamily
  0907.3987}}].

\bibitem{bridgeland2007stability}
T.~Bridgeland, \emph{Stability conditions on triangulated categories},
  {\emph{Annals of Mathematics} (2007) 317}.

\bibitem{li2019stability}
C.~Li, \emph{On stability conditions for the quintic threefold},
  {\emph{Inventiones mathematicae} {\bfseries 218} (2019) 301}.

\bibitem{Sen:2023dps}
A.~Sen, \emph{{Revisiting localization for BPS black hole entropy}},
  \href{https://arxiv.org/abs/2302.13490}{{\ttfamily 2302.13490}}.

\bibitem{Iliesiu:2022onk}
L.V.~Iliesiu, S.~Murthy and G.J.~Turiaci, \emph{{Revisiting the Logarithmic
  Corrections to the Black Hole Entropy}},
  \href{https://arxiv.org/abs/2209.13608}{{\ttfamily 2209.13608}}.

\bibitem{Murthy:2015yfa}
S.~Murthy and V.~Reys, \emph{{Functional determinants, index theorems, and
  exact quantum black hole entropy}},
  \href{https://doi.org/10.1007/JHEP12(2015)028}{\emph{JHEP} {\bfseries 12}
  (2015) 028} [\href{https://arxiv.org/abs/1504.01400}{{\ttfamily
  1504.01400}}].

\bibitem{Jeon:2018kec}
I.~Jeon and S.~Murthy, \emph{{Twisting and localization in supergravity:
  equivariant cohomology of BPS black holes}},
  \href{https://doi.org/10.1007/JHEP03(2019)140}{\emph{JHEP} {\bfseries 03}
  (2019) 140} [\href{https://arxiv.org/abs/1806.04479}{{\ttfamily
  1806.04479}}].

\bibitem{Green:1997as}
M.B.~Green, M.~Gutperle and P.~Vanhove, \emph{{One loop in eleven-dimensions}},
  \href{https://doi.org/10.1016/S0370-2693(97)00931-3}{\emph{Phys. Lett. B}
  {\bfseries 409} (1997) 177}
  [\href{https://arxiv.org/abs/hep-th/9706175}{{\ttfamily hep-th/9706175}}].

\bibitem{Green:1997me}
M.B.~Green, M.~Gutperle and H.-h.~Kwon, \emph{{Sixteen fermion and related
  terms in M theory on T**2}},
  \href{https://doi.org/10.1016/S0370-2693(97)01551-7}{\emph{Phys. Lett. B}
  {\bfseries 421} (1998) 149}
  [\href{https://arxiv.org/abs/hep-th/9710151}{{\ttfamily hep-th/9710151}}].

\end{thebibliography}
\end{document}